\documentclass[twocolumn]{aastex631}

\newcommand\kms{km~s$^{-1}$}
\newcommand\flam{erg~s$^{-1}$~cm$^{-2}$~\AA$^{-1}$}

\newcommand\rlhbetaoptcont{$R_{\mathrm{\hbox{{\rm H}\kern 0.1em{\sc $\beta$}}}}$--$L_{\mathrm{5100}}$}
\newcommand\rlhbetahbeta{$R_{\mathrm{\hbox{{\rm H}\kern 0.1em{\sc $\beta$}}}}$--$L_{\mathrm{\Hbeta}}$}
\newcommand\rlhbetauv{$R_{\mathrm{\hbox{{\rm H}\kern 0.1em{\sc $\beta$}}}}$--$L_{\mathrm{UV}}$}
\newcommand\rlhbetauvcont{$R_{\mathrm{\hbox{{\rm H}\kern 0.1em{\sc $\beta$}}}}$--$L_{\mathrm{1350}}$}
\newcommand\massbhlog{\(\log(M_{\bullet}/M_{\odot})\)}

\newcommand\FeII{\hbox{{\rm Fe~}\kern 0.1em{\sc ii}}}
\newcommand\FeVIIlama{\hbox{{\rm Fe~}\kern 0.1em{\sc vii}~$\lambda 5160$}}
\newcommand\FeVIlam{\hbox{{\rm Fe~}\kern 0.1em{\sc vi}~$\lambda 5177$}}
\newcommand\FeVIIlamb{\hbox{{\rm Fe~}\kern 0.1em{\sc vii}~$\lambda 5278$}}
\newcommand\CaVlam{\hbox{{\rm Ca~}\kern 0.1em{\sc v}~$\lambda 5311$}}
\newcommand\FeVIIlamc{\hbox{{\rm Fe~}\kern 0.1em{\sc vii}~$\lambda 5722$}}
\newcommand\FeVIIlamd{\hbox{{\rm Fe~}\kern 0.1em{\sc vii}~$\lambda 6088$}}
\newcommand\HeIlam{\hbox{{\rm He~}\kern 0.1em{\sc i}~$\lambda 5877$}}
\newcommand\HeI{\hbox{{\rm He~}\kern 0.1em{\sc i}}}
\newcommand\HeIIlamopt{\hbox{{\rm He~}\kern 0.1em{\sc ii}~$\lambda 4687$}}
\newcommand\HeII{\hbox{{\rm He~}\kern 0.1em{\sc ii}}}
\newcommand\Hgammalam{\hbox{{\rm H}\kern 0.1em{\sc $\gamma$}~$\lambda 4342$}}
\newcommand\Lyalpha{\hbox{{\rm Ly}\kern 0.1em{\sc $\alpha$}}}
\newcommand\Hbetalam{\hbox{{\rm H}\kern 0.1em{\sc $\beta$}~$\lambda 4861$}}
\newcommand\Hbeta{\hbox{{\rm H}\kern 0.1em{\sc $\beta$}}}
\newcommand\OIIIlam{\hbox{[{\rm O~}\kern 0.1em{\sc iii}]~$\lambda\lambda 4960, 5008$}}
\newcommand\OIIIlamweak{\hbox{[{\rm O~}\kern 0.1em{\sc iii}]~$\lambda 4960$}}
\newcommand\OIIIlamstrong{\hbox{[{\rm O~}\kern 0.1em{\sc iii}]~$\lambda 5008$}}
\newcommand\OIII{\hbox{[{\rm O~}\kern 0.1em{\sc iii}]}}

\accepted{September 3, 2023}
\submitjournal{ApJ}

\shorttitle{Mrk~142 UV Lag}
\shortauthors{Khatu et al.}

\begin{document}

\title{Supermassive Black Holes with High Accretion Rates in Active Galactic Nuclei. XIII. Ultraviolet Time Lag of \Hbeta\ Emission in Mrk~142}

\author[0000-0002-0581-6506]{Viraja C. Khatu}
\affiliation{Department of Physics and Astronomy \& Institute of Earth and Space Exploration, The University of Western Ontario, 1151 Richmond Street, London, Ontario N6A 3K7, Canada}
\email{vkhatu@uwo.ca}

\author[0000-0001-6217-8101]{Sarah C. Gallagher}
\affiliation{Department of Physics and Astronomy \& Institute of Earth and Space Exploration, The University of Western Ontario, 1151 Richmond Street, London, Ontario N6A 3K7, Canada}

\author[0000-0003-1728-0304]{Keith Horne}
\affiliation{School of Physics and Astronomy, University of St Andrews, North Haugh, St Andrews, KY16 9SS, Scotland, UK}

\author[0000-0002-8294-9281]{Edward M. Cackett}
\affiliation{Department of Physics \& Astronomy, Wayne State University, 666 W Hancock Street, Detroit, Michigan 48201, USA}

\author{Chen Hu}
\affiliation{Key Laboratory for Particle Astrophysics, Institute of High Energy Physics, Chinese Academy of Sciences, 19B Yuquan Road, Beijing 100049, People's Republic of China}

\author{Sofia Pasquini}
\affiliation{Department of Physics and Astronomy, The University of Western Ontario, 1151 Richmond Street, London, Ontario N6A 3K7, Canada}

\author{Patrick Hall}
\affiliation{Department of Physics and Astronomy, York University, 4700 Keele Street, Toronto, Ontario M3J 1P3, Canada}

\author[0000-0001-9449-9268]{Jian-Min Wang}
\affiliation{Key Laboratory for Particle Astrophysics, Institute of High Energy Physics, Chinese Academy of Sciences, 19B Yuquan Road, Beijing 100049, People's Republic of China}

\author[0000-0002-2121-8960]{Wei-Hao Bian}
\affiliation{Physics Department, Nanjing Normal University, Nanjing 210097, People's Republic of China}

\author[0000-0001-5841-9179]{Yan-Rong Li}
\affiliation{Key Laboratory for Particle Astrophysics, Institute of High Energy Physics, Chinese Academy of Sciences, 19B Yuquan Road, Beijing 100049, People's Republic of China}

\author{Jin-Ming Bai}
\affiliation{Yunnan Observatories, The Chinese Academy of Sciences, Kunming 650011, People's Republic of China}

\author{Yong-Jie Chen}
\affiliation{Key Laboratory for Particle Astrophysics, Institute of High Energy Physics, Chinese Academy of Sciences, 19B Yuquan Road, Beijing 100049, People's Republic of China}

\author[0000-0002-5830-3544]{Pu Du}
\affiliation{Key Laboratory for Particle Astrophysics, Institute of High Energy Physics, Chinese Academy of Sciences, 19B Yuquan Road, Beijing 100049, People's Republic of China}

\author{Michael Goad}
\affiliation{School of Physics and Astronomy, University of Leicester, Leicester LE1 7RH, UK}

\author{Bo-Wei Jiang}
\affiliation{Key Laboratory for Particle Astrophysics, Institute of High Energy Physics, Chinese Academy of Sciences, 19B Yuquan Road, Beijing 100049, People's Republic of China}

\author[0000-0003-3823-3419]{Sha-Sha Li}
\affiliation{Yunnan Observatories, The Chinese Academy of Sciences, Kunming 650011, People's Republic of China}

\author{Yu-Yang Songsheng}
\affiliation{Key Laboratory for Particle Astrophysics, Institute of High Energy Physics, Chinese Academy of Sciences, 19B Yuquan Road, Beijing 100049, People's Republic of China}

\author{Chan Wang}
\affiliation{Physics Department, Nanjing Normal University, Nanjing 210097, People's Republic of China}

\author{Ming Xiao}
\affiliation{Key Laboratory for Particle Astrophysics, Institute of High Energy Physics, Chinese Academy of Sciences, 19B Yuquan Road, Beijing 100049, People's Republic of China}

\author{Zhe Yu}
\affiliation{Key Laboratory for Particle Astrophysics, Institute of High Energy Physics, Chinese Academy of Sciences, 19B Yuquan Road, Beijing 100049, People's Republic of China}

\begin{abstract}

We performed a rigorous reverberation-mapping analysis of the broad-line region (BLR) in a highly accreting ($L/L_{\mathrm{Edd}}=0.74$--3.4) active galactic nucleus, Markarian~142 (Mrk~142), for the first time using concurrent observations of the inner accretion disk and the BLR to determine a time lag for the \Hbetalam\ emission relative to the ultraviolet (UV) continuum variations.  We used continuum data taken with the {\em Niel Gehrels Swift Observatory} in the {\em UVW2} band, and the Las Cumbres Observatory, Dan Zowada Memorial Observatory, and Liverpool Telescope in the {\em g} band, as part of the broader Mrk~142 multi-wavelength monitoring campaign in 2019.  We obtained new spectroscopic observations covering the \Hbeta\ broad emission line in the optical from the Gemini North Telescope and the Lijiang 2.4-meter Telescope for a total of 102 epochs (over a period of eight months) contemporaneous to the continuum data.  Our primary result states a UV-to-\Hbeta\ time lag of \(8.68_{-0.72}^{+0.75}\) days in Mrk~142 obtained from light-curve analysis with a Python-based Running Optimal Average algorithm.  We placed our new measurements for Mrk~142 on the optical and UV radius-luminosity relations for NGC~5548 to understand the nature of the continuum driver.  The positions of Mrk~142 on the scaling relations suggest that UV is closer to the ``true’’ driving continuum than the optical.  Furthermore, we obtain \massbhlog~=~\(6.32\pm0.29\) assuming UV as the primary driving continuum.
\newline{{\em Unified Astronomy Thesaurus concepts}: \href{http://astrothesaurus.org/uat/16}{Active galactic nuclei (16)}; \href{http://astrothesaurus.org/uat/14}{Accretion (14)}; \href{http://astrothesaurus.org/uat/1558}{Spectroscopy (1558)}}
\newline{{\em Supporting material}: Machine-readable tables and spectra}

\end{abstract}

\section{Introduction} \label{sec:intro}

Accretion onto supermassive black holes through an accretion disk of ionized gas powers active galactic nuclei (AGN) at the centers of massive galaxies.  AGN accreting at typical rates (a few percent of the Eddington limit) have a geometrically thin but optically thick disk -- the `thin-disk' model \citep{shakura_sunyaev_1973}.  However, theoretical models predict a notably different structure for the AGN with high accretion rates significantly above the Eddington limit -- super-Eddington AGN \citep[e.g.,][]{abramowicz_etal_1988}.  The occurrence of such AGN is likely higher during the peak era of supermassive black hole growth during cosmic noon \citep[redshifts, $z = 1$--3;][]{brandt_alexander_2010, shen_etal_2020}.  Understanding the structure of the accretion system in high-Eddington AGN remains an open issue in accretion physics.

Although models exist for slim-disk systems, observational tests of the structure of the accretion flow in super-Eddington AGN are rare.  At high accretion rates, radiation pressure is expected to dominate, causing the inner disk to inflate vertically -- now called a `slim' (rather than thin) disk -- with a scale height, $H \leq R$, where {\em R} is the disk radius \citep[e.g.,][]{abramowicz_etal_1988}.  Photons are trapped in the fast-flowing matter, eventually falling into the black hole.  Given that not all photons escape, the disks in super-Eddington AGN are underluminous relative to the accretion rates as compared to thin disks \citep{jaroszynski_etal_1980}.  \citet{begelman2002} proposed an alternative scenario where the ``photon-bubble instability'' principle can cause the disks in super-Eddington AGN to become inhomogeneous at scales much smaller than the disk scale height.

Reverberation mapping \citep[RM;][]{blandford_mckee_1982, peterson1993} provides a way to observationally study the slim-disk model and broad-line region (BLR) in super-Eddington AGN.  RM takes advantage of the observed continuum variability of AGN on many time scales \citep[from several days to weeks and years; e.g,][]{peterson_etal_1982}.  The accretion-disk emission illuminates the BLR on larger scales, and sets the ionization structure and thus the location of the gas generating the broad emission lines (e.g., \Hbeta).  An increase in continuum emission from the accretion disk results in an increase in broad emission-line flux after a time lag set by the sum total of the light travel time between the continuum-emitting region and the BLR \citep{peterson2014}, and the recombination timescale, where the latter is much smaller than the former for typical BLR densities (and therefore ignored in the time-lag calculations).  RM converts this time lag into a spatial distance, the size of the BLR.  Thus, applying RM to high-accretion rate AGN gives an observational method to test the structure of the accretion flow and BLR in these systems, and place super-Eddington AGN on the radius-luminosity ({\em R}--{\em L}) relationship for AGN \citep{kaspi_etal_2005, bentz_etal_2013}.

The Narrow-Line Seyfert~1 (NLS1) class of AGN are considered to have high accretion rates, and typically display narrow broad emission lines (e.g., the \Hbeta\ line has a full width at half maximum, FWHM~$\lesssim 2000$~\kms) in comparison to other broad-line objects (and broader than the narrow lines seen in type~2 objects), strong \FeII\ emission lines, and weak \OIII\ lines \citep[e.g.,][]{osterbrock_pogge_1987, boroson_green_1992, boller_etal_1996, veroncetty_etal_2001} in their spectra.  The Super-Eddington Accreting Massive Black Holes (SEAMBH) campaign has been performing photometric and spectroscopic monitoring over the past nine years of high accretion-rate AGN that display spectral characteristics of NLS1s \citep[e.g.,][]{du_etal_2014, wang_etal_2014_2, hu_etal_2015, du_etal_2015, du_etal_2016_1, du_etal_2016_2, du_etal_2018, li_etal_2018, li_etal_2021}.  \citet{du_etal_2016_2} showed that the BLRs in super-Eddington AGN are smaller than those with sub-Eddington accretion rates.  In the context of the slim-disk model, the smaller BLR sizes can be explained as a consequence of the increased scale height of the inner accretion disk that shields the BLR from the central ionizing flux \citep{wang_etal_2014_1}.  \Hbeta, a marker of the hydrogen ionization front in the BLR, can thus exist at smaller radii than in thin accretion-disk systems.  \citet{fonseca_alvarez_etal_2020} offer an alternative explanation.  In their correlation analysis of the physical and spectral properties of the Sloan Digital Sky Survey (SDSS) RM AGN, \citet{fonseca_alvarez_etal_2020} found that the {\em R}--{\em L} offset (defined as the ratio of the measured \Hbeta\ time lag to the expected time lag from the best-fit {\em R}--{\em L} such as that given by \citealt{bentz_etal_2013}) is positively correlated to the \OIIIlamstrong\ to \Hbeta\ luminosity ratio, which is often used as a proxy for the number of ionizing photons \citep[e.g.,][]{baldwin_etal_1981}.  The smaller BLR sizes are therefore likely a result of the changes in the shape of the ultraviolet (UV)/optical spectral energy distribution (SED) of AGN \citep{fonseca_alvarez_etal_2020}.

As the most promising SEAMBH object -- a bright target with an extremely super-Eddington accretion rate \citep[$\dot{M}/\dot{M}_{\mathrm{Edd}}=250$;][]{li_etal_2018} and a well-measured \Hbeta\ lag -- Markarian~142 (Mrk~142 or PG~1022+519, RA~=~10$^\mathrm{h}$25$^\mathrm{m}$31.20$^\mathrm{s}$, Dec~=~+51$^{\circ}$40\arcmin34.87\arcsec, $z=0.045$) is the target of our study to probe the structure of its BLR.  In the 2012 SEAMBH campaign, Mrk~142 was highly variable with a fractional variability amplitude of \(F_{\rm{var}} = 8.1\%\) at 5100~\AA\ over a period of six months.  Its variable nature makes it amenable to RM studies of both accretion-disk structure (from X-ray/UV/optical continuum time-lag studies) and the BLR structure (from continuum-emission line time lags).  Accretion-disk RM applies the same principle as BLR RM to the inner and outer regions of the accretion disk to determine its size and temperature profile \citep{cackett_etal_2007}.  The more energetic X-ray/UV radiation from the inner disk illuminates the disk at larger radii where the optical photons are generated.  Therefore, the lower-energy emission will respond with a positive time lag to changes in the high energy radiation giving rise to correlated continuum light curves.  Mrk~142 has a total \Hbeta\ time lag ($\tau$) with respect to the 5100~\AA\ continuum emission of \(7.9_{-1.1}^{+1.2}\) days \citep{du_etal_2015} and a black hole mass of \massbhlog~=~\(6.23_{-0.45}^{+0.26}\) \citep{li_etal_2018}.

In this paper, we present Mrk~142 time-lag measurements from two ground-based, optical spectroscopic RM campaigns of Mrk~142 concurrent with the photometric monitoring of the target with the {\em Neil Gehrels Swift Observatory} ({\em Swift}) in a UV band; and the Las Cumbres Observatory (LCO), Dan Zowada Memorial Observatory (Zowada), and Liverpool Telescope \citep[Liverpool;][]{steele_etal_2004} in an optical band.  With our joint campaign, we performed, for the first time, simultaneous measurements of the inner accretion disk and BLR size in a super-Eddington AGN.  This paper is organized as follows.  In Section~\ref{sec:observations}, we provide details of the observations, and in Section~\ref{sec:reduction}, we explain the process of data reduction.  In Section~\ref{sec:spectral_analysis}, we describe our spectral modeling followed by light-curve analysis in Section~\ref{sec:light_curve_analysis}.  In Section~\ref{sec:results_discussion}, we outline and discuss our results in the context of previous studies.  Section~\ref{sec:conclusion} provides closing remarks.  Throughout this work, we use the standard cosmology with \(H_0=67\)~\kms~Mpc$^{-1}$, \(\Omega_\Lambda=0.68\), and \(\Omega_\mathrm{M}=0.32\) \citep{ade_etal_2014}.

\section{Observations} \label{sec:observations}

\begin{figure}[ht!]
\epsscale{1.2}
\plotone{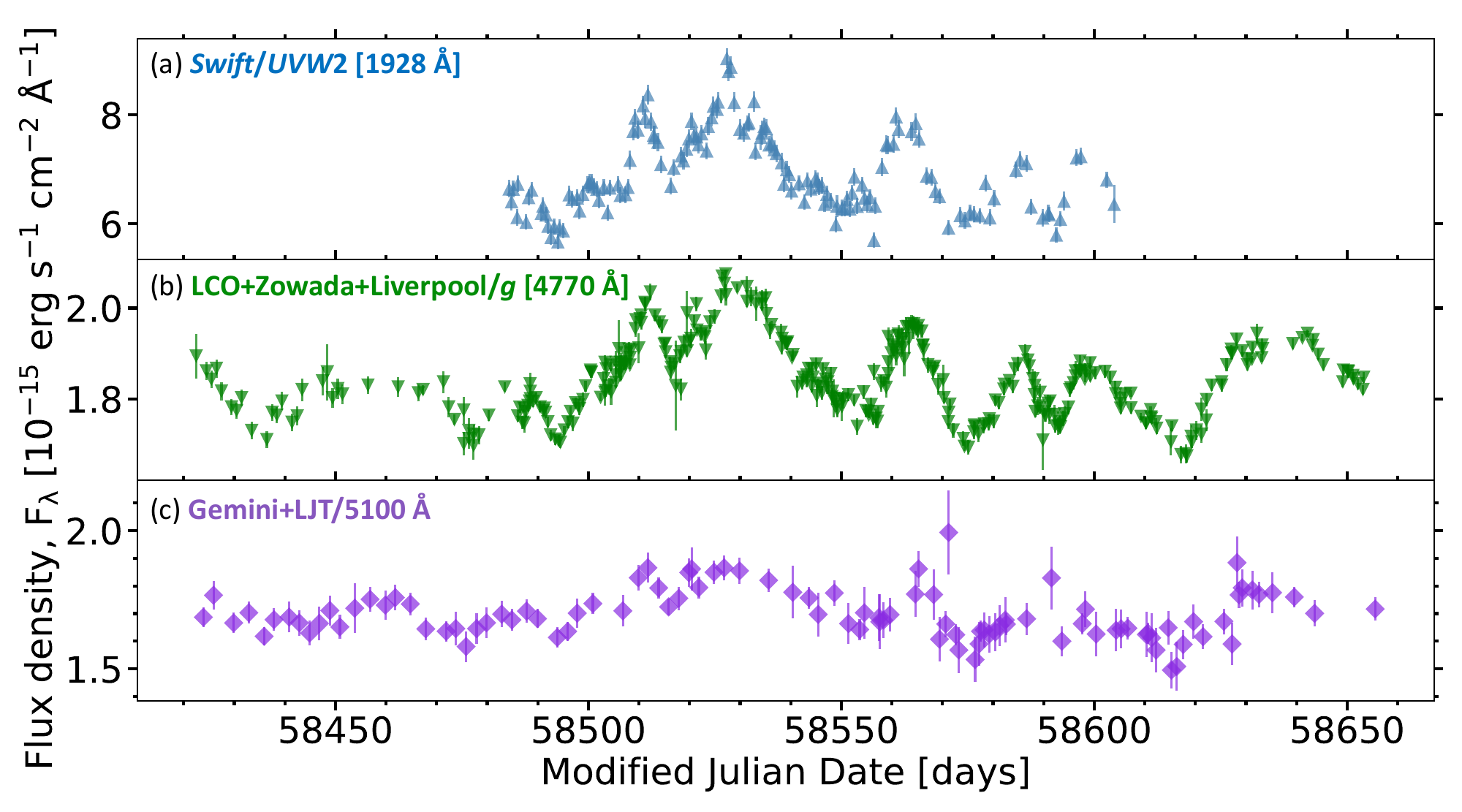}
\caption{Mrk~142 continuum light curves -- {\em UVW2} band with {\em Swift} (blue upright triangles, panel {\em a}) and the {\em g} band with LCO, Zowada, and Liverpool (green flipped triangles, panel {\em b}) from photometric observations; and 5100~\AA\ (purple diamonds, panel {\em c}) inter-calibrated from the spectroscopic observations with the Gemini North Telescope (Gemini) and the Lijiang Telescope (LJT).  The 5100~\AA\ inter-calibrated light curve displays the combined spectroscopic coverage over 102 epochs.  We performed the inter-calibration with a {\tt\string Python}-based Running Optimal Average \citep[{\tt\string PyROA};][]{donnan_etal_2021} technique (see introduction to {\tt\string PyROA} in Section~\ref{subsec:time_lags}).
\label{fig:obs_overlaps}}
\end{figure}

We obtained concurrent observations of Mrk~142 with multiple telescopes to perform RM analysis of the accretion disk and BLR simultaneously.  Figure~\ref{fig:obs_overlaps} showing the continuum light curves of Mrk~142 highlights the simultaneous coverage with different telescopes.

\subsection{Gemini North Telescope} \label{subsec:observations_gemini}

We obtained new observations of Mrk~142 long-slit spectra taken with the Gemini Multi-Object Spectrograph \citep[GMOS;][]{hook_etal_2004} on the 8.1-m Gemini North Telescope (Gemini) on Maunakea, Hawai`i with 33 epochs from February 6 through June 1, 2019.  These spectral observations are concurrent with the Mrk~142 photometric data from the {\em Swift} telescope comprising 180 epochs of 1-kilosecond exposures at X-ray, UV, and optical wavelengths from January 1 through April 30, 2019 (P.I.: E. Cackett) as well as with the photometric {\em g}-band data from LCO (P.I.: R. Edelson), Zowada (P.I.: E.Cackett), and Liverpool (P.I.: M. Goad); the photometric data are presented in \citet{cackett_etal_2020}.  The {\em Swift} observations had a twice-daily cadence until March 19, and the cadence was decreased to daily from March 20 onward.  We required observations from Gemini in early 2019 with considerable overlap with the {\em Swift} campaign to allow, for the first time, simultaneous measurements of the UV-emitting accretion disk and the BLR of a super-Eddington AGN.  The cadence of the Gemini observations was set to {\em one} day.  We obtained data for only two, sparsely separated epochs in February during the beginning of the observing period due to weather interruptions.  However, observations were more frequent in March and May, and the daily cadence was achieved in the first week of April.

The spectra were taken with the GMOS-North Hamamatsu detector and a single grating, B600 with two different slits, 0.75\arcsec\ (narrow slit) and 5.00\arcsec\ (wide slit), in the two-target acquisition mode, where Mrk~142 and a comparison star were observed in the same slit.  The GMOS-North Hamamatsu detector comprises three $\sim$2048$\times$4176~pixel chips (full detector size of 6278$\times$4176~pixels, mosaiced) arranged in a row with pixel size of 0.0807\arcsec\ with two chip gaps 4.88\arcsec\ wide.  The choice of the grating was made to obtain the broad emission line of interest, \Hbetalam, in the spectra.  The narrow 0.75\arcsec\ slit was selected to obtain a spectral resolution of $\sim$1125 (narrow-slit data) required to study the velocity structure of \Hbeta.  Accuracy in spectrophotometric calibration is a key for RM studies, and therefore, we used the wide slit at a resolution of $\sim$170 (wide-slit data) to correct for slit losses due to the narrow slit.  To satisfy this calibration requirement, Mrk~142 and a comparison star for flux calibration (hereafter, calibration star) were placed simultaneously in the same slit.  We achieved this for all observations by fixing the position angle of the slits at 155.20$^{\circ}$ East of North such that Mrk~142 appeared at the center of the slit.  The selected G-type calibration star (RA~=~10$^\mathrm{h}$25$^\mathrm{m}$36.37$^\mathrm{s}$, Dec~=~+51$^{\circ}$38\arcmin52.18\arcsec, {\em r}-band magnitude~=~15.9 from the SDSS catalog) has a well-calibrated spectrum and was used for previous LJT campaigns.  Flatfield images were taken for every object (science target and calibration star) with the Gemini Facility Calibration Unit (GCAL) in the sequence FLAT--OBJECT--OBJECT--FLAT with both slits.  The on-target exposures were 90~s long.  We also took daytime arc lamp spectra with the CuAr lamp, again, for both slits.  Binning of 1 in the spectral (X) direction and 2 in the spatial (Y) direction (1$\times$2) was used for all data except for the wide-slit arc lamp spectra, which used the binning of 1$\times$1.  A summary of the GMOS-North science observations is provided in Table~\ref{tab:obs_summ_gemini}.  The object spectra from all epochs except the narrow-slit spectra from epoch 30 were assigned a Pass (``P'') flag.

\startlongtable
\begin{deluxetable*}{rcllll}
\tablewidth{700pt}
\tablenum{1}
\tablecaption{Summary of Mrk~142 GMOS observations from February to June 2019
\label{tab:obs_summ_gemini}}
\tabletypesize{\scriptsize}
\tablehead{
 \\
\colhead{Epoch} & \colhead{UT\tablenotemark{a} Date} & \multicolumn2c{MJD\tablenotemark{b} Start Time} & \multicolumn2c{Airmass} \\
 & (YYYY-MM-DD) & 0.75$\arcsec$ slit & 5.00$\arcsec$ slit & 0.75$\arcsec$ slit & 5.00$\arcsec$ slit
} 
\startdata
1 & 2019-02-06 & 58520.412 & 58520.418 & 1.275 & 1.258 \\ 
 &  & 58520.414 & 58520.420 & 1.270 & 1.254 \\ 
2 & 2019-02-26 & 58540.415 & 58540.421 & 1.181 & 1.178 \\ 
 &  & 58540.417 & 58540.423 & 1.180 & 1.177 \\ 
3 & 2019-03-03 & 58545.433 & 58545.439 & 1.178 & 1.181 \\ 
 &  & 58545.435 & 58545.441 & 1.178 & 1.182 \\ 
4 & 2019-03-09 & 58551.430 & 58551.437 & 1.187 & 1.193 \\ 
 &  & 58551.432 & 58551.438 & 1.188 & 1.194 \\ 
5 & 2019-03-12 & 58554.550 & 58554.557 & 1.679 & 1.734 \\ 
 &  & 58554.552 & 58554.558 & 1.693 & 1.750 \\ 
6 & 2019-03-15 & 58557.427 & 58557.433 & 1.200 & 1.209 \\ 
 &  & 58557.428 & 58557.442 & 1.202 & 1.225 \\ 
7 & 2019-03-16 & 58558.355 & 58558.362 & 1.189 & 1.184 \\ 
 &  & 58558.357 & 58558.363 & 1.188 & 1.183 \\ 
8 & 2019-03-23 & 58565.281 & 58565.287 & 1.299 & 1.280 \\ 
 &  & 58565.283 & 58565.289 & 1.293 & 1.276 \\ 
9 & 2019-03-26 & 58568.254 & 58568.260 & 1.365 & 1.341 \\ 
 &  & 58568.256 & 58568.262 & 1.359 & 1.335 \\ 
10 & 2019-03-27 & 58569.410 & 58569.416 & 1.225 & 1.237 \\ 
 &  & 58569.411 & 58569.418 & 1.228 & 1.241 \\ 
11 & 2019-03-29 & 58571.233\tablenotemark{c} & 58571.239 & 1.424 & 1.394 \\ 
 &  & 58571.235\tablenotemark{d} & 58571.241 & 1.416 & 1.387 \\ 
12 & 2019-03-31 & 58573.233 & 58573.239 & 1.398 & 1.370 \\ 
 &  & 58573.235 & 58573.241 & 1.390 & 1.364 \\ 
13 & 2019-04-03 & 58576.457 & 58576.463 & 1.451 & 1.486 \\ 
 &  & 58576.459 & 58576.465 & 1.460 & 1.496 \\ 
14 & 2019-04-04 & 58577.234 & 58577.240 & 1.350 & 1.327 \\ 
 &  & 58577.235 & 58577.241 & 1.343 & 1.321 \\ 
15 & 2019-04-05 & 58578.241 & 58578.248 & 1.312 & 1.292 \\ 
 &  & 58578.243 & 58578.249 & 1.306 & 1.287 \\ 
16 & 2019-04-06 & 58579.236 & 58579.242 & 1.322 & 1.302 \\ 
 &  & 58579.237 & 58579.243 & 1.317 & 1.297 \\ 
17 & 2019-04-07 & 58580.461 & 58580.467 & 1.543 & 1.585 \\ 
 &  & 58580.463 & 58580.469 & 1.554 & 1.597 \\ 
18 & 2019-04-08 & 58581.331 & 58581.337 & 1.177 & 1.179 \\ 
 &  & 58581.333 & 58581.339 & 1.177 & 1.180 \\ 
19 & 2019-04-09 & 58582.432 & 58582.438 & 1.408 & 1.439 \\ 
 &  & 58582.434 & 58582.440 & 1.416 & 1.447 \\ 
20 & 2019-04-25 & 58598.274 & 58598.280 & 1.175 & 1.176 \\ 
 &  & 58598.276 & 58598.282 & 1.175 & 1.176 \\ 
21 & 2019-04-27 & 58600.391 & 58600.397\tablenotemark{c} & 1.447 & 1.481 \\ 
 &  & 58600.392 & 58600.398 & 1.456 & 1.491 \\ 
22 & 2019-05-01 & 58604.338 & 58604.344\tablenotemark{c} & 1.282 & 1.300 \\ 
 &  & 58604.340 & 58604.346 & 1.286 & 1.305 \\ 
23 & 2019-05-02 & 58605.327 & 58605.333 & 1.259 & 1.275 \\ 
 &  & 58605.329 & 58605.335 & 1.263 & 1.280 \\ 
24 & 2019-05-07 & 58610.317 & 58610.323 & 1.267 & 1.285 \\ 
 &  & 58610.318 & 58610.324 & 1.272 & 1.289 \\ 
25 & 2019-05-08 & 58611.361 & 58611.367 & 1.450 & 1.485 \\ 
 &  & 58611.363\tablenotemark{c} & 58611.369\tablenotemark{c} & 1.459 & 1.494 \\ 
26 & 2019-05-09 & 58612.246 & 58612.253 & 1.177 & 1.179 \\ 
 &  & 58612.248 & 58612.254 & 1.177 & 1.180 \\ 
27 & 2019-05-12 & 58615.269 & 58615.275 & 1.201 & 1.209 \\ 
 &  & 58615.270 & 58615.276 & 1.203 & 1.212 \\ 
28 & 2019-05-13 & 58616.320\tablenotemark{e} & 58616.326 & 1.329 & 1.353 \\ 
 &  & 58616.322 & 58616.328 & 1.335 & 1.359 \\ 
29 & 2019-05-24 & 58627.266 & 58627.272 & 1.256 & 1.272 \\ 
 &  & 58627.267 & 58627.273 & 1.260 & 1.276 \\ 
30 & 2019-05-25 & 58628.242\tablenotemark{f} & 58628.248 & 1.213 & 1.223 \\ 
 &  & 58628.243\tablenotemark{f} & 58628.249 & 1.216 & 1.227 \\ 
31 & 2019-05-26 & 58629.251 & 58629.257 & 1.235 & 1.249 \\ 
 &  & 58629.253 & 58629.259 & 1.239 & 1.252 \\ 
32 & 2019-05-28 & 58631.251 & 58631.257 & 1.246 & 1.261 \\ 
 &  & 58631.252 & 58631.258 & 1.250 & 1.265 \\ 
33 & 2019-06-01 & 58635.255 & 58635.262 & 1.287 & 1.307 \\ 
 &  & 58635.257 & 58635.263 & 1.292 & 1.312 \\
\enddata
\tablenotetext{a}{UT: Universal Time dates}
\tablenotetext{b}{MJD: Modified Julian Date recorded at the start of the observations for individual exposures.}
\tablenotetext{c}{Science spectrum assigned SPCALF (see $Note$ below)=0.}
\tablenotetext{d}{Science spectrum calibrated with the narrow-slit standard star spectrum from exposure 1 and hence assigned SPCALG (see $Note$ below)=B.}
\tablenotetext{e}{Science spectrum likely had a calibration issue and hence was not used for further analysis (see Appendix~\ref{app:A} for details).}
\tablenotetext{f}{Science spectrum assigned DQF (see $Note$ below)=U.}
\tablecomments{Observations were done with the GMOS-North Hamamatsu detector in the two-target acquisition mode (Mrk~142 and a comparison star in the same slit) positioning the slit at 155.20$^\circ$ East of North, with the B600 grating (covering the broad \Hbeta\ emission line at $\sim$4862\AA) and two slits, 0.75$\arcsec$ (narrow slit) and 5.00$\arcsec$ (wide slit).  Two exposures were taken with every grating/slit combination, each 90 seconds long.  A Data Quality Flag (DQF) of ``P'' or ``U'' that stands for Pass or Usable was assigned to all data at the time of observing.  Unless stated otherwise, all science spectra were assigned a DQF of ``P''.  SpectroPhotometric CALibration Flag (SPCALF) indicates whether the science spectra were calibrated (``1'') or not calibrated (``0'') during spectral reduction (see $\S$\ref{subsubsec:fcalibration} for more details).  SpectroPhotometric CALibration Grade (SPCALG) indicates the grade assigned to the spectrophotometric calibration based on the epoch and exposure of the calibration star spectrum used for calibrating the science spectra (see $\S$\ref{subsubsec:fcalibration} for more details).  All science spectra were assigned an SPCALF of 1 and an SPCALG of ``A'' unless indicated otherwise.}
\end{deluxetable*} 
\subsection{Lijiang Telescope} \label{subsec:observations_ljt}

To complement the short observing period of 33 epochs with Gemini, we incorporated supporting observations of Mrk~142 for our study.  We observed Mrk~142 with the Yunnan Faint Object Spectrograph and Camera on the Lijiang 2.4-m Telescope \citep[LJT;][]{wang_etal_2019} in the two-target acquisition mode with the same calibration star as used for the Gemini observations.  We followed the same observing procedure as for previous SEAMBH campaigns \citep[e.g.,][]{du_etal_2014, du_etal_2015}.  We obtained long-slit spectra of the target at 69 epochs from November 1, 2018 through June 21, 2019, contemporaneously with the {\em Swift}, LCO+Zowada+Liverpool, and Gemini observing campaigns.

Two exposures of 1200 seconds each were taken for each epoch, with Grism~14 and a long slit with a projected width of 2$\farcs$5.  The yielded spectra cover a wavelength rage of 3800~\AA\ to 7200~\AA, with a dispersion of 1.8~\AA~pixel$^{-1}$.  The final instrumental broadening is roughly 695~\kms\ in FWHM.  Bias, dome flats, and arc-lamp spectra were taken each night for calibrations, and spectrophotometric standards were observed in several nights of good weather conditions.

Table~\ref{tab:obs_summ_overall} provides a summary of the overlapping photometric and spectroscopic programs.

\begin{deluxetable*}{ccccc}
\tablewidth{700pt}
\tablenum{2}
\tablecaption{Summary of Overlapping Photometric and Spectroscopic Observations
\label{tab:obs_summ_overall}}
\tabletypesize{\scriptsize}
\tablehead{
 \\
\colhead{Filter/Line} & \colhead{Observatory} & Date Range (MJD\tablenotemark{a}) & \colhead{Number of Epochs} & \colhead{Mean Cadence} \\
} 
\startdata
{\em UVW2} [1928~\AA]\tablenotemark{b} & {\em Swift} & 58484.349--58603.941 & 149 & $\sim$twice-daily \\ 
{\em g} [4770~\AA]\tablenotemark{b} & LCO, Zowada, Liverpool & 58422.973--58653.260 & 361 & 2 days \\ 
\Hbetalam \tablenotemark{c} & Gemini & 58520.414--58635.257 & 33 & $\sim$3.6 days \\ 
\Hbetalam \tablenotemark{c} & LJT & 58423.899--58655.549 & 69 & $\sim$3.4 days \\ 
\enddata
\tablenotetext{a}{MJD: Modified Julian Date}
\tablenotetext{b}{Photometry}
\tablenotetext{c}{Spectroscopy}
\end{deluxetable*}

\section{Spectral Reduction} \label{sec:reduction}

\subsection{Gemini Spectral Reduction} \label{subsec:reduction_gemini}

The spectral reduction process for all Gemini epochs included four stages (in the order of appearance below) with the Gemini Image Reduction and Analysis Facility (Gemini {\tt\string IRAF}\footnote{Gemini {\tt\string IRAF} is an external package that makes use of {\tt\string IRAF} (a software system used for the reduction and analysis of astronomical data, created and supported by the National Optical Astronomy Observatory in Tucson, Arizona).  See more at \url{https://www.gemini.edu/sciops/data-and-results/processing-software/description}.}) reduction package: (1) baseline calibrations with GCAL flats, two-dimensional (2D) arc lamp spectra, and bias frames; (2) cleaning of 2D spectra followed by the wavelength calibration and extraction of one-dimensional (1D) science and calibration-star spectra (in the same slit); (3) preparing 1D spectra for analysis with {\tt\string PrepSpec} (see introduction to {\tt\string PrepSpec} in Section~\ref{subsec:prepspec_modeling}); and (4) flux calibration of the 1D science spectra.  For each epoch, we first sorted the data into lists of bias frames, GCAL flats, arc lamp spectra, and object spectra for both the narrow and the wide slits.  We then used the the same reduction script with different parameter settings for processing the data taken with the two slits.

\subsubsection{Baseline Calibrations} \label{subsubsec:base_calibrations}

Baseline calibrations comprised creating a masterbias image, generating a dispersion solution with the narrow-slit arc lamp spectra, and constructing masterflat images with both the narrow- and wide-slit flatfield images.  For individual observing nights, we used bias frames with the binning of 1$\times$2 and a full-frame readout from the \href{https://archive.gemini.edu}{Gemini Observatory Archive}.  We applied an overscan noise correction to all bias images, for a given night, before combining them into a masterbias image.  We then reduced the narrow-slit arc-lamp spectra with bias subtraction turned off and used them to generate 2D dispersion solutions with the task {\tt\string gswavelength}.  Generating dispersion solutions was a two-step process -- fitting the 1D wavelength solution in the spectral direction and fitting any distortions in the spatial direction.  The reference wavelengths for the arc-lamp spectra were used from the Gemini {\tt\string IRAF} package.  Because the re-binned, wide-slit arc-lamp spectra -- binned from 1$\times$1 to 1$\times$2 to match the binning of the corresponding GCAL flats and object spectra -- were unable to provide a non-distorted wide-slit dispersion solution, we used the narrow-slit solution to wavelength calibrate the wide-slit data.  For a given epoch, we combined the two GCAL flats (including a quantum efficiency correction for each) taken with the two slits to create a masterflat corrected for the uneven illumination along the GMOS detector in the long-slit mode.

\subsubsection{Cleaning, Wavelength Calibration, and Extraction} \label{subsubsec:cleaning_wcalibration_extraction}

We corrected the 2D object spectra affected by cosmic-ray hits and performed their wavelength calibration to then extract the 1D science and calibration-star spectra.  With the task {\tt\string gscrrej}, we first selected a fixed square region surrounding the cosmic-ray affected pixels above a specified threshold and then replaced them with interpolated values from local noise levels.  However, this method did not correct for all cosmic rays.  We applied an additional correction to the affected pixels that remained uncorrected in the next stage of the reduction process.  We applied the derived narrow-slit dispersion solutions to both the narrow- and the wide-slit object spectra.

For a given epoch, we extracted 1D science and calibration-star spectra separately from individual exposures with the task {\tt\string gsextract}.  We selected a considerable swath of background for subtraction from both sides of each trace during the extraction process.  The subtraction of bright skylines from the extracted 1D spectra resulted in some sharp spikes in the spectra owing to residual noise.  We applied additional correction to remove the sharp features in the 1D spectra in the next stage of reduction (see Section~\ref{subsubsec:prep_data_ps} for details).  

A few of the extracted science and calibration-star spectra showed flat regions (zero flux values) on the shorter-wavelength (or blue) end ($\sim$3355~\AA\ to $\sim$4325~\AA) that do not match the true shape of the continuum, while some spectra showed bump-like features.  The flat regions were a consequence of the slit position angle not aligned along the parallactic angle, whereas the bump-like features likely resulted from the flat-fielding process, where a higher order spline was used to create the masterflat to appropriately trace detector sensitivity near the chip-gap regions and avoid discontinuities in the calibrated spectra near the chip edges.  We corrected the spectra containing flat blue ends or bumpy features individually before attempting flux calibration (see Section~\ref{subsubsec:prep_data_ps} for details).

\subsubsection{Additional Corrections to 1D Spectra -- Preparing Data for {\tt\string PrepSpec}} \label{subsubsec:prep_data_ps}

To prepare the spectra for PrepSpec, it was important that each spectrum have no gaps.  Before the flux calibration stage, we trimmed the blue ends of the spectra shorter than $\sim$4325~\AA\ in the rest frame because they were very noisy and not required for the purposes of this study.  We further processed the 1D spectra for: (1) flat blue ends (due to the slit position angle) or bump-like features (from the flat-fielding process) appearing in some spectra; and (2) spectral regions affected by artefacts from cosmic-ray removal and sky subtraction as well as chip gaps with no flux.  This additional processing was important for the initial stage of modeling spectra with {\tt\string PrepSpec}, the software tool that corrects spectra for relative calibration differences (see Section~\ref{subsec:prepspec_modeling} for details).

We developed a script to correct flat and bump-like regions in the 1D spectra in {\tt\string Python}\footnote{Visit \url{https://www.python.org/} for full documentation on {\tt\string Python}.}~v3.6.5.  A spline function fit to a reference spectrum modeled the true shape of the affected region.  We then modeled the flux over the affected pixels assuming a Gaussian distribution of data points with standard deviation equal to the measured standard deviation at the same location in the reference spectrum.  The reference spectrum used for recovery was typically the spectrum from another exposure taken on the same night (see Appendix~\ref{app:A} for exceptions).  

To correct for spectral regions affected by artefacts and chip-gaps, we developed another {\tt\string Python} script to replace the regions with affected data points by local median values or interpolated and simulated data.  In a given window of affected points: (1) if the number of pixels was $<$5, the algorithm replaced every data point by the median value of a range of 5~pixels on either side of that point with the noise equal to the local median noise; and (2) if the number of pixels was $\geq$5, the algorithm first linearly interpolated across the affected region and then replaced the interpolated points with simulated data assuming a Gaussian distribution with a standard deviation equal to twice the noise in the interpolated data.  The uncertainties for the corrected pixel regions were assigned to be twice as much as the standard deviation of the unaffected individual pixel values in the region.

We used the wide-slit science and calibration-star spectra to correct for the wavelength-dependent slit losses in the narrow-slit spectra with a {\tt\string PyRAF} ({\tt\string IRAF} with {\tt\string Python} wrapper) script.  We employed the {\tt\string IRAF} task {\tt\string curfit} to fit a spline function to the ratios of the narrow-slit to the reference spectra.  We used a single reference spectrum: the mean of the bright, wide-slit spectra.  Finally, we updated the starting pixel value of the wavelength scale in the FITS file headers of the slit-loss corrected spectra to generate the appropriate wavelength grid for the trimmed spectra.

The {\tt\string Python} scripts for performing the above corrections to prepare spectra for {\tt\string PrepSpec} analysis are publicly available on {\tt\string GitHub}\footnote{Please contact the corresponding author for further details.}.

\subsubsection{Flux Calibration} \label{subsubsec:fcalibration}

The flux calibration process included two steps -- fitting a sensitivity curve of the detector response to the flux standard with {\tt\string gsstandard}, and applying the sensitivity solution to the science spectra with the task {\tt\string gscalibrate}.  For flux calibration, we used the calibration star captured in the same slit as the science target except for a handful of spectra for which we used the star from another exposure of the same epoch (see Appendix~\ref{app:A} for details).  Accordingly, we assigned a SpectroPhotometric CALibration Flag (SPCALF) of 1 (0) for calibrated (non-calibrated) science spectra (stated in Table~\ref{tab:obs_summ_gemini}).  Based on the epoch and exposure of the standard star spectrum used for calibration, we further assigned a SpectroPhotometric CALibration Grade (SPCALG; see Table~\ref{tab:obs_summ_gemini}) to the science spectra as follows.

\begin{itemize}
    \item SPCALG ``A'': Science spectrum calibrated with the standard star spectrum from the same exposure.
    \item SPCALG ``B'': Science spectrum calibrated with the standard star spectrum from the same epoch but different exposure.
\end{itemize}

\noindent Appendix~\ref{app:A} outlines special cases of spectral reduction that were treated separately. \\

\subsection{LJT Spectral Reduction} \label{subsec:reduction_ljt}

We first reduced the LJT spectra with {\tt\string IRAF}, following the standard procedures for bias subtraction, flat-field correction, and wavelength calibration.  The spectra of both the target and the calibration star were extracted in a uniform aperture of 8$\farcs$5.  For those nights with good weather conditions, the spectra of the calibration star were flux-calibrated using the spectrophotometric standards.  We combined these flux-calibrated spectra to generate a fiducial spectrum of the calibration star.  Then, for each exposure, a sensitivity function was obtained by fitting the fiducial spectrum to the extracted spectrum of the comparison star.  Finally, we performed flux-calibration of the target spectrum (see \citealt{li_etal_2021} for more details).

\subsection{Comparison Between Gemini and LJT Spectra} \label{subsec:gemini_ljt_mean_comparison}

\begin{figure*}[ht!]
\epsscale{1.15}
\plotone{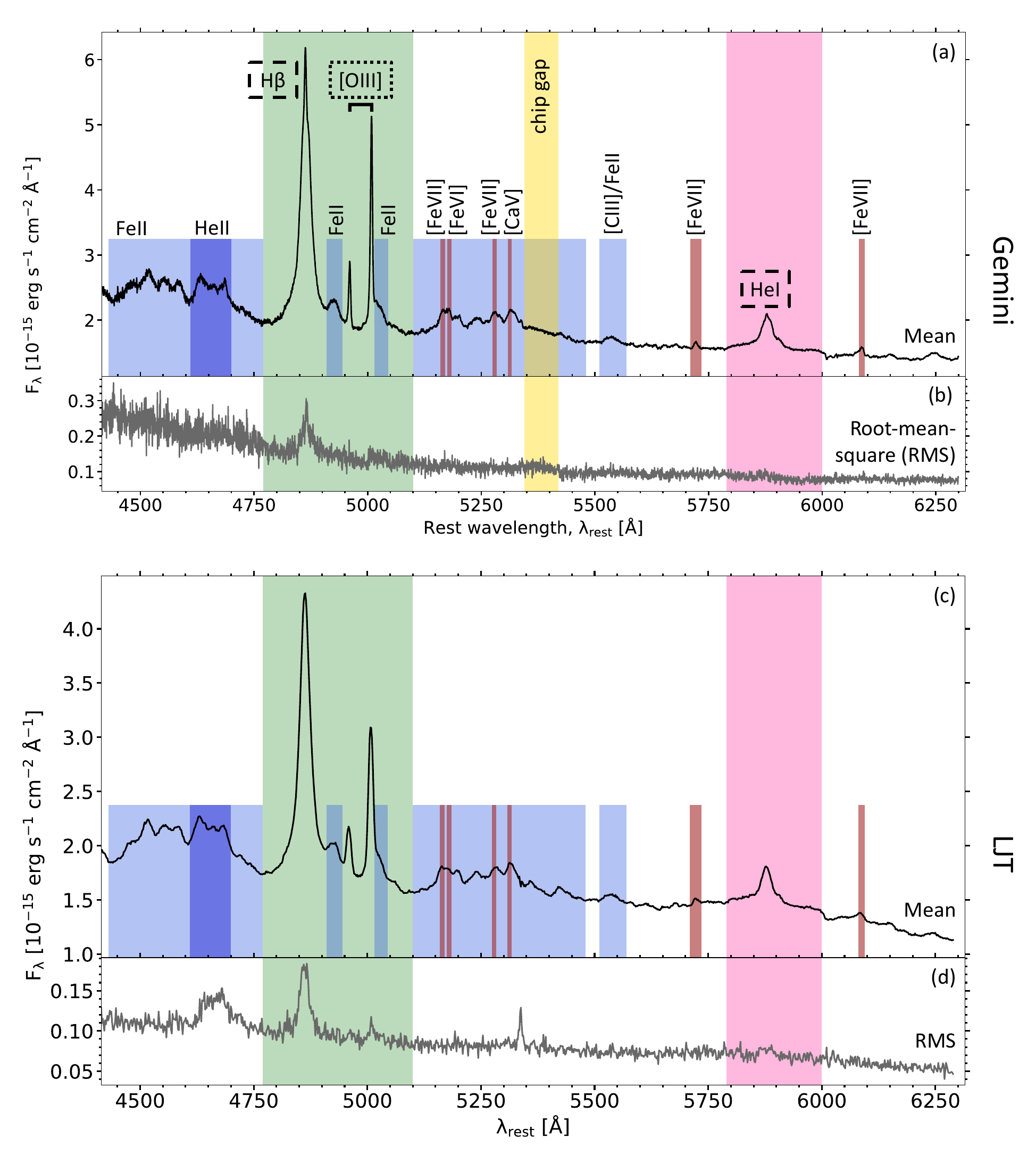}
\caption{Rest-frame mean (black; panels {\em a} and {\em c}) and root-mean-square (RMS; gray; panels {\em b} and {\em d}) of Mrk~142 Gemini ({\em Top}; velocity resolution of $185.6\pm10.2$~\kms) and LJT ({\em Bottom}; velocity resolution of $695.2\pm3.9$~\kms) spectra highlighting the {\em Regions of Interest} -- \Hbetalam\ and \OIIIlam\ (region shaded in green), and \HeIlam\ (region shaded in pink).  Labels enclosed in dashed (dotted) boxes indicate broad (narrow) lines.  The Gemini mean spectrum shows sharp \FeII\ features (shaded in faint blue), which appear blended with the red wings of \Hbeta\ and \OIIIlamstrong\ (blue shaded bars) in the LJT mean.  Owing to the high signal-to-noise of the Gemini spectra, the peculiar shape of the \HeIlam\ line is clearly evident.  The high-ionization coronal lines (shaded in brown) also appear sharp in contrast to the LJT spectrum, as a result.  The LJT RMS spectrum shows clear variability in \HeII\ (shaded in blue).  Both RMS spectra indicate variability in the broad \Hbeta.  However, no significant variability is evident in \HeI\ over the timescale of Gemini+LJT observations.  The yellow-shaded region in the Gemini spectra indicates the GMOS chip gap from $\sim$5350~\AA\ to $\sim$5410~\AA. \\
\noindent (The reduced and calibrated Mrk~142 Gemini spectra are available in a machine-readable form.)
\label{fig:mean_rms_gemini_ljt}}
\end{figure*}

Gemini spectra from 33 epochs and LJT spectra from 69 epochs provided 102 epochs of Mrk~142 spectral observations overlapping with the {\em Swift} and LCO+Zowada+Liverpool photometric campaigns.  A mean spectrum allows us to visualize spectral features in high signal-to-noise (S/N) from the combined observations, while a root-mean-square (RMS) spectrum signifies the variability in the spectral features.  Figure~\ref{fig:mean_rms_gemini_ljt} displays the mean and RMS of the Gemini ({\em Top}) and LJT ({\em Bottom}) spectra.  The higher-resolution Gemini mean spectrum shows sharper emission-line profiles (\Hbeta, \OIII, and \HeI) as compared to the LJT mean.  At lower resolution, LJT spectra are affected by instrumental broadening which results in the narrow emission lines, e.g., \OIII, appearing broader than in the Gemini mean.  The instrumental broadening effect also blurs the \FeII\ emission (shaded in faint blue) and the coronal lines (high-ionization forbidden transitions shaded in brown) in the LJT mean spectrum.  In contrast to the Gemini mean, the \FeII\ features at $\sim$4925~\AA\ and $\sim$5030~\AA\ in the LJT mean appear blended with the \Hbeta\ wings on the longer-wavelength (red) side and \OIIIlamstrong, respectively.  The RMS of the Gemini spectra shows a noisy region blueward of 4750~\AA\ likely dominated by calibration noise.  It is worth noting, however, that the finer wavelength sampling of the Gemini spectra (owing to the narrow-slit observations) makes that region appear even noisier.  On the other hand, the region towards the blue end of the LJT RMS spectrum shows clear evidence of variability in the \HeIIlamopt\ line although it is heavily contaminated with \FeII\ in the surrounding region.  Variability in \Hbeta\ is revealed by both the Gemini and the LJT RMS spectra.  Although no variability in \HeI\ is evident from the Gemini RMS, the LJT RMS shows a weak signature of variability in broad \HeI.  A very low, broad wave appears from $\sim$5250~\AA\ to $\sim$5450~\AA\ and from $\sim$5650~\AA\ to $\sim$5950~\AA\ in the LJT RMS spectrum likely resulting from calibration.  The GMOS chip gap region in the Gemini spectrum extends from $\sim$5350~\AA\ to $\sim$5410~\AA, which also appears as a low bump in the RMS spectrum.

\section{Spectral Analysis} \label{sec:spectral_analysis}

To measure the \Hbeta\ and \HeI\ emission lines in the calibrated spectra, we first corrected any discrepancies in the calibrations of the Gemini and LJT spectra, independently with {\tt\string PrepSpec}, and then modeled their spectral features with {\tt\string Sherpa}.  For {\tt\string PrepSpec} modeling of Gemini spectra, we used the spectral region from $\sim$4430~\AA\ to $\sim$6300~\AA.  For LJT spectra, we kept the spectral region from $\sim$3390~\AA\ to $\sim$6300~\AA.

\subsection{{\tt\string PrepSpec} Modeling} \label{subsec:prepspec_modeling}

We independently modeled the 64 narrow-slit Gemini spectra and the 69 LJT spectra with {\tt\string PrepSpec}\footnote{Find current version of {\tt\string PrepSpec} at \url{http://star-www.st-andrews.ac.uk/~kdh1/lib/prepspec/prepspec.tar.gz}.} (developer: K. Horne) to correct for any relative deviations in the calibrated wavelength and flux scales.  {\tt\string PrepSpec} models spectra by fitting the continuum and emission lines with a composite model through an iterative process.  We included the following model components for fitting the Mrk~142 spectra: (1) [A]verage spectrum (specified by ``A'') -- mean of the input spectra; (2) [C]ontinuum -- variations in the continuum emission from the accretion disk modeled as a polynomial defined by \(\log{\lambda}\) with time-dependent coefficients; (3) [W]avelength jitter -- inter-spectra shifts in the wavelength scales; (4) [F]lux jitter -- time-dependent photometric corrections to minimize the scatter of narrow emission-line fluxes relative to their median; and (4) [B]road-line variations -- variability in the broad emission-line features.  Modeling emission lines in {\tt\string PrepSpec} takes into account the velocity window half-widths of the broad as well as the narrow lines, whose initial values were set to 3000~\kms\ and 500~\kms, respectively.  We set the broad \Hbetalam\ and \HeIlam\ as variable lines for Gemini spectra, and \Hgammalam, \HeIIlamopt, \Hbeta, and \HeIlam\ as variable for LJT spectra.  The software uses the I~Zwicky~1 (I~Zw~1) template model \citep{veroncetty_etal_2001} to fit \FeII\ emission in the mean spectrum.  {\tt\string PrepSpec} is not designed to handle gaps in spectra or extremely large flux values, e.g., from cosmic-ray hits.  Therefore, chip gaps and artefacts from cosmic-ray correction or sky subtraction in the Gemini spectra were replaced by median or simulated data (see Section~\ref{subsubsec:prep_data_ps} for details) during spectral reduction.

\begin{figure}[ht!]
\epsscale{1.2}
\plotone{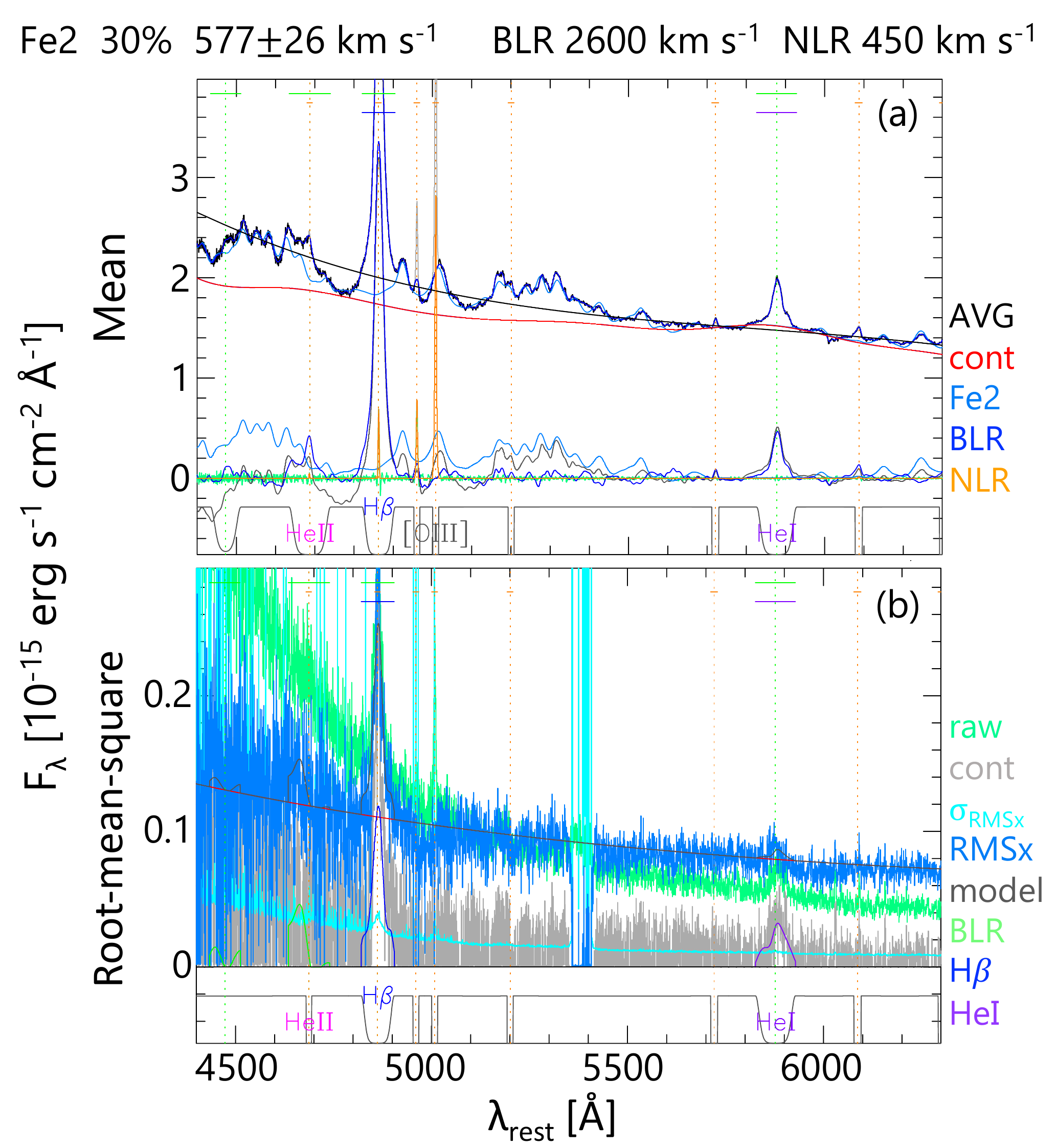}
\caption{Mean (panel {\em a}) and the root-mean-square (RMS) spectra for 64 narrow-slit Mrk~142 Gemini spectra processed through {\tt\string PrepSpec} showing model fits with individual components.  In panel {\em a}, the composite model (dark blue curve), including the components shown at the bottom of the plot -- average spectrum (AVG; black), continuum (cont; red), \FeII\ (Fe2; faint blue), broad-line region (BLR; dark blue), and narrow-line region (NLR; orange), is overlaid on the mean spectrum (black curve).  The broad (narrow) emission lines are indicated with green (orange) dotted vertical lines as well as with green (orange) solid horizontal dashes.  The broad lines of \Hbeta\ (blue label) and \HeI\ (purple label) are marked with solid horizontal dashes.  In panel {\em b}, model (model; dark gray curve) fit to the residual RMS spectrum (RMSx; blue curve) includes the components: continuum (cont; red curve) and BLR (green curve at the bottom of the plot).  The raw RMS spectrum is the upper green curve.  The BLR component comprises the broad lines of \Hbeta\ (blue bump around $\sim$4862~\AA) and \HeI\ (purple bump around $\sim$5877~\AA) shown at the bottom of the plot.  The broad \HeII\ emission feature at $\sim$4687~\AA\ (box-like feature in the model) is contaminated with \FeII\ and hence difficult to fit given the noise in the region.  The deviation in the residual RMS spectrum ($\sigma_{\mathrm{RMSx}}$; cyan curve) shows large values in the region of the GMOS detector chip gap from $\sim$5345~\AA\ to $\sim$5420~\AA.  In both panels, the active broad-line windows of \Hbeta\ and \HeI\ used for {\tt\string PrepSpec} modeling are shown in negative-value space.}
\label{fig:ps_mean_rms_gemini}
\end{figure}

In the {\tt\string PrepSpec} modeling stage, we first corrected the Gemini and LJT spectra for pixel shifts relative to the \OIIIlamstrong\ line and then modeled the spectra with a composite model.  We observed small pixel shifts ($<$6~pixels) while aligning the spectra along the wavelength axis.  The model components were jointly fit starting with a single component and then adding components up to the ACWFB composite model for both the Gemini and the LJT spectra.  {\tt\string PrepSpec} determines the best-fitting model by accessing the Bayesian Information Criterion and reduced $\chi^2$ ($\chi_{\nu}^2$, where $\nu$ stands for degrees of freedom) statistics.  The goal of the fitting process is to use the fewest possible parameters to describe the data while penalizing the model for the number of parameters used.  A good model yields $\mathrm{\chi_{\nu}^2\sim1}$.  Figure~\ref{fig:ps_mean_rms_gemini} displays the final model (dark blue curve) passing through the black mean spectrum (panel {\em a}) and the model (dark gray curve) to the residual root-mean-square (RMSx) spectrum (panel {\em b}) for the 64 narrow-slit Gemini spectra.  The RMS spectrum shows that the spectra are noisier at the bluer end.

\begin{figure}[ht!]
\epsscale{1.2}
\plotone{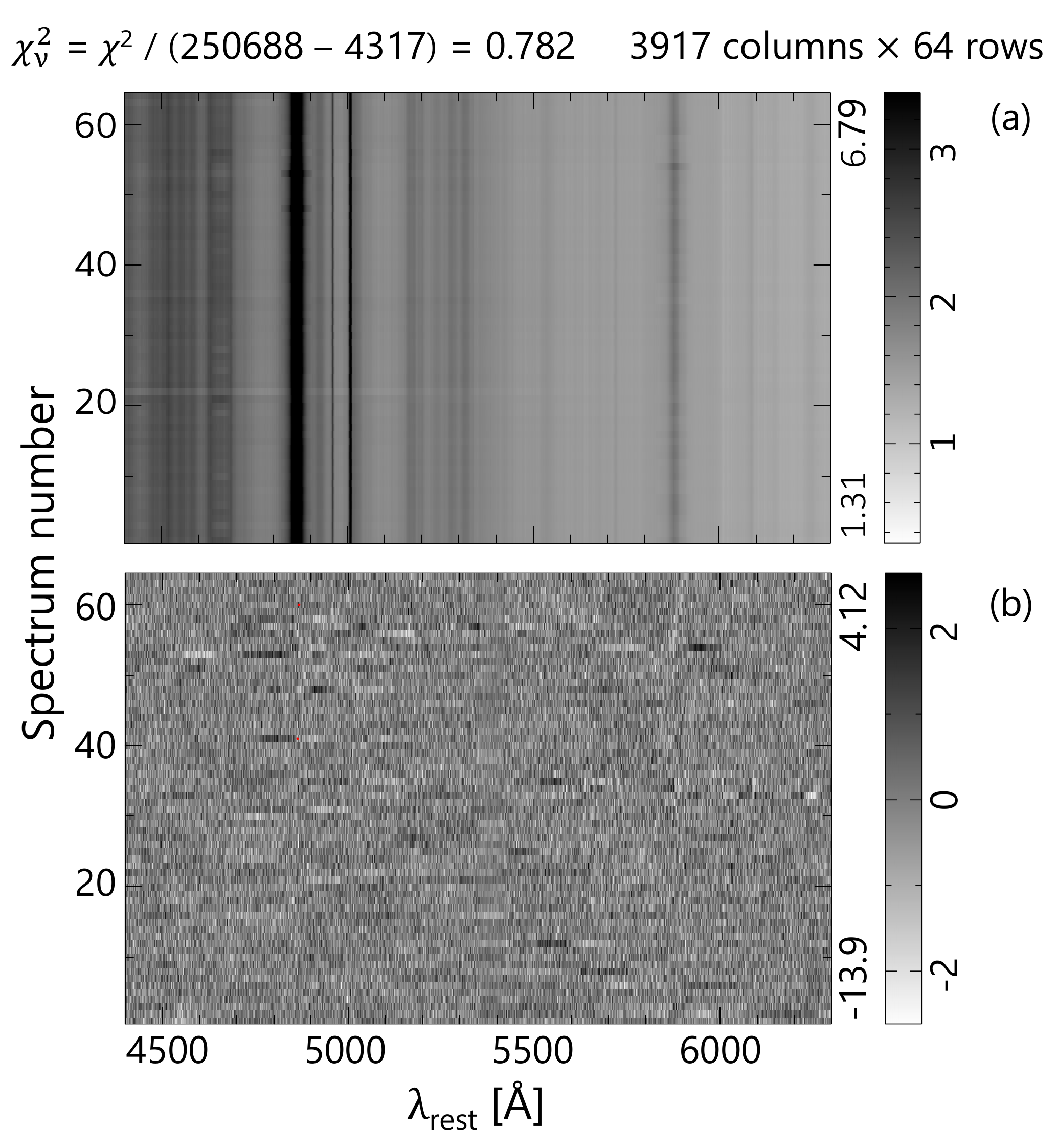}
\caption{{\tt\string PrepSpec} model including all components (panel {\em a}) and residuals (\(data - model\) in units of standard deviation, $\sigma$; panel {\em b}) with a reduced $\chi^2$ ($\chi_{\nu}^2$) of 0.782 for 64 narrow-slit Mrk~142 spectra.  In panel {\em a}, dark regions indicate strong emission lines of \Hbeta\ and \OIII, whereas the weaker \HeI\ lines and \FeII\ emission appears as less prominent features. 
 In panel {\em b}, each row represents a single exposure spectrum (where multiple exposures at a given epoch are not yet combined).  The horizontal wiggles strongly evident in some spectra are likely the result of either replacing values with simulated data in those regions or using a higher order function during flat-fielding (see text for more details). 
The smeared region from $\sim$5345~\AA\ to $\sim$5420~\AA\ is one of the chip gaps of the GMOS detector where simulated data was added during reduction.
\label{fig:ps_gray_model_residuals_gemini}}
\end{figure}

Figure~\ref{fig:ps_gray_model_residuals_gemini} shows the final model (panel {\em a}) along with the residuals (in the units of $\sigma$; panel {\em b}) in grayscale for the Gemini spectra.  The best-fit model yielded a $\chi_{\nu}^2$ value of 0.782, which indicates overfitting of the data, possibly indicating inaccurate error bars larger than the scatter in the data.  The dark regions in the model highlight the prominent emission-line features of \Hbetalam\ and \OIIIlam.  The weak fluctuations blueward of $\sim$4700~\AA\ indicate more noise in that region as compared to the red end of the spectra.  The residuals in grayscale display horizontal wiggles that are strongly evident in some spectra.  We noted that the wiggles appear in the spectral regions replaced by simulated data to correct for residual features either from cosmic-ray correction or sky subtraction.  The replacement with simulated data may have resulted in a lower performance of the model in those regions.  Another probable reason for the wiggles is the use of a higher-order spline during flat-fielding in the spectral reduction process (refer Section~\ref{subsubsec:prep_data_ps} for details).  However, we visually inspected all spectra processed through {\tt\string PrepSpec} and observed no anomalous behavior in the regions with wiggles.  Therefore, the spectra were considered valid for further analysis.

{\tt\string PrepSpec} modeling of LJT spectra yielded nearly even residuals with a $\chi_{\nu}^2$ value of 0.791.  The region redward of 6300~\AA\ in LJT spectra comprises several blended narrow-line features, which resulted in a sub-optimal performance of the {\tt\string PrepSpec} model.  Therefore, we excluded the red side of the LJT spectra ($\lambda > 6300$~\AA) during {\tt\string PrepSpec} processing.

\subsection{Spectral Modeling in {\tt\string Sherpa}} \label{subsec:sherpa_modeling}

We modeled the continuum and emission lines in the Gemini and LJT spectra in {\tt\string Sherpa}\footnote{{\tt\string Sherpa} is a software application for modeling and fitting astronomical images and spectra.  In this work, the {\tt\string Sherpa}~v4.10.0 application was used within Coronagraphic Imager with Adaptive Optics ({\tt\string CIAO}) v4.10.0, the X-ray Data Analysis Software designed by the Chandra X-ray Center.  For full documentation of {\tt\string CIAO}-{\tt\string Sherpa}, see \url{https://cxc.harvard.edu/sherpa4.14/}.} \citep{freeman_etal_2001, doug_burke_2018_1245678} v4.10.0 with a {\tt\string Python} wrapper script.  We first corrected the Gemini and LJT spectra for Galactic reddening using $E(B-V)=0.015$ \citep{schlafly_finkbeiner_2011}.  Averaging the two narrow-slit Gemini exposures from every night into a single spectrum per epoch (with exceptions for spectra from epochs 11, 25, and 28, where we only used single exposures) yielded a total of 33 Gemini spectra.  Together with the 69 LJT spectra, we modeled a total of 102 Mrk~142 spectra.

We developed a composite model with a goal of performing a clean extraction of the \Hbeta\ and \HeI\ emission lines from the Gemini and LJT spectra.  We included a power-law fit to the continuum, three Gaussians to model each of the \Hbeta, \HeI\, and \HeII\ emission lines, and a single Gaussian for each of the \OIII\ doublet lines.  We adopted the I~Zw~1 template model from \citet{boroson_green_1992} as a pseudo-continuum to trace the \FeII\ emission-line features.  We also experimented with the \FeII\ template from \citet{veroncetty_etal_2001}.  However, it failed to suitably trace the sharp \FeII\ features in Mrk~142.  With the \citet{boroson_green_1992} \FeII\ template model, the fits yielded lower (${\chi_{\nu}}^2$) values than with the \citet{veroncetty_etal_2001} template.  Following the procedure in \citet{hu_etal_2015}, we added single Gaussian profiles for each of the six coronal lines (\FeVIIlama, \FeVIlam, \CaVlam, \FeVIIlamb, \FeVIIlamc, \FeVIIlamd; see Figure~\ref{fig:mean_rms_gemini_ljt}).  In addition, we included the host-galaxy template with 11~Gyr at $z=0.05$ from the 2013 updated version of \citet{bruzual_charlot_2003} galaxy templates.  The host-galaxy template, affecting the redder part  of the spectrum more than the bluer, contributed greatly in producing a good fit to the \HeI\ emission-line region.  The fit in the \Hbeta\ region was less sensitive to host-galaxy emission.  We referred to the \citet{vandenberk_etal_2001} rest wavelengths for setting the positions of all emission lines.

\subsubsection{Gemini Spectral Analysis} \label{subsubsec:sherpa_modeling_gemini}

\begin{figure*}[ht!]
\epsscale{1.15}
\plotone{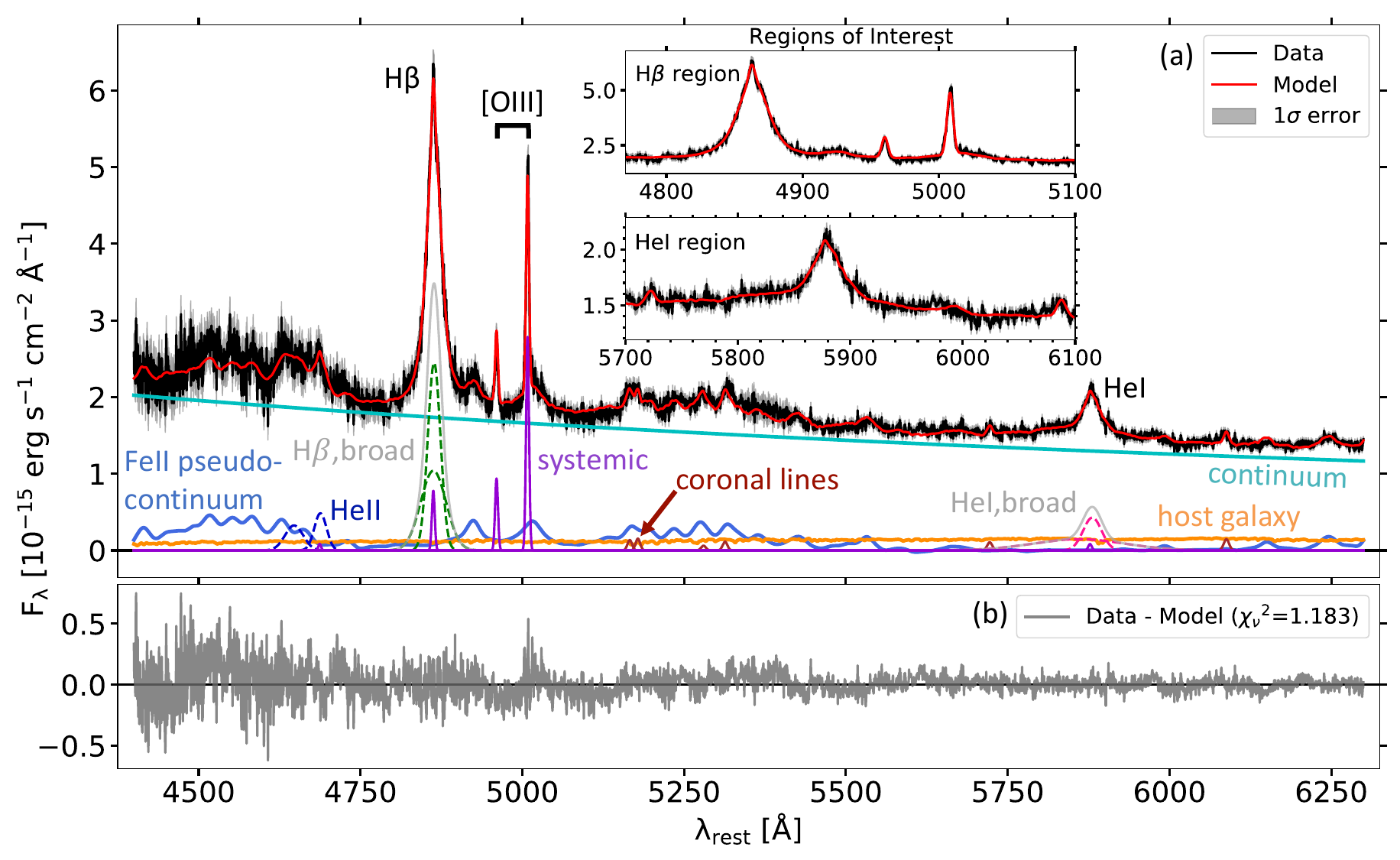}
\caption{Composite model fit to epoch 24 of the Mrk~142 Gemini data displaying individual components of the model.  Panel {\em a}: Composite model (red solid curve) fit to the data (black solid curve) from 4430~\AA\ to 6300~\AA\ is shown in the main panel, and the \Hbeta\ and \HeI\ {\em Regions of Interest} are shown in the {\em Inset} panels.  The individual components of the model are displayed at the bottom of the panel: continuum (cyan solid curve); \FeII\ I~Zw~1 template as a pseudo-continuum (faint blue solid curve); host-galaxy template (orange solid curve); \HeII\ broad components (blue dashed Gaussians); \Hbeta\ broad components (green dashed Gaussians); \HeI\ broad components (pink dashed Gaussians); narrow-line components of \Hbeta, \HeI, and \OIII\ (purple solid Gaussians); and high-ionization coronal lines (brown solid Gaussians).  The total broad \Hbeta\ and \HeI\ profiles are also overplotted (gray solid curves).  Panel {\em b}: Residuals of the model with ${\chi_{\nu}}^2=1.183$.  The model shows larger residuals around the \OIIIlamstrong\ line indicating a sub-optimal fit in that region.  The noisier blue end of the spectrum affects the overall fit in that region, thus resulting in larger residuals compared to the red end of the spectrum.
\label{fig:spectral_model_single_epoch_gemini}}
\end{figure*}

Our goal of spectral fitting was to accurately estimate the \Hbeta, \OIII, and \HeI\ profiles in the Gemini spectra.  We aimed at finding a robust and flexible set of parameters that fit the structure in the spectra over all epochs.  Figure~\ref{fig:spectral_model_single_epoch_gemini} shows the composite model fit to a single-epoch Gemini spectrum.  

We describe the fitting process as follows.  The Gaussian used for each of the \OIII\ doublet lines traces the systemic narrow-line emission peaks.  While fitting the Gemini spectra, the position, FWHM, and flux of the \OIIIlamstrong\ emission line were freed.  However, we fixed the position of the \OIIIlamweak\ line relative to the \OIIIlamstrong\ line and the flux in the \OIIIlamweak\ line to a factor of 1/3 compared to the \OIIIlamstrong\ flux.  We chose Gaussians over Lorentzians to appropriately trace the narrower wings of the \OIII\ lines.  We fixed the positions of the narrow components of \Hbeta, \HeI, and \HeII\ relative to the position of the \OIIIlamstrong\ line, and their widths equal to the \OIIIlamstrong\ FWHM.  Spectral fitting in {\tt\string Sherpa} takes into account the scaling of the FWHM for a given line relative to the calibrator line, which is the \OIIIlamstrong\ line for this work.  The position and flux parameters of the two broad components of both \Hbeta\ and \HeI\ were freed.  The width of only one of the \Hbeta\ Gaussians was allowed to vary while the second Gaussian was fixed at twice the width of the first.  On the other hand, the widths of the \HeI\ broad components were fixed at factors of 1 and 6 of the FWHM of the flexible, broad \Hbeta\ component.  We determined the above FWHM ratios for the broad \Hbeta\ and \HeI\ emission lines from the mean spectrum.  The insets in Figure~\ref{fig:spectral_model_single_epoch_gemini} show a closer view of the two {\em Regions of Interest} -- \Hbeta\ and \HeI.  A single broad component of \HeII\ with position fixed relative to the flexible, broad \Hbeta\ component and width equal to the FWHM of the same component proved insufficient to trace the broader emission around $\sim$4650~\AA, indicating a plausible blueshifted broad component of \HeII.  Adding another blueshifted Gaussian 1.5 times the width of the flexible, broad \Hbeta\ component with  flux equal to that of the first broad \HeII\ component significantly improved the fit in that region.  Table~\ref{tab:fitting_params} lists the emission-line parameters along with their settings as used during spectral fitting.

\begin{deluxetable*}{ccccc}
\tablewidth{700pt}
\tablenum{3}
\tablecaption{Emission-Line Fitting Parameters for Gemini and LJT Spectra
\label{tab:fitting_params}}
\tabletypesize{\scriptsize}
\tablehead{
 \\
\colhead{Line} & \colhead{Parameter} & \colhead{Fixed} & \multicolumn2c{Ratio relative to the fixed line} \\
 &  & relative to line\tablenotemark{a} & Gemini & LJT \\
}
\startdata
 & Position & \nodata & \nodata & \nodata \\
\OIIIlamstrong\ & FWHM & \nodata & \nodata & \nodata \\
 & Flux & \nodata & \nodata & \nodata \\
\hline
 & Position & \OIIIlamstrong\ & 0.990 & 0.990 \\
\OIIIlamweak\ & FWHM & \OIIIlamstrong\ & 1.000 & 1.000 \\
 & Flux & \OIIIlamstrong\ & 0.333 & 0.333 \\
\hline
 & Position & \OIIIlamstrong\ & 0.971 & 0.971  \\
Narrow \Hbetalam\ & FWHM & \OIIIlamstrong\ & 1.000 & 1.000 \\
 & Flux & \OIIIlamstrong\ & \nodata & 0.293 \\
\hline
 & Position & \nodata & \nodata & \nodata \\
Narrower broad \Hbetalam\ & FWHM & \nodata & \nodata & \nodata \\
 & Flux & \nodata & \nodata & \nodata \\
\hline
 & Position & \nodata & \nodata & \nodata \\
Broader broad \Hbetalam\ & FWHM & Narrower broad \Hbetalam\ & 2.000 & 2.500 \\
 & Flux & \nodata & \nodata & \nodata \\
\hline
 & Position & \OIIIlamstrong\ & 1.174 & 1.174 \\
Narrow \HeIlam\ & FWHM & \OIIIlamstrong\ & 1.000 & 1.000 \\
 & Flux & \OIIIlamstrong\ & \nodata & 0.034 \\
\hline
 & Position & \nodata & \nodata & \nodata \\
Narrower broad \HeIlam\ & FWHM & Narrower broad \Hbetalam\ & 1.200 & 1.000 \\
 & Flux & \nodata & \nodata & \nodata\\
\hline
 & Position & \nodata & \nodata & \nodata \\
Broader broad \HeIlam\ & FWHM & Narrower broad \Hbetalam\ & 6.000 & 6.000 \\
 & Flux & \nodata & \nodata & \nodata \\
\hline
 & Position & \OIIIlamstrong\ & 0.936 & 0.936 \\
Narrow \HeIIlamopt\ & FWHM & \OIIIlamstrong\ & 1.000 & 1.000 \\
 & Flux & \nodata & \nodata & \nodata \\
\hline
 & Position & Narrower broad \Hbetalam\ & 0.964 & 0.964 \\
Narrower broad \HeIIlamopt\ & FWHM & Narrower broad \Hbetalam\ & 1.000 & 1.000 \\
 & Flux & \nodata & \nodata & \nodata \\
\hline
 & Position & Narrower broad \Hbetalam\ & 0.956 & 0.956 \\
Blue-shifted broad \HeIIlamopt\ & FWHM & Narrower broad \Hbetalam\ & 1.500 & 1.500 \\
 & Flux & Narrower broad \HeIIlamopt\ & 1.000 & 1.000 \\
\enddata
\tablenotetext{a}{Parameter settings with no data indicate that the parameter was kept flexible during spectral fitting.}
\end{deluxetable*}
 
We followed the \FeII\ template fitting procedure described in \citet{hu_etal_2015}, where the \FeII\ emission is defined by a convolution of the \citet{boroson_green_1992} template with a Gaussian.  We applied the Gaussian as a 1D Point Spread Function (PSF) with a fixed FWHM, and the amplitude of the convolved \FeII\ model was set as a flexible parameter while fitting.  Although the \FeII\ model successfully traces the sharp \FeII\ features in most parts of the spectrum, it performs sub-optimally near the \FeII\ emission at the red wing of the \OIIIlamstrong\ line thus resulting in larger residuals in that region.  Further, the model overestimates the emission between the two \OIII\ lines due to the broader wing of \FeII\ from the template model.  

Modeling individual coronal lines in the spectra considerably improved the fit in the \FeII\ emission region from $\sim$5150~\AA\ to $\sim$5350~\AA.  In this region, the Fe coronal lines appeared to be slightly redshifted ($\leq$0.003) with respect to their rest wavelengths.  We set the coronal-line widths to 1.5 times the \OIIIlamstrong\ line width and their flux values to specific fractions of the \OIIIlamstrong\ flux, as determined from the fit to the Gemini mean spectrum.

From spectral modeling, we derived, at each epoch, the total FWHM and flux values of the \Hbeta, \HeI, and \OIII\ lines.  We measured the FWHM of the broad and total (including both the broad and the narrow components) \Hbeta\ and \HeI\ lines empirically by subtracting all other model components from the spectra including the narrow lines.  To calculate the contribution from the broad-line and total (again, including both the broad- and the narrow-line) flux in the \Hbeta\ and \HeI\ emission profiles, we simply added the contribution from each of their components.  Tables~\ref{tab:spectral_meas_oiii_gemini}, \ref{tab:spectral_meas_hbeta_gemini}, and \ref{tab:spectral_meas_hei_gemini} provide emission-line measurements for the Gemini spectra from 33 epochs.  For epoch 11, the model failed to constrain the broad \Hbeta\ emission as the region blueward of the \Hbeta\ line appeared noisier compared to the other epochs.  We therefore excluded epoch 11 from further analysis.  Also, due to improper flux calibration at the location of the \Hbeta\ line in epoch 25, the line appeared unusually broader and brighter than at the other epochs.  We therefore excluded the spectrum from epoch 25 as well.

\begin{deluxetable*}{rcccc}
\tablewidth{700pt}
\tablenum{4}
\tablecaption{\OIIIlamstrong\ Emission-Line Measurements for Gemini Spectra
\label{tab:spectral_meas_oiii_gemini}}
\tabletypesize{\scriptsize}
\tablehead{
 \\
\colhead{Epoch} & \colhead{Position$_{\OIII\lambda 5008}$} & \colhead{FWHM$_{\OIII\lambda 5008}$} & \colhead{$F_{\OIII\lambda 5008}$} & \colhead{$\chi_{\nu}^2$\tablenotemark{{\rm [}a{\rm ]}}} \\
 & [\AA] & [\kms] & [10$^{-15}$~erg~s$^{-1}$~cm$^{-2}$] & 
} 
\startdata
1 & \(5008.21\pm0.04\) & \(313\pm5\) & \(15.7\pm0.2\) & 1.07 \\ 
2 & \(5008.19\pm0.05\) & \(318\pm9\) & \(15.8\pm0.3\) & 1.06 \\ 
3 & \(5008.12\pm0.04\) & \(325\pm6\) & \(16.1\pm0.2\) & 1.23 \\ 
4 & \(5008.19\pm0.04\) & \(303\pm9\) & \(15.7\pm0.3\) & 1.21 \\ 
5 & \(5008.18\pm0.05\) & \(320\pm5\) & \(16.2\pm0.3\) & 0.99 \\ 
6 & \(5008.20\pm0.04\) & \(325\pm3\) & \(16.1\pm0.2\) & 1.08 \\ 
7 & \(5008.23\pm0.03\) & \(311\pm4\) & \(15.5\pm0.2\) & 1.18 \\ 
8 & \(5008.15\pm0.04\) & \(322\pm5\) & \(16.1\pm0.2\) & 1.22 \\ 
9 & \(5008.15\pm0.05\) & \(336\pm7\) & \(16.1\pm0.3\) & 0.97 \\ 
10 & \(5008.11\pm0.04\) & \(326\pm5\) & \(16.0\pm0.3\) & 1.04 \\ 
11 & \(5008.11\pm0.07\) & \(317\pm9\) & \(15.6\pm0.5\) & 0.99 \\ 
12 & \(5008.16\pm0.06\) & \(333\pm10\) & \(16.3\pm0.4\) & 1.08 \\ 
13 & \(5008.19\pm0.04\) & \(304\pm6\) & \(15.6\pm0.3\) & 1.12 \\ 
14 & \(5008.16\pm0.04\) & \(317\pm5\) & \(15.8\pm0.2\) & 1.24 \\ 
15 & \(5008.20\pm0.04\) & \(319\pm4\) & \(15.9\pm0.2\) & 1.20 \\ 
16 & \(5008.19\pm0.04\) & \(311\pm5\) & \(15.6\pm0.2\) & 1.26 \\ 
17 & \(5008.22\pm0.04\) & \(315\pm4\) & \(16.0\pm0.2\) & 1.27 \\ 
18 & \(5008.14\pm0.04\) & \(338\pm6\) & \(16.1\pm0.2\) & 1.33 \\ 
19 & \(5008.22\pm0.04\) & \(312\pm2\) & \(15.8\pm0.2\) & 1.10 \\ 
20 & \(5008.21\pm0.03\) & \(292\pm6\) & \(15.5\pm0.2\) & 1.17 \\ 
21 & \(5008.23\pm0.04\) & \(313\pm6\) & \(16.0\pm0.2\) & 1.28 \\ 
22 & \(5008.18\pm0.03\) & \(308\pm6\) & \(15.8\pm0.2\) & 1.21 \\ 
23 & \(5008.20\pm0.03\) & \(304\pm4\) & \(15.6\pm0.2\) & 1.24 \\ 
24 & \(5008.15\pm0.04\) & \(320\pm3\) & \(15.8\pm0.2\) & 1.18 \\ 
25 & \(5008.20\pm0.05\) & \(307\pm5\) & \(15.9\pm0.3\) & 1.11 \\ 
26 & \(5008.20\pm0.05\) & \(310\pm7\) & \(15.9\pm0.3\) & 1.19 \\ 
27 & \(5008.21\pm0.04\) & \(302\pm5\) & \(15.5\pm0.2\) & 1.24 \\ 
28 & \(5008.34\pm0.05\) & \(301\pm2\) & \(15.3\pm0.3\) & 1.07 \\ 
29 & \(5008.18\pm0.04\) & \(323\pm7\) & \(16.1\pm0.3\) & 1.19 \\ 
30 & \(5008.24\pm0.04\) & \(321\pm4\) & \(15.9\pm0.2\) & 1.35 \\ 
31 & \(5008.27\pm0.03\) & \(306\pm5\) & \(15.7\pm0.2\) & 1.16 \\ 
32 & \(5008.25\pm0.04\) & \(321\pm6\) & \(16.0\pm0.3\) & 1.10 \\ 
33 & \(5008.22\pm0.05\) & \(328\pm4\) & \(16.4\pm0.2\) & 1.05 \\
\enddata
\tablenotetext{a}{Reduced $\chi^2$, \(\chi_{\nu}^2 = \chi^2/\nu\), where $\nu$ indicates 3897 degrees of freedom, gives the model statistic for individual epochs.}
\tablecomments{This table is available in machine-readable form.}
\end{deluxetable*} 
\begin{deluxetable*}{rcccccc}
\tablewidth{700pt}
\tablenum{5}
\tablecaption{\Hbetalam\ Emission-Line Measurements for Gemini Spectra
\label{tab:spectral_meas_hbeta_gemini}}
\tabletypesize{\scriptsize}
\tablehead{
\colhead{Epoch} & \colhead{FWHM$_{\Hbeta,b}$} & \colhead{$F_{\Hbeta,b}$} & \colhead{FWHM$_{\Hbeta,n}$} & \colhead{$F_{\Hbeta,n}$} & \colhead{FWHM$_{\Hbeta,t}$} &   \colhead{$F_{\Hbeta,t}$} \\
 & \colhead{[\kms]} & \colhead{[10$^{-15}$~erg~s$^{-1}$~cm$^{-2}$]} & \colhead{[\kms]} & \colhead{[10$^{-15}$~erg~s$^{-1}$~cm$^{-2}$]} & \colhead{[\kms]} & \colhead{[10$^{-15}$~erg~s$^{-1}$~cm$^{-2}$]}
}
\startdata
1 & \(1777\pm72\) & \(112.2\pm2.4\) & \(312\pm5\) & \(5.3\pm0.4\) & \(1445\pm52\) & \(117.5\pm2.4\) \\ 
2 & \(1681\pm109\) & \(106.3\pm4.3\) & \(317\pm8\) & \(5.5\pm0.6\) & \(1324\pm83\) & \(111.8\pm4.3\) \\ 
3 & \(1703\pm69\) & \(115.8\pm5.1\) & \(325\pm5\) & \(5.1\pm0.4\) & \(1412\pm73\) & \(120.9\pm5.1\) \\ 
4 & \(1745\pm83\) & \(110.5\pm2.7\) & \(303\pm9\) & \(4.5\pm0.4\) & \(1478\pm67\) & \(115.0\pm2.7\) \\ 
5 & \(1555\pm80\) & \(107.5\pm3.4\) & \(319\pm4\) & \(3.1\pm0.6\) & \(1466\pm96\) & \(110.6\pm3.4\) \\ 
6 & \(1830\pm77\) & \(109.3\pm3.7\) & \(324\pm3\) & \(5.1\pm0.5\) & \(1524\pm80\) & \(114.4\pm3.7\) \\ 
7 & \(1797\pm69\) & \(112.2\pm2.2\) & \(310\pm3\) & \(4.5\pm0.4\) & \(1525\pm56\) & \(116.7\pm2.2\) \\ 
8 & \(1697\pm62\) & \(106.5\pm2.3\) & \(321\pm4\) & \(3.5\pm0.4\) & \(1447\pm62\) & \(110.0\pm2.4\) \\ 
9 & \(1499\pm83\) & \(100.2\pm3.1\) & \(336\pm6\) & \(5.1\pm0.6\) & \(1379\pm58\) & \(105.4\pm3.2\) \\ 
10 & \(1593\pm77\) & \(103.4\pm2.7\) & \(326\pm4\) & \(5.0\pm0.4\) & \(1328\pm55\) & \(108.4\pm2.8\) \\ 
11\tablenotemark{a} & \(1637\pm132\) & \(104.3\pm3.3\) & \(317\pm9\) & \(5.7\pm0.8\) & \(1235\pm165\) & \(110.1\pm3.4\) \\ 
12 & \(1732\pm120\) & \(101.5\pm5.8\) & \(332\pm10\) & \(5.9\pm0.7\) & \(1322\pm81\) & \(107.3\pm5.8\) \\ 
13 & \(1628\pm88\) & \(101.3\pm2.5\) & \(303\pm6\) & \(3.7\pm0.5\) & \(1354\pm77\) & \(105.0\pm2.5\) \\ 
14 & \(1707\pm69\) & \(109.1\pm2.6\) & \(316\pm5\) & \(4.2\pm0.4\) & \(1384\pm56\) & \(113.3\pm2.7\) \\ 
15 & \(1672\pm104\) & \(108.8\pm2.7\) & \(319\pm3\) & \(4.9\pm0.4\) & \(1341\pm30\) & \(113.7\pm2.7\) \\ 
16 & \(1780\pm76\) & \(103.2\pm2.3\) & \(311\pm5\) & \(4.4\pm0.4\) & \(1412\pm73\) & \(107.6\pm2.3\) \\ 
17 & \(1623\pm65\) & \(108.4\pm3.0\) & \(315\pm4\) & \(3.8\pm0.4\) & \(1402\pm60\) & \(112.2\pm3.0\) \\ 
18 & \(1712\pm72\) & \(105.6\pm2.4\) & \(337\pm5\) & \(4.6\pm0.5\) & \(1477\pm64\) & \(110.2\pm2.4\) \\ 
19 & \(1579\pm91\) & \(106.1\pm4.2\) & \(311\pm1\) & \(6.1\pm0.5\) & \(1348\pm64\) & \(112.2\pm4.2\) \\ 
20 & \(1558\pm62\) & \(115.2\pm2.1\) & \(291\pm5\) & \(3.4\pm0.4\) & \(1473\pm42\) & \(118.7\pm2.1\) \\ 
21 & \(1604\pm43\) & \(112.6\pm2.1\) & \(312\pm5\) & \(3.6\pm0.4\) & \(1446\pm71\) & \(116.2\pm2.1\) \\ 
22 & \(1698\pm62\) & \(109.3\pm2.8\) & \(307\pm5\) & \(4.4\pm0.4\) & \(1442\pm55\) & \(113.7\pm2.8\) \\ 
23 & \(1590\pm57\) & \(108.3\pm2.1\) & \(304\pm3\) & \(3.7\pm0.4\) & \(1416\pm52\) & \(112.0\pm2.2\) \\ 
24 & \(1668\pm77\) & \(110.9\pm2.7\) & \(319\pm3\) & \(4.3\pm0.5\) & \(1391\pm54\) & \(115.2\pm2.7\) \\ 
25\tablenotemark{b} & \(1672\pm73\) & \(140.4\pm3.6\) & \(307\pm5\) & \(5.9\pm0.7\) & \(1603\pm83\) & \(146.3\pm3.7\) \\ 
26 & \(1655\pm95\) & \(107.0\pm2.6\) & \(309\pm7\) & \(4.9\pm0.4\) & \(1297\pm50\) & \(111.9\pm2.6\) \\ 
27 & \(1649\pm48\) & \(103.4\pm3.9\) & \(301\pm5\) & \(4.6\pm0.4\) & \(1356\pm59\) & \(108.0\pm3.9\) \\ 
28 & \(1678\pm96\) & \(102.6\pm3.2\) & \(300\pm2\) & \(3.9\pm0.6\) & \(1459\pm148\) & \(106.5\pm3.3\) \\ 
29 & \(1713\pm76\) & \(94.9\pm3.1\) & \(322\pm7\) & \(4.3\pm0.4\) & \(1410\pm66\) & \(99.2\pm3.1\) \\ 
30 & \(1754\pm96\) & \(102.7\pm3.7\) & \(321\pm4\) & \(5.5\pm0.5\) & \(1395\pm92\) & \(108.2\pm3.8\) \\ 
31 & \(1816\pm81\) & \(109.4\pm4.3\) & \(306\pm5\) & \(6.1\pm0.4\) & \(1489\pm96\) & \(115.5\pm4.3\) \\ 
32 & \(1682\pm79\) & \(109.6\pm2.3\) & \(321\pm5\) & \(4.9\pm0.4\) & \(1423\pm47\) & \(114.5\pm2.4\) \\ 
33 & \(1699\pm58\) & \(110.0\pm5.0\) & \(328\pm4\) & \(3.8\pm0.5\) & \(1515\pm78\) & \(113.8\pm5.0\) \\
\enddata
\tablenotetext{a}{Because the noisy region blueward of the \Hbeta\ emission line was unable to well constrain the broad, blue wing of \Hbeta, this epoch was excluded from further analysis.} 
\tablenotetext{b}{Due to a calibration issue at the location of the \Hbeta\ emission line, the \Hbeta\ profile appeared unusually broader and brighter than in other spectra.  Therefore, this epoch was excluded from further analysis.}
\tablecomments{The second and the third columns providing the FWHM and flux values, respectively, for the broad (`{\em b}') \Hbeta\ component include contributions from both the broad Gaussians defined for the line. The FWHM of the narrow (`{\em n}') \Hbeta\ is equal to the FWHM of the \OIIIlamstrong\ (see Table~\ref{tab:spectral_meas_oiii_gemini}, third column).  The total (`{\em t}') FWHM and flux include contributions from both the broad and the narrow components ({\em t} = {\em b} + {\em n}).}
\vspace{4pt}
\tablecomments{This table is available in machine-readable form.}
\end{deluxetable*}

\begin{deluxetable*}{rcccccc}
\tablewidth{700pt}
\tablenum{6}
\tablecaption{\HeIlam\ Emission-Line Measurements for Gemini Spectra
\label{tab:spectral_meas_hei_gemini}}
\tabletypesize{\scriptsize}
\tablehead{
\colhead{Epoch} & \colhead{FWHM$_{\HeI,b}$} & \colhead{$F_{\HeI,b}$} & \colhead{FWHM$_{\HeI,n}$} & \colhead{$F_{\HeI,n}$} & \colhead{FWHM$_{\HeI,t}$} & \colhead{$F_{\HeI,t}$} \\
 & \colhead{[\kms]} & \colhead{[10$^{-15}$~erg~s$^{-1}$~cm$^{-2}$]} & \colhead{[\kms]} & \colhead{[10$^{-15}$~erg~s$^{-1}$~cm$^{-2}$]} & \colhead{[\kms]} & \colhead{[10$^{-15}$~erg~s$^{-1}$~cm$^{-2}$]}
}
\startdata
1 & \(47\pm37\) & \(34.4\pm0.8\) & \(312\pm5\) & \(0.7\pm0.1\) & \(47\pm95\) & \(35.1\pm0.8\) \\ 
2 & \(53\pm37\) & \(33.4\pm1.0\) & \(317\pm8\) & \(0.6\pm0.2\) & \(53\pm36\) & \(34.0\pm1.0\) \\ 
3 & \(90\pm44\) & \(41.3\pm0.9\) & \(325\pm5\) & \(0.7\pm0.1\) & \(90\pm41\) & \(42.0\pm0.9\) \\ 
4 & \(36\pm25\) & \(36.0\pm0.9\) & \(303\pm9\) & \(0.8\pm0.1\) & \(36\pm20\) & \(36.8\pm0.9\) \\ 
5 & \(335\pm70\) & \(35.2\pm1.1\) & \(319\pm4\) & \(0.3\pm0.2\) & \(335\pm71\) & \(35.5\pm1.1\) \\ 
6 & \(46\pm36\) & \(35.1\pm0.9\) & \(324\pm3\) & \(0.6\pm0.1\) & \(46\pm35\) & \(35.7\pm0.9\) \\ 
7 & \(136\pm170\) & \(39.7\pm0.8\) & \(310\pm3\) & \(0.6\pm0.1\) & \(1962\pm312\) & \(40.3\pm0.8\) \\ 
8 & \(49\pm65\) & \(39.1\pm0.7\) & \(321\pm4\) & \(0.5\pm0.1\) & \(49\pm59\) & \(39.6\pm0.7\) \\ 
9 & \(34\pm43\) & \(33.1\pm0.9\) & \(336\pm6\) & \(0.7\pm0.2\) & \(34\pm46\) & \(33.7\pm0.9\) \\ 
10 & \(92\pm53\) & \(36.3\pm0.8\) & \(326\pm4\) & \(0.7\pm0.2\) & \(92\pm52\) & \(37.0\pm0.8\) \\ 
11\tablenotemark{a} & \(153\pm39\) & \(36.6\pm1.4\) & \(317\pm9\) & \(0.4\pm0.2\) & \(153\pm38\) & \(37.1\pm1.4\) \\ 
12 & \(92\pm51\) & \(31.4\pm1.3\) & \(332\pm10\) & \(0.5\pm0.2\) & \(92\pm52\) & \(32.0\pm1.3\) \\ 
13 & \(1419\pm90\) & \(35.9\pm0.9\) & \(303\pm6\) & \(0.0\pm0.2\) & \(1419\pm108\) & \(35.9\pm0.9\) \\ 
14 & \(83\pm51\) & \(41.0\pm0.8\) & \(316\pm5\) & \(0.8\pm0.1\) & \(83\pm55\) & \(41.8\pm0.8\) \\ 
15 & \(66\pm41\) & \(38.4\pm0.8\) & \(319\pm3\) & \(0.5\pm0.1\) & \(66\pm41\) & \(38.9\pm0.8\) \\ 
16 & \(152\pm225\) & \(38.8\pm0.8\) & \(311\pm5\) & \(0.6\pm0.1\) & \(1758\pm497\) & \(39.4\pm0.8\) \\ 
17 & \(1825\pm303\) & \(38.8\pm0.9\) & \(315\pm4\) & \(0.3\pm0.1\) & \(1528\pm415\) & \(39.1\pm0.9\) \\ 
18 & \(37\pm32\) & \(37.2\pm0.8\) & \(337\pm5\) & \(0.4\pm0.1\) & \(37\pm181\) & \(37.6\pm0.9\) \\ 
19 & \(1828\pm42\) & \(35.4\pm0.9\) & \(311\pm1\) & \(0.7\pm0.1\) & \(1806\pm41\) & \(36.1\pm0.9\) \\ 
20 & \(1862\pm357\) & \(42.0\pm0.7\) & \(291\pm5\) & \(0.4\pm0.1\) & \(1829\pm631\) & \(42.4\pm0.7\) \\ 
21 & \(1798\pm399\) & \(38.4\pm0.8\) & \(312\pm5\) & \(0.9\pm0.1\) & \(1651\pm647\) & \(39.3\pm0.8\) \\ 
22 & \(1870\pm654\) & \(41.9\pm0.8\) & \(307\pm5\) & \(0.4\pm0.1\) & \(1870\pm682\) & \(42.2\pm0.8\) \\ 
23 & \(38\pm189\) & \(40.2\pm0.8\) & \(304\pm3\) & \(0.5\pm0.1\) & \(1873\pm336\) & \(40.7\pm0.8\) \\ 
24 & \(54\pm84\) & \(41.1\pm0.9\) & \(319\pm3\) & \(0.5\pm0.1\) & \(54\pm132\) & \(41.7\pm0.9\) \\ 
25\tablenotemark{b} & \(137\pm43\) & \(38.3\pm1.3\) & \(307\pm5\) & \(0.6\pm0.2\) & \(137\pm45\) & \(38.9\pm1.3\) \\ 
26 & \(61\pm53\) & \(36.9\pm0.8\) & \(309\pm7\) & \(0.5\pm0.1\) & \(61\pm110\) & \(37.4\pm0.8\) \\ 
27 & \(114\pm180\) & \(38.7\pm0.9\) & \(301\pm5\) & \(0.5\pm0.1\) & \(1935\pm389\) & \(39.3\pm0.9\) \\ 
28 & \(81\pm60\) & \(55.6\pm1.1\) & \(300\pm2\) & \(0.1\pm0.2\) & \(81\pm52\) & \(55.7\pm1.1\) \\ 
29 & \(181\pm100\) & \(36.0\pm0.8\) & \(322\pm7\) & \(0.4\pm0.1\) & \(181\pm96\) & \(36.4\pm0.8\) \\ 
30 & \(34\pm86\) & \(40.0\pm1.0\) & \(321\pm4\) & \(0.5\pm0.1\) & \(34\pm125\) & \(40.5\pm1.0\) \\ 
31 & \(66\pm268\) & \(44.6\pm0.9\) & \(306\pm5\) & \(0.9\pm0.1\) & \(66\pm627\) & \(45.5\pm0.9\) \\ 
32 & \(294\pm50\) & \(41.2\pm0.9\) & \(321\pm5\) & \(0.9\pm0.1\) & \(1631\pm160\) & \(42.1\pm0.9\) \\ 
33 & \(1537\pm161\) & \(35.3\pm0.9\) & \(328\pm4\) & \(0.3\pm0.1\) & \(1528\pm267\) & \(35.6\pm0.9\) \\
\enddata
\tablenotetext{a}{Excluded from further analysis.  See corresponding note in Table~\ref{tab:spectral_meas_hbeta_gemini}.}
\tablenotetext{b}{Excluded from further analysis.  See corresponding note in Table~\ref{tab:spectral_meas_hbeta_gemini}.}
\tablecomments{The second and the third columns providing the FWHM and flux values, respectively, for the broad (`{\em b}') \HeI\ component include contributions from both the broad Gaussians defined for the line. The FWHM of the narrow (`{\em n}') \HeI\ is equal to the FWHM of the \OIIIlamstrong\ (see Table~\ref{tab:spectral_meas_oiii_gemini}, third column).  The total (`{\em t}') FWHM and flux include contributions from both the broad and the narrow components ({\em t} = {\em b} + {\em n}).}
\tablecomments{This table is available in machine-readable form.}
\end{deluxetable*}

\subsubsection{LJT Spectral Analysis} \label{subsubsec:sherpa_modeling_ljt}

\begin{figure*}[ht!]
\epsscale{1.1}
\plotone{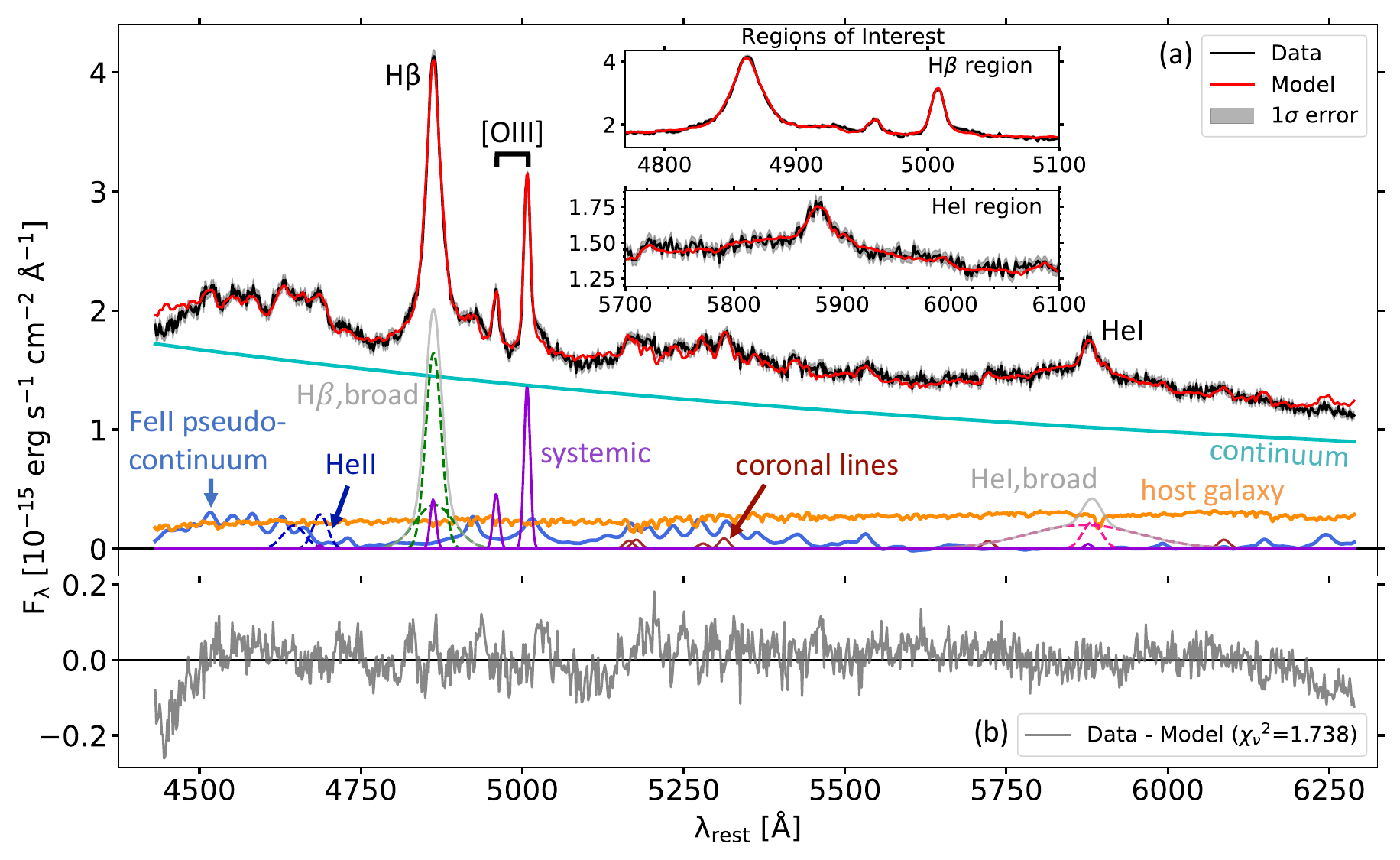}
\caption{Composite model fit to epoch 24 of the Mrk~142 LJT data displaying individual components of the model.  See caption of Figure~\ref{fig:spectral_model_single_epoch_gemini} for a description of the individual model components in Panel {\em a}.  The red side of the broad \Hbeta\ emission line shows contamination with the \FeII\ emission at $\sim$4923~\AA.  Similarly, the \OIIIlamstrong\ line shows considerable blending with the \FeII\ feature in its red wing, thus affecting a reliable measurement of the \OIIIlamstrong\ line.  Panel {\em b} shows the residuals of the model with ${\chi_{\nu}}^2=1.711$.  The smaller residuals indicate an overall good fit to the spectrum.  The model performance drops significantly at both the end of the spectrum although it does not impact measurements in the {\em Regions of Interest}.
\label{fig:spectral_model_single_epoch_ljt}}
\end{figure*}

We fit the 69 LJT spectra with the same goal of modeling the \Hbeta, \OIII, and \HeI\ lines accurately.  Figure~\ref{fig:spectral_model_single_epoch_ljt} shows the fit to a single-epoch LJT spectrum.

We adopted the same model as for the Gemini spectra with small modifications to the ratios of certain fixed parameters.  We determined the flux ratios for the coronal lines relative to the \OIIIlamstrong\ line flux from the fit to the mean LJT spectrum.  The width of the fixed broad \Hbeta\ Gaussian was fixed at a factor of 2.5 (instead of 2 for the Gemini spectra).  A factor of 2 for this second \Hbeta\ Gaussian in LJT spectra was insufficient to trace the broad wing of \Hbeta, which also affected the fit to the blended \FeII\ feature at $\sim$4923~\AA.  Therefore, a broader \Hbeta\ component was required to generate a good fit in that region.  This indicates an interplay between the broad \Hbeta\ and \FeII\ line emission in the fitting process.  Similarly, the \OIIIlamstrong\ appears to be blended with the \FeII\ emission feature at its red wing.  This is caused by the instrumental broadening in LJT spectra that further resulted in wider \OIII\ FWHM measurements than typically expected for \OIII\ in NLS1 objects.  To contain the effect of the ``broader'' \OIII\ FWHM measurements on the \Hbeta\ and \HeI\ line measurements, we fixed the narrow-line flux ratios of \Hbeta\ to \OIIIlamstrong\ and \HeI\ to \OIIIlamstrong\ from the Gemini spectral measurements (see Table~\ref{tab:fitting_params}).  Tables~\ref{tab:spectral_meas_oiii_ljt}, \ref{tab:spectral_meas_hbeta_ljt}, and \ref{tab:spectral_meas_hei_ljt} provide emission-line measurements for the LJT spectra from 69 epochs.

\clearpage

\startlongtable
\begin{deluxetable*}{rcccc}
\tablewidth{700pt}
\tablenum{7}
\tablecaption{\OIIIlamstrong\ Emission-Line Measurements for LJT Spectra
\label{tab:spectral_meas_oiii_ljt}}
\tabletypesize{\scriptsize}
\tablehead{
 \\
\colhead{Epoch} & \colhead{Position$_{\OIII\lambda 5008}$} & \colhead{FWHM$_{\OIII\lambda 5008}$} & \colhead{$F_{\OIII\lambda 5008}$} & \colhead{$\chi_{\nu}^2$\tablenotemark{{\rm [}a{\rm ]}}} \\
 & [\AA] & [\kms] & [10$^{-15}$~erg~s$^{-1}$~cm$^{-2}$] & 
} 
\startdata
1 & \(5007.25\pm0.09\) & \(718\pm9\) & \(16.6\pm0.2\) & 2.27 \\ 
2 & \(5007.22\pm0.13\) & \(754\pm22\) & \(17.5\pm0.4\) & 1.45 \\ 
3 & \(5007.25\pm0.13\) & \(741\pm13\) & \(17.2\pm0.3\) & 2.00 \\ 
4 & \(5007.40\pm0.12\) & \(787\pm19\) & \(17.8\pm0.4\) & 2.03 \\ 
5 & \(5007.33\pm0.10\) & \(716\pm18\) & \(16.1\pm0.4\) & 2.07 \\ 
6 & \(5007.43\pm0.13\) & \(740\pm21\) & \(16.4\pm0.4\) & 1.71 \\ 
7 & \(5007.44\pm0.16\) & \(880\pm6\) & \(19.6\pm0.4\) & 1.60 \\ 
8 & \(5007.46\pm0.06\) & \(691\pm13\) & \(16.9\pm0.3\) & 2.56 \\ 
9 & \(5007.42\pm0.10\) & \(679\pm19\) & \(17.1\pm0.3\) & 1.68 \\ 
10 & \(5007.51\pm0.17\) & \(707\pm27\) & \(17.4\pm0.6\) & 1.28 \\ 
11 & \(5007.43\pm0.14\) & \(742\pm\cdots\) & \(17.6\pm0.3\) & 1.51 \\ 
12 & \(5007.26\pm0.09\) & \(713\pm10\) & \(16.7\pm0.3\) & 2.19 \\ 
13 & \(5007.41\pm0.20\) & \(716\pm27\) & \(17.2\pm0.6\) & 1.51 \\ 
14 & \(5007.49\pm0.12\) & \(757\pm13\) & \(16.8\pm0.3\) & 1.67 \\ 
15 & \(5007.57\pm0.14\) & \(737\pm22\) & \(16.8\pm0.4\) & 1.42 \\ 
16 & \(5007.53\pm0.10\) & \(682\pm16\) & \(17.4\pm0.3\) & 1.66 \\ 
17 & \(5007.44\pm0.10\) & \(682\pm2\) & \(15.7\pm0.2\) & 1.98 \\ 
18 & \(5007.46\pm0.15\) & \(736\pm27\) & \(16.3\pm0.5\) & 1.20 \\ 
19 & \(5007.65\pm0.14\) & \(752\pm15\) & \(16.6\pm0.3\) & 1.93 \\ 
20 & \(5007.45\pm0.15\) & \(727\pm28\) & \(17.2\pm0.6\) & 1.24 \\ 
21 & \(5007.53\pm0.14\) & \(758\pm17\) & \(17.1\pm0.4\) & 1.40 \\ 
22 & \(5007.67\pm0.17\) & \(802\pm16\) & \(17.1\pm0.4\) & 1.47 \\ 
23 & \(5007.16\pm0.23\) & \(772\pm26\) & \(17.7\pm0.7\) & 1.21 \\ 
24 & \(5007.56\pm0.10\) & \(674\pm17\) & \(16.5\pm0.3\) & 1.74 \\ 
25 & \(5007.57\pm0.11\) & \(700\pm26\) & \(16.9\pm0.4\) & 1.66 \\ 
26 & \(5007.55\pm0.11\) & \(733\pm14\) & \(17.1\pm0.3\) & 1.77 \\ 
27 & \(5007.45\pm0.09\) & \(680\pm4\) & \(16.8\pm0.3\) & 2.02 \\ 
28 & \(5007.66\pm0.14\) & \(801\pm6\) & \(16.9\pm0.3\) & 1.45 \\ 
29 & \(5007.48\pm0.10\) & \(785\pm16\) & \(18.6\pm0.3\) & 1.96 \\ 
30 & \(5007.60\pm0.14\) & \(742\pm22\) & \(17.7\pm0.5\) & 1.34 \\ 
31 & \(5007.46\pm0.10\) & \(707\pm11\) & \(16.1\pm0.3\) & 2.25 \\ 
32 & \(5007.38\pm0.08\) & \(699\pm5\) & \(16.3\pm0.4\) & 1.32 \\ 
33 & \(5007.48\pm0.13\) & \(684\pm17\) & \(16.5\pm0.4\) & 1.80 \\ 
34 & \(5007.52\pm0.11\) & \(683\pm17\) & \(17.1\pm0.4\) & 1.76 \\ 
35 & \(5007.38\pm0.11\) & \(776\pm15\) & \(16.8\pm0.3\) & 2.82 \\ 
36 & \(5007.39\pm0.06\) & \(681\pm2\) & \(16.5\pm0.2\) & 4.36 \\ 
37 & \(5007.44\pm0.09\) & \(682\pm0\) & \(16.5\pm0.2\) & 2.13 \\ 
38 & \(5007.66\pm0.13\) & \(810\pm9\) & \(17.3\pm0.4\) & 1.47 \\ 
39 & \(5007.65\pm0.24\) & \(719\pm49\) & \(17.2\pm1.0\) & 2.02 \\ 
40 & \(5007.63\pm0.12\) & \(719\pm16\) & \(16.9\pm0.4\) & 1.71 \\ 
41 & \(5007.56\pm0.08\) & \(728\pm9\) & \(16.9\pm0.2\) & 4.65 \\ 
42 & \(5007.61\pm0.09\) & \(688\pm2\) & \(16.9\pm0.2\) & 3.74 \\ 
43 & \(5007.41\pm0.14\) & \(713\pm27\) & \(17.2\pm0.5\) & 1.52 \\ 
44 & \(5007.48\pm0.16\) & \(712\pm45\) & \(17.4\pm0.9\) & 1.63 \\ 
45 & \(5007.51\pm0.13\) & \(745\pm18\) & \(18.3\pm0.4\) & 1.53 \\ 
46 & \(5007.47\pm0.10\) & \(731\pm14\) & \(16.8\pm0.3\) & 1.98 \\ 
47 & \(5007.57\pm0.22\) & \(744\pm36\) & \(16.5\pm0.7\) & 1.05 \\ 
48 & \(5007.60\pm0.12\) & \(685\pm26\) & \(16.1\pm0.5\) & 2.26 \\ 
49 & \(5008.33\pm0.21\) & \(857\pm30\) & \(18.8\pm0.6\) & 2.06 \\ 
50 & \(5007.51\pm0.09\) & \(691\pm31\) & \(17.1\pm0.5\) & 3.21 \\ 
51 & \(5007.76\pm0.08\) & \(726\pm11\) & \(16.7\pm0.2\) & 3.28 \\ 
52 & \(5007.64\pm0.12\) & \(686\pm31\) & \(16.1\pm0.5\) & 1.61 \\ 
53 & \(5007.44\pm0.11\) & \(707\pm19\) & \(17.1\pm0.4\) & 1.82 \\ 
54 & \(5007.64\pm0.11\) & \(679\pm15\) & \(16.6\pm0.4\) & 1.59 \\ 
55\tablenotemark{b} & \(5007.50\pm0.23\) & \(670\pm26\) & \(14.5\pm0.6\) & 1.97 \\ 
56 & \(5007.40\pm0.12\) & \(646\pm31\) & \(15.8\pm0.4\) & 1.53 \\ 
57 & \(5007.39\pm0.08\) & \(634\pm1\) & \(16.1\pm0.2\) & 2.55 \\ 
58 & \(5007.55\pm0.08\) & \(711\pm8\) & \(17.4\pm0.2\) & 3.54 \\ 
59 & \(5007.62\pm0.08\) & \(724\pm12\) & \(17.7\pm0.3\) & 3.37 \\ 
60 & \(5007.57\pm0.16\) & \(726\pm21\) & \(17.5\pm0.5\) & 1.48 \\ 
61 & \(5007.39\pm0.14\) & \(668\pm27\) & \(17.2\pm0.6\) & 1.50 \\ 
62 & \(5007.60\pm0.13\) & \(750\pm14\) & \(18.6\pm0.3\) & 2.58 \\ 
63 & \(5007.68\pm0.09\) & \(659\pm14\) & \(16.4\pm0.3\) & 2.02 \\ 
64 & \(5007.32\pm0.08\) & \(660\pm11\) & \(16.4\pm0.3\) & 3.39 \\ 
65 & \(5007.70\pm0.08\) & \(727\pm12\) & \(17.4\pm0.2\) & 3.46 \\ 
66 & \(5007.55\pm0.09\) & \(681\pm12\) & \(17.1\pm0.3\) & 2.27 \\ 
67 & \(5007.63\pm0.10\) & \(709\pm14\) & \(17.0\pm0.3\) & 2.03 \\ 
68 & \(5007.70\pm0.10\) & \(756\pm3\) & \(17.4\pm0.3\) & 1.73 \\ 
69 & \(5007.59\pm0.17\) & \(752\pm25\) & \(17.3\pm0.4\) & 1.78 \\
\enddata
\tablenotetext{a}{Reduced $\chi^2$, \(\chi_{\nu}^2 = \chi^2/\nu\), where $\nu$ indicates 1071 degrees of freedom, gives the model statistic for individual epochs.}
\tablenotetext{b}{This spectrum appeared very noisy likely due to some disturbance in the field of view at the time of observation.  Therefore, we excluded this epoch from further analysis.}
\tablecomments{This table is available in machine-readable form.}
\end{deluxetable*} 
\startlongtable
\begin{deluxetable*}{rcccccc}
\tablewidth{700pt}
\tablenum{8}
\tablecaption{\Hbetalam\ Emission-Line Measurements for LJT Spectra
\label{tab:spectral_meas_hbeta_ljt}}
\tabletypesize{\scriptsize}
\tablehead{
\colhead{Epoch} & \colhead{FWHM$_{\Hbeta,b}$} & \colhead{$F_{\Hbeta,b}$} & \colhead{FWHM$_{\Hbeta,n}$} & \colhead{$F_{\Hbeta,n}$} & \colhead{FWHM$_{\Hbeta,t}$} & \colhead{$F_{\Hbeta,t}$} \\
 & \colhead{[\kms]} & \colhead{[10$^{-15}$~erg~s$^{-1}$~cm$^{-2}$]} & \colhead{[\kms]} & \colhead{[10$^{-15}$~erg~s$^{-1}$~cm$^{-2}$]} & \colhead{[\kms]} & \colhead{[10$^{-15}$~erg~s$^{-1}$~cm$^{-2}$]}
}
\startdata
1 & \(1912\pm52\) & \(83.5\pm1.5\) & \(718\pm9\) & \(4.85\pm0.07\) & \(1660\pm42\) & \(88.3\pm1.5\) \\ 
2 & \(1933\pm73\) & \(82.7\pm3.5\) & \(754\pm22\) & \(5.13\pm0.13\) & \(1702\pm46\) & \(87.8\pm3.5\) \\ 
3 & \(1951\pm38\) & \(83.3\pm1.6\) & \(741\pm13\) & \(5.03\pm0.09\) & \(1748\pm48\) & \(88.3\pm1.6\) \\ 
4 & \(1901\pm52\) & \(84.9\pm2.8\) & \(787\pm19\) & \(5.22\pm0.11\) & \(1700\pm69\) & \(90.1\pm2.8\) \\ 
5 & \(1881\pm46\) & \(80.3\pm3.8\) & \(716\pm18\) & \(4.72\pm0.10\) & \(1706\pm46\) & \(85.0\pm3.8\) \\ 
6 & \(1958\pm85\) & \(82.2\pm3.1\) & \(740\pm21\) & \(4.80\pm0.12\) & \(1708\pm47\) & \(87.0\pm3.1\) \\ 
7 & \(2055\pm86\) & \(82.6\pm2.3\) & \(880\pm6\) & \(5.74\pm0.11\) & \(1750\pm69\) & \(88.3\pm2.3\) \\ 
8 & \(1956\pm54\) & \(81.3\pm2.4\) & \(691\pm13\) & \(4.95\pm0.09\) & \(1804\pm71\) & \(86.3\pm2.4\) \\ 
9 & \(1884\pm63\) & \(78.3\pm1.5\) & \(679\pm19\) & \(5.02\pm0.10\) & \(1618\pm33\) & \(83.4\pm1.5\) \\ 
10 & \(1988\pm85\) & \(85.2\pm4.7\) & \(707\pm27\) & \(5.10\pm0.17\) & \(1742\pm82\) & \(90.3\pm4.7\) \\ 
11 & \(1985\pm102\) & \(78.2\pm4.9\) & \(742\pm\cdots\) & \(5.17\pm0.09\) & \(1647\pm54\) & \(83.3\pm4.9\) \\ 
12 & \(1924\pm44\) & \(82.1\pm1.8\) & \(713\pm10\) & \(4.90\pm0.08\) & \(1703\pm46\) & \(87.0\pm1.8\) \\ 
13 & \(1981\pm87\) & \(84.8\pm7.0\) & \(716\pm27\) & \(5.03\pm0.18\) & \(1848\pm48\) & \(89.8\pm7.0\) \\ 
14 & \(2041\pm64\) & \(80.2\pm1.5\) & \(757\pm13\) & \(4.93\pm0.10\) & \(1779\pm58\) & \(85.1\pm1.5\) \\ 
15 & \(2004\pm57\) & \(78.5\pm4.3\) & \(737\pm22\) & \(4.94\pm0.12\) & \(1684\pm77\) & \(83.4\pm4.3\) \\ 
16 & \(2017\pm83\) & \(82.3\pm1.5\) & \(682\pm16\) & \(5.11\pm0.10\) & \(1630\pm61\) & \(87.4\pm1.5\) \\ 
17 & \(1928\pm64\) & \(81.8\pm1.9\) & \(682\pm2\) & \(4.61\pm0.07\) & \(1702\pm41\) & \(86.4\pm1.9\) \\ 
18 & \(1904\pm79\) & \(82.9\pm4.6\) & \(736\pm27\) & \(4.77\pm0.14\) & \(1678\pm55\) & \(87.7\pm4.6\) \\ 
19 & \(2024\pm63\) & \(83.0\pm2.2\) & \(752\pm15\) & \(4.87\pm0.09\) & \(1782\pm52\) & \(87.9\pm2.2\) \\ 
20 & \(1961\pm91\) & \(81.3\pm4.0\) & \(727\pm28\) & \(5.05\pm0.17\) & \(1657\pm70\) & \(86.4\pm4.0\) \\ 
21 & \(1999\pm68\) & \(77.6\pm2.0\) & \(758\pm17\) & \(5.02\pm0.12\) & \(1696\pm92\) & \(82.7\pm2.0\) \\ 
22 & \(1878\pm92\) & \(76.6\pm2.2\) & \(802\pm16\) & \(5.01\pm0.12\) & \(1685\pm55\) & \(81.6\pm2.2\) \\ 
23 & \(1904\pm60\) & \(77.1\pm13.0\) & \(772\pm26\) & \(5.18\pm0.21\) & \(1755\pm59\) & \(82.3\pm13.0\) \\ 
24 & \(1933\pm44\) & \(77.7\pm1.5\) & \(674\pm17\) & \(4.83\pm0.09\) & \(1653\pm57\) & \(82.6\pm1.5\) \\ 
25 & \(1885\pm73\) & \(76.8\pm2.5\) & \(700\pm26\) & \(4.96\pm0.11\) & \(1647\pm61\) & \(81.8\pm2.5\) \\ 
26 & \(2037\pm49\) & \(78.9\pm1.8\) & \(733\pm14\) & \(5.02\pm0.09\) & \(1816\pm70\) & \(83.9\pm1.8\) \\ 
27 & \(1968\pm76\) & \(79.7\pm1.6\) & \(680\pm4\) & \(4.92\pm0.08\) & \(1735\pm41\) & \(84.6\pm1.6\) \\ 
28 & \(1976\pm77\) & \(76.7\pm2.0\) & \(801\pm6\) & \(4.94\pm0.08\) & \(1764\pm58\) & \(81.7\pm2.0\) \\ 
29 & \(1984\pm45\) & \(78.0\pm3.0\) & \(785\pm16\) & \(5.46\pm0.10\) & \(1742\pm47\) & \(83.4\pm3.0\) \\ 
30 & \(1862\pm89\) & \(75.4\pm2.7\) & \(742\pm22\) & \(5.18\pm0.14\) & \(1698\pm62\) & \(80.6\pm2.7\) \\ 
31 & \(2007\pm39\) & \(80.2\pm1.9\) & \(707\pm11\) & \(4.71\pm0.08\) & \(1744\pm42\) & \(84.9\pm1.9\) \\ 
32 & \(2058\pm96\) & \(81.3\pm2.3\) & \(699\pm5\) & \(4.77\pm0.11\) & \(1741\pm63\) & \(86.0\pm2.3\) \\ 
33 & \(1912\pm56\) & \(82.9\pm2.1\) & \(684\pm17\) & \(4.84\pm0.11\) & \(1688\pm73\) & \(87.7\pm2.1\) \\ 
34 & \(1936\pm39\) & \(84.7\pm2.9\) & \(683\pm17\) & \(5.00\pm0.10\) & \(1726\pm50\) & \(89.7\pm2.9\) \\ 
35 & \(1964\pm29\) & \(81.9\pm1.2\) & \(776\pm15\) & \(4.91\pm0.09\) & \(1730\pm67\) & \(86.9\pm1.2\) \\ 
36 & \(1919\pm32\) & \(84.6\pm1.0\) & \(681\pm2\) & \(4.85\pm0.05\) & \(1702\pm29\) & \(89.5\pm1.0\) \\ 
37 & \(1895\pm76\) & \(85.5\pm1.7\) & \(682\pm0\) & \(4.82\pm0.07\) & \(1654\pm34\) & \(90.3\pm1.7\) \\ 
38 & \(1922\pm72\) & \(84.5\pm1.7\) & \(810\pm9\) & \(5.08\pm0.11\) & \(1708\pm50\) & \(89.6\pm1.7\) \\ 
39 & \(1869\pm69\) & \(85.1\pm21.5\) & \(719\pm49\) & \(5.05\pm0.29\) & \(1655\pm33\) & \(90.1\pm21.5\) \\ 
40 & \(1911\pm50\) & \(87.1\pm3.9\) & \(719\pm16\) & \(4.94\pm0.11\) & \(1670\pm68\) & \(92.1\pm3.9\) \\ 
41 & \(1866\pm45\) & \(88.7\pm0.7\) & \(728\pm9\) & \(4.96\pm0.06\) & \(1641\pm31\) & \(93.6\pm0.7\) \\ 
42 & \(1933\pm47\) & \(87.3\pm1.9\) & \(688\pm2\) & \(4.96\pm0.06\) & \(1682\pm45\) & \(92.2\pm1.9\) \\ 
43 & \(1836\pm74\) & \(91.0\pm6.8\) & \(713\pm27\) & \(5.03\pm0.16\) & \(1700\pm34\) & \(96.0\pm6.8\) \\ 
44 & \(1919\pm58\) & \(87.2\pm8.9\) & \(712\pm45\) & \(5.10\pm0.27\) & \(1686\pm45\) & \(92.3\pm8.9\) \\ 
45 & \(1888\pm66\) & \(86.5\pm3.0\) & \(745\pm18\) & \(5.36\pm0.12\) & \(1676\pm57\) & \(91.8\pm3.0\) \\ 
46 & \(1943\pm46\) & \(85.2\pm2.8\) & \(731\pm14\) & \(4.94\pm0.09\) & \(1663\pm43\) & \(90.1\pm2.8\) \\ 
47 & \(1809\pm133\) & \(80.0\pm4.8\) & \(744\pm36\) & \(4.85\pm0.21\) & \(1606\pm79\) & \(84.9\pm4.8\) \\ 
48 & \(1937\pm54\) & \(82.2\pm3.2\) & \(685\pm26\) & \(4.71\pm0.13\) & \(1700\pm62\) & \(86.9\pm3.2\) \\ 
49 & \(2129\pm115\) & \(80.4\pm2.7\) & \(857\pm30\) & \(5.51\pm0.17\) & \(1702\pm127\) & \(85.9\pm2.7\) \\ 
50 & \(2032\pm39\) & \(87.9\pm1.9\) & \(691\pm31\) & \(5.02\pm0.14\) & \(1773\pm45\) & \(92.9\pm1.9\) \\ 
51 & \(1996\pm36\) & \(85.7\pm1.8\) & \(726\pm11\) & \(4.89\pm0.07\) & \(1749\pm40\) & \(90.6\pm1.8\) \\ 
52 & \(1921\pm51\) & \(81.7\pm2.6\) & \(686\pm31\) & \(4.73\pm0.15\) & \(1665\pm56\) & \(86.5\pm2.6\) \\ 
53 & \(1857\pm43\) & \(81.6\pm1.9\) & \(707\pm19\) & \(5.00\pm0.11\) & \(1667\pm48\) & \(86.6\pm1.9\) \\ 
54 & \(1846\pm71\) & \(80.3\pm3.6\) & \(679\pm15\) & \(4.87\pm0.11\) & \(1627\pm53\) & \(85.1\pm3.6\) \\ 
55\tablenotemark{a} & \(2051\pm101\) & \(98.8\pm3.2\) & \(670\pm26\) & \(4.24\pm0.17\) & \(1837\pm91\) & \(103.0\pm3.2\) \\ 
56 & \(1900\pm60\) & \(84.4\pm3.3\) & \(646\pm31\) & \(4.64\pm0.12\) & \(1657\pm35\) & \(89.1\pm3.3\) \\ 
57 & \(1855\pm32\) & \(84.3\pm1.5\) & \(634\pm1\) & \(4.71\pm0.06\) & \(1673\pm36\) & \(89.0\pm1.5\) \\ 
58 & \(1873\pm45\) & \(85.1\pm1.1\) & \(711\pm8\) & \(5.11\pm0.06\) & \(1648\pm35\) & \(90.3\pm1.1\) \\ 
59 & \(1880\pm33\) & \(83.1\pm1.2\) & \(724\pm12\) & \(5.18\pm0.07\) & \(1683\pm33\) & \(88.3\pm1.2\) \\ 
60 & \(1831\pm79\) & \(82.8\pm4.3\) & \(726\pm21\) & \(5.13\pm0.14\) & \(1674\pm49\) & \(87.9\pm4.3\) \\ 
61 & \(1802\pm71\) & \(79.5\pm7.3\) & \(668\pm27\) & \(5.04\pm0.17\) & \(1536\pm60\) & \(84.5\pm7.3\) \\ 
62 & \(1943\pm57\) & \(78.1\pm2.9\) & \(750\pm14\) & \(5.44\pm0.09\) & \(1724\pm43\) & \(83.6\pm2.9\) \\ 
63 & \(1934\pm51\) & \(79.9\pm2.3\) & \(659\pm14\) & \(4.81\pm0.09\) & \(1639\pm39\) & \(84.7\pm2.3\) \\ 
64 & \(1852\pm36\) & \(77.7\pm1.7\) & \(660\pm11\) & \(4.82\pm0.07\) & \(1653\pm37\) & \(82.5\pm1.7\) \\ 
65 & \(1893\pm33\) & \(81.0\pm1.1\) & \(727\pm12\) & \(5.11\pm0.07\) & \(1682\pm46\) & \(86.2\pm1.1\) \\ 
66 & \(1930\pm47\) & \(82.6\pm1.5\) & \(681\pm12\) & \(5.01\pm0.09\) & \(1706\pm32\) & \(87.6\pm1.5\) \\ 
67 & \(1877\pm35\) & \(83.7\pm2.2\) & \(709\pm14\) & \(4.98\pm0.09\) & \(1695\pm56\) & \(88.7\pm2.2\) \\ 
68 & \(1838\pm47\) & \(81.5\pm2.4\) & \(756\pm3\) & \(5.11\pm0.09\) & \(1651\pm43\) & \(86.6\pm2.4\) \\ 
69 & \(1893\pm67\) & \(81.3\pm3.0\) & \(752\pm25\) & \(5.05\pm0.11\) & \(1696\pm35\) & \(86.3\pm3.0\) \\
\enddata
\tablenotetext{a}{Excluded from further analysis.  See note {\em a} in Table~\ref{tab:spectral_meas_oiii_ljt}.}
\tablecomments{The second and the third columns providing the FWHM and flux values, respectively, for the broad (`{\em b}') \Hbeta\ component include contributions from both the broad Gaussians defined for the line. The FWHM of the narrow (`{\em n}') \Hbeta\ is equal to the FWHM of the \OIIIlamstrong\ (see Table~\ref{tab:spectral_meas_oiii_ljt}, third column).  The total (`{\em t}') FWHM and flux include contributions from both the broad and the narrow components ({\em t} = {\em b} + {\em n}).} 
\vspace{4pt}
\tablecomments{This table is available in machine-readable form.}
\end{deluxetable*}

\startlongtable
\begin{deluxetable*}{rcccccc}
\tablewidth{700pt}
\tablenum{9}
\tablecaption{\HeIlam\ Emission-Line Measurements for LJT Spectra
\label{tab:spectral_meas_hei_ljt}}
\tabletypesize{\scriptsize}
\tablehead{
\colhead{Epoch} & \colhead{FWHM$_{\HeI,b}$} & \colhead{$F_{\HeI,b}$} & \colhead{FWHM$_{\HeI,n}$} & \colhead{$F_{\HeI,n}$} & \colhead{FWHM$_{\HeI,t}$} & \colhead{$F_{\HeI,t}$} \\
 & \colhead{[\kms]} & \colhead{[10$^{-15}$~erg~s$^{-1}$~cm$^{-2}$]} & \colhead{[\kms]} & \colhead{[10$^{-15}$~erg~s$^{-1}$~cm$^{-2}$]} & \colhead{[\kms]} & \colhead{[10$^{-15}$~erg~s$^{-1}$~cm$^{-2}$]}
}
\startdata
1 & \(3321\pm340\) & \(56.5\pm1.3\) & \(718\pm9\) & \(0.56\pm0.01\) & \(3233\pm310\) & \(57.0\pm1.3\) \\ 
2 & \(4215\pm487\) & \(54.0\pm1.9\) & \(754\pm22\) & \(0.59\pm0.01\) & \(3719\pm432\) & \(54.6\pm1.9\) \\ 
3 & \(2677\pm393\) & \(53.4\pm1.5\) & \(741\pm13\) & \(0.58\pm0.01\) & \(2640\pm335\) & \(54.0\pm1.5\) \\ 
4 & \(3526\pm460\) & \(51.2\pm1.6\) & \(787\pm19\) & \(0.61\pm0.01\) & \(3472\pm439\) & \(51.8\pm1.6\) \\ 
5 & \(3672\pm430\) & \(50.7\pm1.5\) & \(716\pm18\) & \(0.55\pm0.01\) & \(2955\pm380\) & \(51.2\pm1.5\) \\ 
6 & \(4021\pm559\) & \(53.0\pm1.7\) & \(740\pm21\) & \(0.56\pm0.01\) & \(3358\pm509\) & \(53.6\pm1.7\) \\ 
7 & \(2800\pm413\) & \(56.6\pm1.9\) & \(880\pm6\) & \(0.67\pm0.01\) & \(2652\pm370\) & \(57.3\pm1.9\) \\ 
8 & \(2393\pm429\) & \(59.9\pm1.8\) & \(691\pm13\) & \(0.57\pm0.01\) & \(2390\pm356\) & \(60.5\pm1.8\) \\ 
9 & \(3007\pm395\) & \(51.5\pm1.5\) & \(679\pm19\) & \(0.58\pm0.01\) & \(3002\pm377\) & \(52.1\pm1.5\) \\ 
10 & \(3440\pm567\) & \(57.7\pm2.4\) & \(707\pm27\) & \(0.59\pm0.02\) & \(3384\pm534\) & \(58.3\pm2.4\) \\ 
11 & \(3742\pm464\) & \(50.2\pm2.0\) & \(742\pm\cdots\) & \(0.60\pm0.01\) & \(3336\pm414\) & \(50.8\pm2.0\) \\ 
12 & \(3707\pm394\) & \(52.6\pm1.3\) & \(713\pm10\) & \(0.57\pm0.01\) & \(3234\pm349\) & \(53.1\pm1.3\) \\ 
13 & \(2031\pm523\) & \(58.0\pm3.7\) & \(716\pm27\) & \(0.58\pm0.02\) & \(2020\pm440\) & \(58.6\pm3.7\) \\ 
14 & \(4636\pm499\) & \(54.8\pm1.7\) & \(757\pm13\) & \(0.57\pm0.01\) & \(3448\pm477\) & \(55.3\pm1.7\) \\ 
15 & \(3570\pm482\) & \(51.3\pm2.1\) & \(737\pm22\) & \(0.57\pm0.01\) & \(3392\pm451\) & \(51.9\pm2.1\) \\ 
16 & \(3777\pm455\) & \(52.6\pm1.6\) & \(682\pm16\) & \(0.59\pm0.01\) & \(2817\pm387\) & \(53.2\pm1.6\) \\ 
17 & \(3086\pm433\) & \(55.4\pm1.5\) & \(682\pm2\) & \(0.53\pm0.01\) & \(3061\pm378\) & \(55.9\pm1.5\) \\ 
18 & \(2644\pm473\) & \(49.8\pm2.2\) & \(736\pm27\) & \(0.55\pm0.02\) & \(2584\pm432\) & \(50.4\pm2.2\) \\ 
19 & \(3681\pm401\) & \(54.9\pm1.5\) & \(752\pm15\) & \(0.56\pm0.01\) & \(3607\pm366\) & \(55.4\pm1.5\) \\ 
20 & \(3500\pm533\) & \(52.3\pm2.2\) & \(727\pm28\) & \(0.59\pm0.02\) & \(3260\pm504\) & \(52.9\pm2.2\) \\ 
21 & \(2511\pm515\) & \(45.0\pm1.9\) & \(758\pm17\) & \(0.58\pm0.01\) & \(2479\pm447\) & \(45.6\pm1.9\) \\ 
22 & \(4378\pm683\) & \(57.8\pm2.1\) & \(802\pm16\) & \(0.58\pm0.01\) & \(3636\pm586\) & \(58.4\pm2.1\) \\ 
23 & \(3628\pm556\) & \(53.8\pm3.7\) & \(772\pm26\) & \(0.60\pm0.02\) & \(3230\pm523\) & \(54.4\pm3.7\) \\ 
24 & \(3526\pm399\) & \(52.0\pm1.4\) & \(674\pm17\) & \(0.56\pm0.01\) & \(3148\pm367\) & \(52.6\pm1.4\) \\ 
25 & \(3227\pm518\) & \(51.5\pm1.7\) & \(700\pm26\) & \(0.58\pm0.01\) & \(3101\pm485\) & \(52.1\pm1.7\) \\ 
26 & \(3575\pm504\) & \(53.0\pm1.6\) & \(733\pm14\) & \(0.58\pm0.01\) & \(3365\pm459\) & \(53.5\pm1.6\) \\ 
27 & \(4302\pm565\) & \(53.1\pm1.4\) & \(680\pm4\) & \(0.57\pm0.01\) & \(4133\pm484\) & \(53.6\pm1.4\) \\ 
28 & \(3591\pm602\) & \(52.1\pm1.7\) & \(801\pm6\) & \(0.57\pm0.01\) & \(3578\pm529\) & \(52.6\pm1.7\) \\ 
29 & \(3724\pm551\) & \(49.4\pm1.5\) & \(785\pm16\) & \(0.63\pm0.01\) & \(3287\pm540\) & \(50.0\pm1.5\) \\ 
30 & \(2501\pm565\) & \(53.2\pm2.1\) & \(742\pm22\) & \(0.60\pm0.02\) & \(2342\pm479\) & \(53.8\pm2.1\) \\ 
31 & \(4400\pm518\) & \(50.7\pm1.4\) & \(707\pm11\) & \(0.55\pm0.01\) & \(4224\pm445\) & \(51.3\pm1.4\) \\ 
32 & \(3186\pm487\) & \(57.9\pm2.0\) & \(699\pm5\) & \(0.55\pm0.01\) & \(3123\pm437\) & \(58.4\pm2.0\) \\ 
33 & \(3456\pm555\) & \(58.8\pm1.8\) & \(684\pm17\) & \(0.56\pm0.01\) & \(3362\pm519\) & \(59.3\pm1.8\) \\ 
34 & \(2880\pm373\) & \(57.8\pm1.9\) & \(683\pm17\) & \(0.58\pm0.01\) & \(2793\pm302\) & \(58.4\pm1.9\) \\ 
35 & \(3257\pm410\) & \(55.0\pm1.3\) & \(776\pm15\) & \(0.57\pm0.01\) & \(3217\pm374\) & \(55.6\pm1.3\) \\ 
36 & \(3778\pm482\) & \(55.3\pm0.9\) & \(681\pm2\) & \(0.56\pm0.01\) & \(3648\pm431\) & \(55.8\pm0.9\) \\ 
37 & \(3369\pm351\) & \(57.0\pm1.4\) & \(682\pm0\) & \(0.56\pm0.01\) & \(3007\pm302\) & \(57.5\pm1.4\) \\ 
38 & \(3024\pm428\) & \(52.6\pm2.0\) & \(810\pm9\) & \(0.59\pm0.01\) & \(3009\pm400\) & \(53.2\pm2.0\) \\ 
39 & \(2901\pm389\) & \(53.0\pm4.1\) & \(719\pm49\) & \(0.59\pm0.03\) & \(2809\pm349\) & \(53.5\pm4.1\) \\ 
40 & \(3124\pm489\) & \(57.3\pm1.8\) & \(719\pm16\) & \(0.57\pm0.01\) & \(3069\pm431\) & \(57.9\pm1.8\) \\ 
41 & \(4761\pm506\) & \(57.8\pm1.0\) & \(728\pm9\) & \(0.58\pm0.01\) & \(3452\pm380\) & \(58.4\pm1.0\) \\ 
42 & \(3699\pm399\) & \(61.1\pm1.2\) & \(688\pm2\) & \(0.58\pm0.01\) & \(3664\pm382\) & \(61.7\pm1.2\) \\ 
43 & \(2720\pm318\) & \(56.2\pm2.4\) & \(713\pm27\) & \(0.58\pm0.02\) & \(2658\pm296\) & \(56.8\pm2.4\) \\ 
44 & \(3220\pm447\) & \(51.5\pm2.4\) & \(712\pm45\) & \(0.59\pm0.03\) & \(3184\pm440\) & \(52.1\pm2.4\) \\ 
45 & \(3481\pm491\) & \(53.9\pm2.0\) & \(745\pm18\) & \(0.62\pm0.01\) & \(3422\pm444\) & \(54.5\pm2.0\) \\ 
46 & \(3367\pm428\) & \(50.9\pm1.4\) & \(731\pm14\) & \(0.57\pm0.01\) & \(3317\pm417\) & \(51.5\pm1.4\) \\ 
47 & \(3618\pm841\) & \(54.1\pm3.2\) & \(744\pm36\) & \(0.56\pm0.02\) & \(3612\pm762\) & \(54.6\pm3.2\) \\ 
48 & \(2302\pm454\) & \(51.5\pm1.7\) & \(685\pm26\) & \(0.55\pm0.02\) & \(2300\pm404\) & \(52.0\pm1.7\) \\ 
49 & \(3621\pm1008\) & \(56.9\pm2.6\) & \(857\pm30\) & \(0.64\pm0.02\) & \(3598\pm876\) & \(57.6\pm2.6\) \\ 
50 & \(3405\pm461\) & \(55.3\pm1.4\) & \(691\pm31\) & \(0.58\pm0.02\) & \(2868\pm423\) & \(55.9\pm1.4\) \\ 
51 & \(2974\pm379\) & \(50.9\pm1.1\) & \(726\pm11\) & \(0.57\pm0.01\) & \(2512\pm286\) & \(51.4\pm1.1\) \\ 
52 & \(3095\pm465\) & \(49.0\pm1.6\) & \(686\pm31\) & \(0.55\pm0.02\) & \(3061\pm412\) & \(49.6\pm1.6\) \\ 
53 & \(3560\pm386\) & \(55.2\pm1.5\) & \(707\pm19\) & \(0.58\pm0.01\) & \(2853\pm361\) & \(55.8\pm1.5\) \\ 
54 & \(4011\pm529\) & \(54.9\pm1.8\) & \(679\pm15\) & \(0.56\pm0.01\) & \(3867\pm475\) & \(55.4\pm1.8\) \\ 
55\tablenotemark{a} & \(1713\pm519\) & \(56.2\pm2.9\) & \(670\pm26\) & \(0.49\pm0.02\) & \(1708\pm510\) & \(56.6\pm2.9\) \\ 
56 & \(3026\pm343\) & \(50.5\pm1.7\) & \(646\pm31\) & \(0.54\pm0.01\) & \(2986\pm363\) & \(51.0\pm1.7\) \\ 
57 & \(3362\pm424\) & \(49.0\pm1.2\) & \(634\pm1\) & \(0.55\pm0.01\) & \(2803\pm397\) & \(49.6\pm1.2\) \\ 
58 & \(3533\pm224\) & \(52.5\pm1.1\) & \(711\pm8\) & \(0.59\pm0.01\) & \(3401\pm203\) & \(53.1\pm1.1\) \\ 
59 & \(4223\pm565\) & \(51.4\pm1.1\) & \(724\pm12\) & \(0.60\pm0.01\) & \(2775\pm415\) & \(52.0\pm1.1\) \\ 
60 & \(2515\pm545\) & \(50.3\pm2.4\) & \(726\pm21\) & \(0.60\pm0.02\) & \(2515\pm480\) & \(50.8\pm2.4\) \\ 
61 & \(3274\pm521\) & \(50.4\pm2.2\) & \(668\pm27\) & \(0.59\pm0.02\) & \(3129\pm497\) & \(51.0\pm2.2\) \\ 
62 & \(3352\pm421\) & \(51.9\pm1.7\) & \(750\pm14\) & \(0.63\pm0.01\) & \(3337\pm419\) & \(52.5\pm1.7\) \\ 
63 & \(3756\pm487\) & \(53.0\pm1.4\) & \(659\pm14\) & \(0.56\pm0.01\) & \(3257\pm447\) & \(53.6\pm1.4\) \\ 
64 & \(4127\pm1067\) & \(49.8\pm1.2\) & \(660\pm11\) & \(0.56\pm0.01\) & \(3030\pm761\) & \(50.3\pm1.2\) \\ 
65 & \(4321\pm452\) & \(57.7\pm1.2\) & \(727\pm12\) & \(0.59\pm0.01\) & \(3535\pm420\) & \(58.3\pm1.2\) \\ 
66 & \(5320\pm693\) & \(57.0\pm1.4\) & \(681\pm12\) & \(0.58\pm0.01\) & \(3285\pm608\) & \(57.6\pm1.4\) \\ 
67 & \(3695\pm507\) & \(56.9\pm1.5\) & \(709\pm14\) & \(0.58\pm0.01\) & \(3549\pm504\) & \(57.5\pm1.5\) \\ 
68 & \(3375\pm481\) & \(52.8\pm1.7\) & \(756\pm3\) & \(0.59\pm0.01\) & \(2925\pm432\) & \(53.4\pm1.7\) \\ 
69 & \(3431\pm465\) & \(53.3\pm1.7\) & \(752\pm25\) & \(0.59\pm0.01\) & \(2660\pm385\) & \(53.9\pm1.7\) \\
\enddata
\tablenotetext{a}{Excluded from further analysis.  See note {\em a} in Table~\ref{tab:spectral_meas_oiii_ljt}.}
\tablecomments{The second and the third columns providing the FWHM and flux values, respectively, for the broad (`{\em b}') \HeI\ component include contributions from both the broad Gaussians defined for the line. The FWHM of the narrow (`{\em n}') \HeI\ is equal to the FWHM of the \OIIIlamstrong\ (see Table~\ref{tab:spectral_meas_oiii_ljt}, third column).  The total (`{\em t}') FWHM and flux include contributions from both the broad and the narrow components ({\em t} = {\em b} + {\em n}).}
\tablecomments{This table is available in machine-readable form.}
\end{deluxetable*}

\section{Light Curve Analysis} \label{sec:light_curve_analysis}

We used the Mrk~142 Gemini and LJT spectral measurements to generate light curves for the broad \Hbeta\ and \HeI\ emission-line profiles.  We obtained the total broad-line light curves by integrating the flux under the two broad components (see green dashed Gaussians for \Hbeta\ and pink dashed Gaussians for \HeI\ in Figures~\ref{fig:spectral_model_single_epoch_gemini} and \ref{fig:spectral_model_single_epoch_ljt}) for the two emission lines.

\begin{figure}[ht!]
\epsscale{1.2}
\plotone{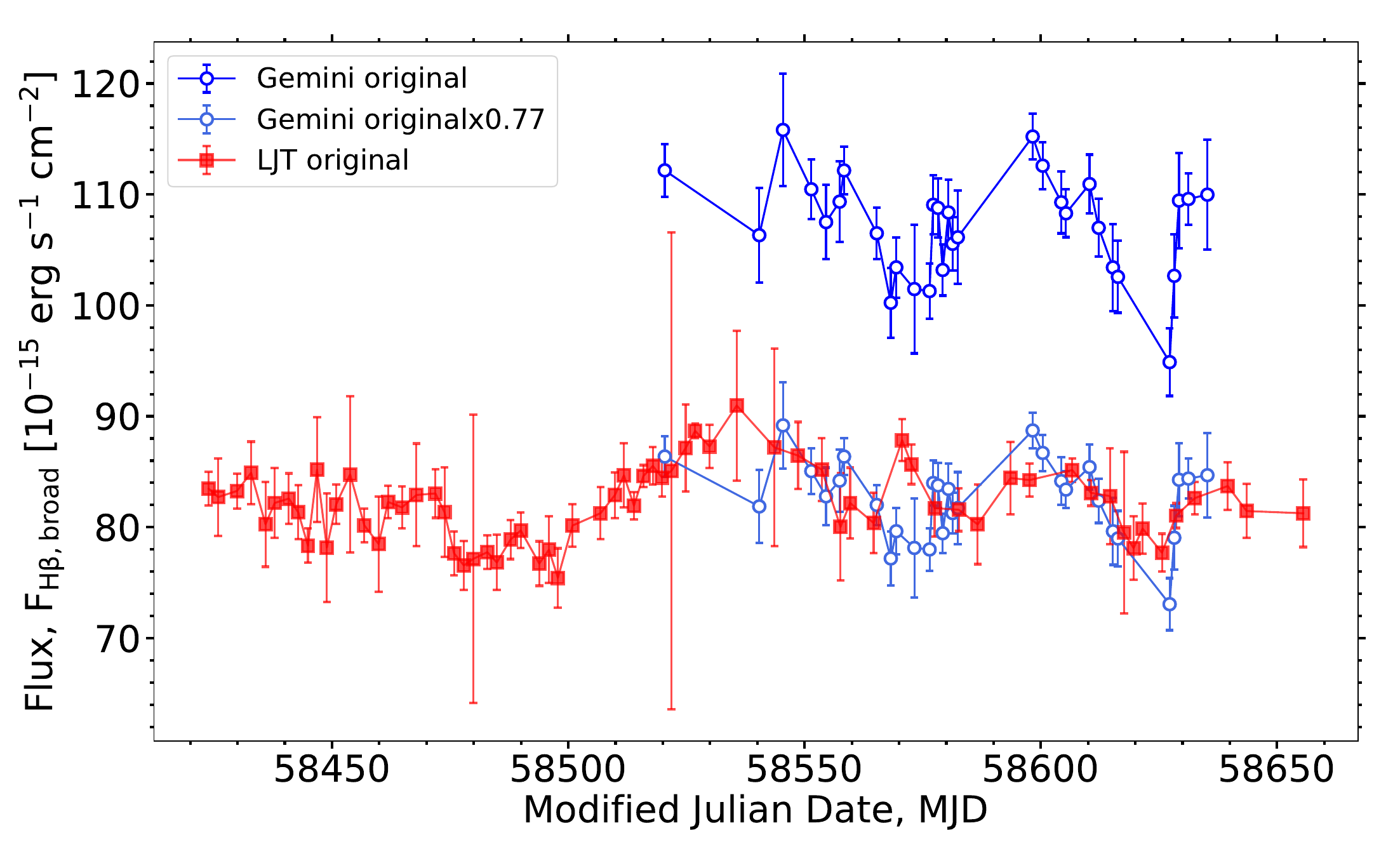}
\caption{Broad \Hbeta\ light curve from Gemini (dark blue open circles) scaled down (faint blue open circles) to the broad \Hbeta\ light curve from LJT (red solid squares).  The scaling factor of 0.77 was determined such that the Gemini data points are distributed evenly above and below the LJT light curve.  The two light curves show similar behavior except at Modified Julian Date $\sim$58540, where the LJT light curve increases in flux while the Gemini light curve appears fainter.
\label{fig:lightcurves_hbeta_b_scaled}}
\end{figure}

We scaled the broad \Hbeta\ light curve from Gemini to the broad \Hbeta\ light curve from LJT to generate an inter-calibrated light curve.  The \Hbeta\ light curves from Gemini and LJT were offset by $\sim$25\% from each other although they displayed similar fluctuations in their patterns.  The offset can be attributed to various factors -- different seeing conditions at Gemini and LJT or even the difference in the calibrations from the two telescopes.  Because we are interested in measuring a time shift in the pattern with reference to the continuum variations, scaling and combining the light curves is valid for our purpose.  Figure~\ref{fig:lightcurves_hbeta_b_scaled} shows the scaled broad \Hbeta\ light curve plotted with the original Gemini and LJT light curves.  We then inter-calibrated the original \Hbeta\ light curve from LJT and the scaled \Hbeta\ light curve from Gemini with {\tt\string PyROA} (see Section~\ref{subsec:time_lags}) to use the combined light curve to determine the time lag between the continuum and emission-line variability.

\subsection{Cross Correlation Time Lags} \label{subsec:time_lags}

We cross-correlated the broad \Hbeta\ Gemini+LJT inter-calibrated light curve with the {\em UVW2} light curve from {\em Swift} to measure the reverberation lag of the BLR response to the continuum variability from the accretion disk, which is at smaller size scales than the BLR.  Following \citet{cackett_etal_2020}, we chose the {\em UVW2} because we aim to measure the response of \Hbeta\ line-emitting gas to the UV continuum and the {\em UVW2} was the shortest wavelength available from the photometric monitoring of Mrk~142.  During cross-correlation, we also included the LCO+Zowada+Liverpool/{\em g}, the inter-calibrated 5100~\AA\ continuum from Gemini and LJT data, and the LJT broad \Hbeta\ light curves to use maximum available information for a reliable measurement of the variability pattern.

\begin{figure*}[ht!]
\epsscale{1.15}
\plotone{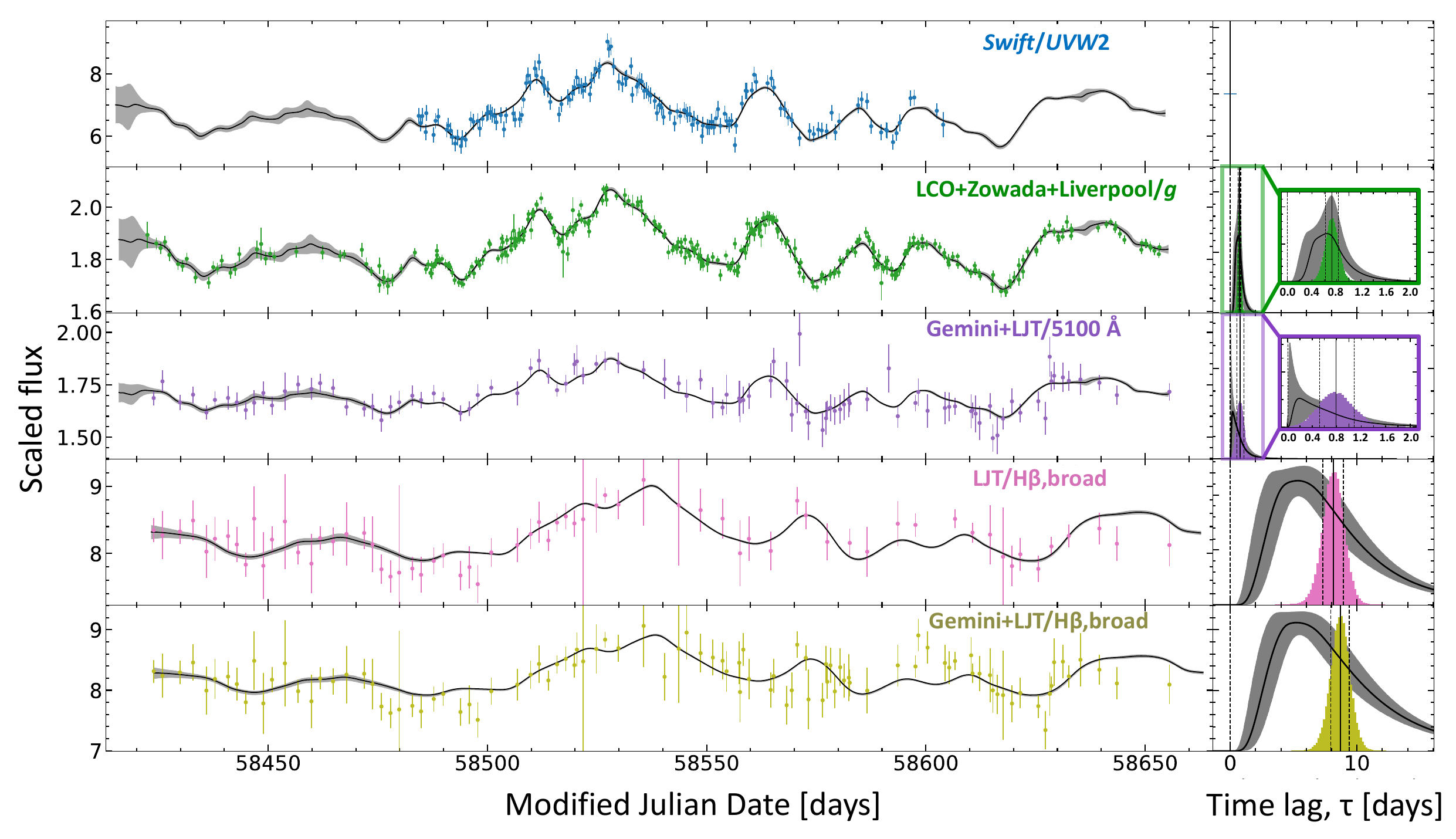}
\caption{Time-lag measurements with reference to the {\em Swift}/{\em UVW2} band (top row) with lag distributions modeled as Log-Gaussians.  {\em Left}: Top three panels show continuum light curves -- {\em Swift}/{\em UVW2} (blue), LCO+Zowada+Liverpool/{\em g} (green), and inter-calibrated Gemini+LJT/5100~\AA\ -- in units of 10$^{-15}$~\flam.  Bottom two panels show the broad \Hbeta\ light curves -- LJT only (pink) and Gemini+LJT inter-calibrated (olive) -- in units of 10$^{-14}$~\flam.  {\em Right}: Time-lag distributions (colored histograms for the mean lag and gray shaded uncertainty estimates from a Log-Gaussian lag distribution) of the light curves on the left with reference to the {\em Swift}/{\em UVW2} band, which has a fixed lag of 0.00 days.  Insets in the second and the third panels show a close view of the time-lag distributions for LCO+Zowada+Liverpool/{\em g} and Gemini+LJT/5100~\AA\ light curves, respectively.  For each of the echo light curves, the black solid vertical line marks the lag measurement, given by the median lag of the colored distribution} and the black dashed vertical lines (on either sides of the solid line) mark the corresponding uncertainty.
\label{fig:time_lag_distributions}
\end{figure*}

We employed {\tt\string Python}-based Running Optimal Average \citep[{\tt\string PyROA};][]{donnan_etal_2021} to calculate cross-correlation time lags.  {\tt\string PyROA}\footnote{See {\tt\string PyROA} code and documentation at \url{https://github.com/FergusDonnan/PyROA}.} uses a running optimal average (ROA) calculated with a window function (defined by a Gaussian by default) of a certain width to estimate light-curve behavior while fitting all input light curves simultaneously.  The width of the window function controls the flexibility of the model in deriving the driving light curve -- a narrower window function traces the fluctuating pattern of a highly variable light curve more appropriately than a wider window, which behaves more rigidly.  The code uses priors to initiate the modeling process, and the performance of the model is evaluated using Bayesian Information Criterion.  {\tt\string PyROA} offers a robust treatment for outliers with an extra variance parameter and a standard deviation threshold.  The extra variance adds in quadrature to the nominal uncertainties of the input light curves, and the threshold set by the user allows further inflation of the uncertainties to mitigate the influence of large outliers.  Figure~\ref{fig:time_lag_distributions} shows the cross-correlation results with reference to the {\em Swift}/{\em UVW2} band.

In addition to the \Hbeta\ time lag for the Gemini+LJT inter-calibrated light curve (\(8.68_{-0.72}^{+0.75}\) days), {\tt\string PyROA} provided lag measurements for other input light curves (\(0.73_{-0.10}^{+0.10}\) days for LCO+Zowada+Liverpool/{\em g}, \(0.79_{-0.29}^{+0.27}\) days for Gemini+LJT/5100~\AA, and \(8.14_{-0.80}^{+0.82}\) days for LJT/\Hbeta) with reference to the {\em Swift}/{\em UVW2} band.  With respect to the shorter-wavelength {\em UVW2} emission, we expect to measure positive lags for the longer-wavelength emission in the {\em g}-band, at 5100~\AA, and for the \Hbeta\ emission line.  We thus modeled the distribution of time lags as a Log-Gaussian function that imposes positive lags with reference to {\em UVW2}, whose lag is fixed at 0.00 days (see Figure~\ref{fig:time_lag_distributions}).  In addition to measuring the time shift in the light-curve pattern, the width of the Log-Gaussian model also accounts for the amount of blurring applied to the reference light curve (here, {\em UVW2}) to match the response in the echo light curves.  This becomes important for BLR RM, where the emission-line variations, emerging farther away from the central engine and from a more spatially extended structure (size scale $\sim$1~parsec) than the accretion disk, are smoother compared to the continuum variations closer to the center.  In {\tt\string PyROA}, the width of the time-lag distribution quantifies the blurring determined for each of the echo light curves: \(0.28_{-0.18}^{+0.19}\) days for LCO+Zowada+Liverpool/{\em g}, \(0.88_{-0.48}^{+0.62}\) days for Gemini+LJT/5100~\AA, \(4.88_{-0.90}^{+1.16}\) days for LJT/\Hbeta), and \(5.47_{-0.89}^{+1.06}\) days for Gemini+LJT/\Hbeta.  We also performed light-curve analysis with the Interpolated Cross-Correlation Function \citep[{\tt\string ICCF};][]{gaskell_sparke_1986, gaskell_peterson_1987} and Just Another Vehicle for Estimating Lags In Nuclei \citep[{\tt\string JAVELIN};][using a top-hat time-lag distribution function]{zu_etal_2011, zu_etal_2013} for comparison with the {\tt\string PyROA} results.  Table~\ref{tab:time_lag_measurements} displays the time-lag measurements with all three methods.

\begin{deluxetable*}{lccc}
\tablewidth{700pt}
\tablenum{10}
\tablecaption{Time-Lag Measurements
\label{tab:time_lag_measurements}}
\tabletypesize{\scriptsize}
\tablehead{
 \\
\colhead{Time Lag [days]} & \colhead{{\tt\string PyROA}$^*$} & \colhead{{\tt\string ICCF}} & \colhead{{\tt\string JAVELIN}} \\
} 
\startdata
{\em UVW2}-to-{\em g} & \(0.73_{-0.10}^{+0.10}\) & \(0.7_{-0.2}^{+0.2}\) & \(0.54_{-0.08}^{+0.08}\) \\
{\em UVW2}-to-5100~\AA & \(0.79_{-0.29}^{+0.27}\) & \(1.7_{-1.2}^{+1.8}\) & \(0.78_{-0.38}^{+0.42}\) \\
{\em UVW2}-to-\Hbeta,LJT & \(8.14_{-0.80}^{+0.82}\) & \(8.8_{-3.5}^{+2.5}\) & \(6.95_{-0.46}^{+0.69}\) \\
{\em UVW2}-to-\Hbeta,Gemini+LJT & \(8.68_{-0.72}^{+0.75}\) & \(8.7_{-9.1}^{+4.2}\) & \(11.38_{-4.49}^{+0.51}\) \\
\enddata
\tablenotetext{*}{We consider these as the most robust time-lag measurements of the three methods.  See Section~\ref{sec:results_discussion} for further details.}
\end{deluxetable*}
 
\subsection{\HeI\ Light Curves} \label{subsec:hei_light_curves}

The peculiar, asymmetrical shape of the \HeI\ line -- narrow-line emission and a broad, asymmetrical component (modeled by a Gaussian six times the width of the broad \Hbeta\ line in our spectra fitting procedure) -- is clearly evident in the high S/N Gemini mean spectrum.  The asymmetry in the broad component due to the stronger blueshifted emission feature likely indicates a wind component in the BLR.  However, the cause of such a disk-wind component is not clear.  \citet{leighly2004} performed {\tt\string CLOUDY} simulations to model 10 high- and low-ionization emission lines observed in NLS1s.  She suggested that the blueshifted emission evident in the high-ionization lines in NLS1s arises from a wind moving towards us.  Interestingly, the plausible broad, blueshifted component for \HeII\ in the spectral model may also be a result of such a wind emission.  Further analysis of the \HeII\ line is needed to draw firm inferences in this regard.  In addition to the blueshifted wind, \citet{leighly2004} found that the high-ionization \Lyalpha\ was dominated by emission in the accretion-disk atmosphere or at the low-velocity base of the broad-line wind.  The very broad, flattened emission feature in \HeI\ may be indicative of a disk-wind feature as noted for \Lyalpha.  Furthermore, \citet{leighly2004} derived a small covering fraction for the BLR.  She argued that in an object with a small black hole mass, as in the case of Mrk~142 ($M_{\bullet}=3.89~\times~10^6~M_\odot$$^[$\footnote{1~$M_\odot$~=~1~Solar mass}$^]$), a small covering fraction can result from an emission-line region closer to the plane of the disk. 
 \citet{li_etal_2018} performed velocity-resolved RM of Mrk~142, where the authors concluded that the two-zone BLR model \citep{wang_etal_2014_1} best fit the Mrk~142 BLR with an opening angle ($\theta$) of 10--30$^\circ$ (representing a disk-like BLR).  Estimating the covering fraction from the opening angle as $\theta/90^\circ$ yields a value of 0.1--0.3 for the covering fraction; however, these small values do not confirm the presence of a disk wind in Mrk~142.  To understand the components that form the atypical Mrk~142 BLR system or the kind of BLR geometry in a super-Eddington that can show a broad, blueshifted emission feature similar to the \HeI\ line feature, we would need further investigation of BLR models and data for super-Eddington AGN.

\begin{figure}[ht!]
\epsscale{1.2}
\plotone{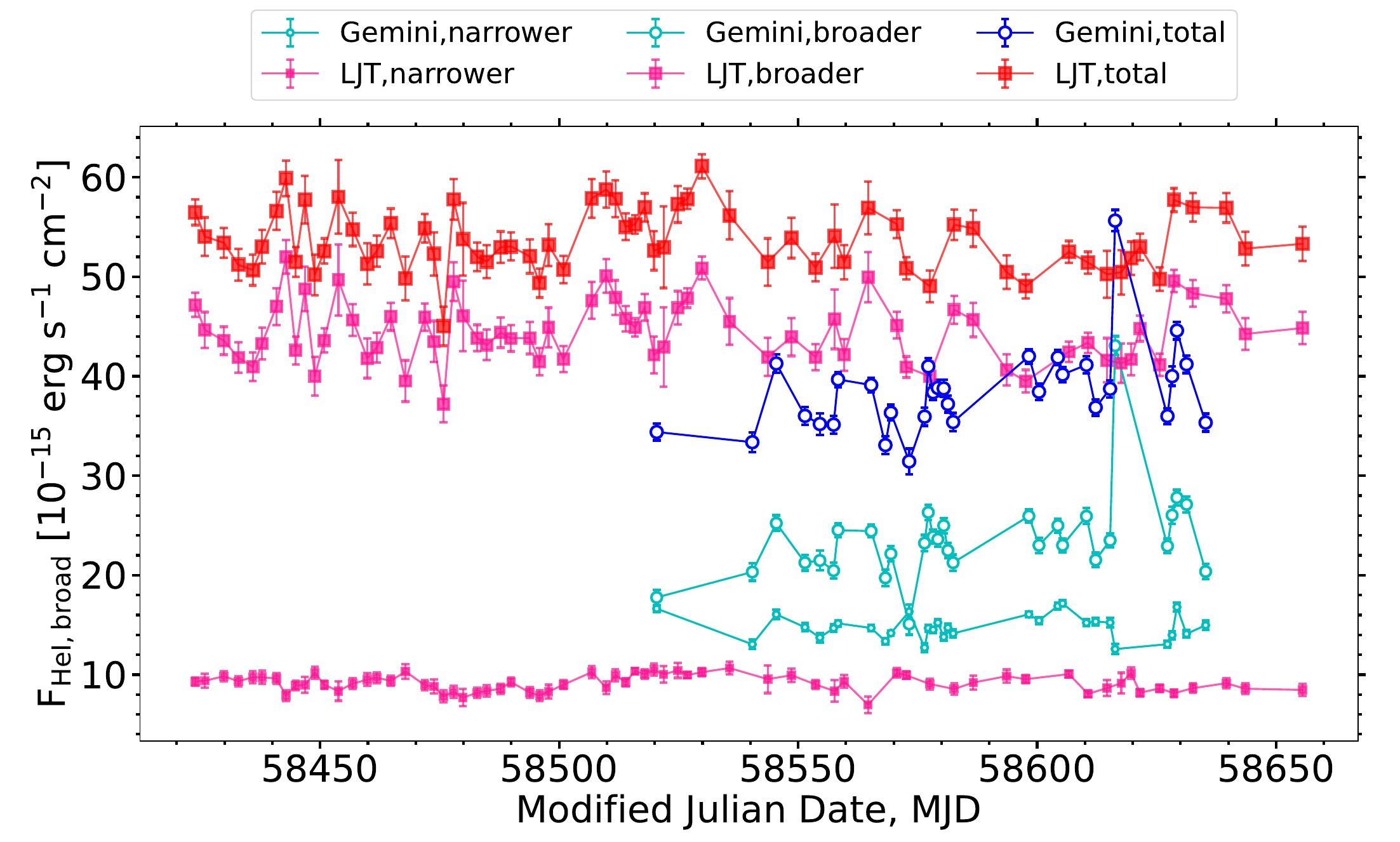}
\caption{\HeI\ light curves highlighting the contributions from the individual broad components (referred to as broader and narrower) to the total broad components from Gemini (open circles) and LJT (solid squares) data.  Smaller (larger) size of the markers represents narrower (broader) components.  It is interesting to note that the broader \HeI\ light curve from the LJT spectra is brighter than the broader \HeI\ component from the Gemini spectra likely due to the blue wing of \HeI\ (traced primarily by the broader component) in the LJT data affected by instrumental broadening.
\label{fig:lightcurves_hei_b}}
\end{figure}

Although the LJT RMS spectrum of Mrk~142 shows a weak variable feature for \HeI, the variability is not usefully quantifiable given the timescale and S/N of our current Gemini+LJT spectroscopic campaigns.  Figure~\ref{fig:lightcurves_hei_b} displays the individual and total broad-line components of \HeI\ emission from both the Gemini and the LJT observations.  The offset observed in the narrower broad component is similar to that observed in the broad \Hbeta, where the light curve from Gemini appears at higher flux values than the LJT light curve.  Interestingly, the broader broad component is brighter in LJT than in Gemini likely resulting from the blueshifted disk-wind component broadened due to the wider slit used for LJT data.

\section{Results and Discussion} \label{sec:results_discussion}

We present the first results on the lag of the broad \Hbeta\ line with respect to the UV continuum in Mrk~142 from optical spectroscopic observations from Gemini+LJT with simultaneous monitoring in the photometric {\em Swift}/{\em UVW2} and LCO+Zowada+Liverpool/{\em g} bands.

We applied a spectral model with same number of components but different parameter settings to the Gemini and LJT spectra to derive \Hbeta\ and \HeI\ light curves.  We noted that a profile composed of one narrow~+~two broad Gaussians sufficiently traced both the \Hbeta\ and the \HeI\ lines, where the widths and positions of the narrow components were tied to those of the \OIIIlamstrong\ line.  The use of Gaussians for fitting broad lines such as \Hbeta\ in our data does not comply with the \citet{wang_etal_2014_1} line-profile predictions for a super-Eddington AGN, where self-shadowing effects due to a slim-disk structure in super-Eddington AGN could result in the broad-line profiles appearing more Lorentzian than Gaussian.  We recognize that our Mrk~142 spectral fitting model does not align the \citet{wang_etal_2014_1}; however, our results are not sufficient evidence to rule out the self-shadowing hypothesis.  A detailed analysis of spectral line profiles for a larger sample of super-Eddington AGN is required to build a robust understanding of how the accretion-disk behavior affects the observed broad-line profiles in these objects.  We employed fixed narrow-line flux ratios $F_{\mathrm{\Hbeta}}/F_{\mathrm{\OIIIlamstrong}}$ and $F_{\mathrm{\HeI}}/F_{\mathrm{\OIIIlamstrong}}$ for the LJT spectral model (determined from flexible flux ratios for the Gemini spectra) due to the instrumental broadening in the LJT spectra affecting emission-line measurements.  Although the LJT RMS spectrum shows a weak feature at the \HeI\ location, {\tt\string PrepSpec} modeling and light-curve analysis suggested that there is no adequate variability in the line that is measurable with the current Gemini+LJT data.  However, we acknowledge that the \HeI\ emission line shows a peculiar profile evident from the high S/N of the Gemini spectra.  The broader Gaussian used for the line indicates stronger blueshifted emission than the redshifted side of the line.  Also, the \HeII\ line specifically required a broad, blueshifted component to accurately trace the emission in that region.  Followed by spectral analysis, we empirically measured the FWHM values as well as calculated the narrow- and broad-line flux values of the \Hbeta\ and \HeI\ lines to obtain their light curves.

Applying {\tt\string PyROA}, we performed cross-correlation analysis with continuum ({\em Swift}/{\em UVW2}, LCO+Zowada+Liverpool/{\em g}, and Gemini+LJT/5100~\AA) and broad \Hbeta\ light curves (LJT and Gemini+LJT inter-calibrated) with a goal of determining reverberation time lag for the Gemini+LJT inter-calibrated \Hbeta\ light curve.  {\tt\string PyROA} provided an improvement in quantifying the uncertainties compared to previous studies.  Most early RM studies have extensively applied the {\tt\string ICCF} method for time-lag measurements which makes it a good comparison standard.  However, {\tt\string ICCF} struggles with non-uniformly sampled data, which is true for our Gemini+LJT campaigns similar to most other studies, and uses linear interpolation to estimate the light-curve behavior in the regions with data gaps.  Consequently, the uncertainties reported for {\tt\string ICCF}-based measurements are typically conservative compared to {\tt\string JAVELIN} and {\tt\string PyROA}, as noted in this work for the 5100~\AA\ continuum as well as the \Hbeta\ emission from LJT and Gemini+LJT inter-calibrated data (see Table~\ref{tab:time_lag_measurements}).  In the context of the uncertainties on time-lag measurements, {\tt\string JAVELIN} has been shown to perform better.  For instance, \citet{edelson_etal_2019} reported uncertainties from {\tt\string ICCF} to be twice as large as those from {\tt\string JAVELIN} which is also evident from the results in this work (see Table~\ref{tab:time_lag_measurements}).  {\tt\string JAVELIN} uses damped random walk (DRW) to estimate the light-curve pattern in the regions where data are not available.  DRW closely characterizes the variability observed in AGN, plausibly leading to smaller uncertainties in the final lag measurements.  However, {\tt\string JAVELIN} requires a good estimation of uncertainties in data.  For sub-optimally calibrated uncertainties, {\tt\string JAVELIN} can sometimes fail to deliver reliable lag measurements \citep{donnan_etal_2021}.  This is likely the reason the LJT/\Hbeta\ and Gemini+LJT/\Hbeta\ emission-line light curves, which have larger calibrated uncertainties than the continuum light curves, show time-lag measurements differing from those reported by {\tt\string PyROA}.  {\tt\string PyROA} offers an improvement over {\tt\string JAVELIN} -- the ROA along with a robust error treatment not only prevents the outlier points from disrupting the estimation of the driving light curve but also applies a valid algorithm for resolving data gaps.

We measured a time lag of \(8.68_{-0.72}^{+0.75}\) days for the Gemini+LJT inter-calibrated \Hbeta\ emission with reference to the {\em UVW2} continuum.  We also obtained a lag of \(0.79_{-0.29}^{+0.27}\) days for the 5100~\AA\ continuum with reference to the {\em UVW2} band \citep[that is consistent, within uncertainties, with the time lag--wavelength relationship in][]{cackett_etal_2020}, and a lag of \(7.89\pm0.80\) days for the \Hbeta\ emission with respect to the 5100~\AA\ continuum.  From here, we report a black hole mass of \massbhlog~=~\(6.28\pm0.29\) derived using Equation~\ref{eqn:mass_bh},

\begin{equation}
    M_{\bullet} = \frac{f~c~\tau_{\text{\Hbeta}}~V_{\text{FWHM,\Hbeta}}^2}{G}
\label{eqn:mass_bh}
\end{equation}

\noindent where we used a \(\log(f)\) value of \(-0.36_{-0.54}^{+0.33}\) from the dynamical modeling of the Mrk~142 BLR by \citet{li_etal_2018}.  It is possible that the value of \(\log(f)\) has a large uncertainty, which depends on the BLR's dynamical model or the calibration using the $M-\sigma_*$ relation for classical bulges and pseudobulges \citep[e.g.][]{ho_kim_2014, li_etal_2018, yu_etal_2020}.  For $V_\mathrm{FWHM}$, we used the mean \Hbeta\ FWHM of \(1680\pm14\)~\kms\ from the Gemini spectra.  Here, we chose the Gemini spectra due to their higher resolution providing a more reliable measurement of the narrower \Hbeta\ broad-line profile than the LJT spectra.  We also considered the mean FWHM from the spectra as against the \Hbeta\ RMS profiles as the RMS spectra from Gemini were noisier blueward of the blue wing of the \Hbeta\ line.  We further measured mean luminosities of \(\log(L_{\text{UVW2}})=43.832\pm0.001\), \(\log(L_{5100})=43.643\pm0.002\), and \(\log(L_{\mathrm{\Hbeta}})=41.621\pm0.002\).  For the $L_{5100}$ and $L_{\mathrm{\Hbeta}}$ measurements, we adopted the mean flux value from our Gemini+LJT inter-calibrated light curves as their respective Gemini light curves alone were insufficient to provide a reliable flux scale due to the shorter observing timescale.

Our results agree with previously published measurements of Mrk~142 \citep{du_etal_2015, li_etal_2018}.  From the previous six-month SEAMBH campaign, \citet{du_etal_2015} reported a time lag of \(7.9_{-1.1}^{+1.2}\) days for \Hbeta\ with reference to 5100~\AA.  We measured an optical lag of \(7.89\pm0.80\) days for \Hbeta\ in agreement with the \citet{du_etal_2015} value within uncertainties.  Furthermore, the derived black hole mass for Mrk~142 in this work, \massbhlog~=~\(6.28\pm0.29\), agrees with the value reported in the recent velocity-resolved RM analysis by \citet{li_etal_2018}, \(6.23_{-0.45}^{+0.26}\), within uncertainty limits.  Table~\ref{tab:measurements_comparison} summarizes the measured quantities in this work and shows their comparison with the values from previous studies.

\begin{deluxetable*}{lcc}
\tablewidth{700pt}
\tablenum{11}
\tablecaption{Comparison of Measurements to Previous SEAMBH Studies
\label{tab:measurements_comparison}}
\tabletypesize{\scriptsize}
\tablehead{
 \\
\colhead{Measured Quantity} & \colhead{This Work} & \colhead{Value From Previous SEAMBH Studies} \\
} 
\startdata
$\tau_{\mathrm{UVW2-to-\Hbeta}}^*$ & \(8.68_{-0.72}^{+0.75}\) days & \nodata \\
$\tau_{\mathrm{5100~\AA-to-\Hbeta}}$ & \(7.89\pm0.80\) days & \(7.9_{-1.1}^{+1.2}\) days \citep{du_etal_2015} \\
\massbhlog & \(6.28\pm0.29\) & \(6.23_{-0.45}^{+0.26}\) \citep{li_etal_2018} \\
\(\log(L_{\mathrm{UVW2}})\)$^{*}$ & \(43.832\pm0.001\) & \nodata \\
\(\log(L_{\mathrm{5100}})\) & \(43.643\pm0.002\) & \(43.56\pm0.06\) \citep{du_etal_2015} \\
\(\log(L_{\mathrm{\Hbeta}})\) & \(41.621\pm0.002\) & \(43.56\pm0.06\) \citep{du_etal_2015} \\
\enddata
\tablenotetext{*}{New measurements}
\end{deluxetable*}
 
To visualize our results in the broader context of reverberation-mapped AGN, we placed the measured size of the \Hbeta\ line-emitting region on various {\em R}--{\em L} scaling relations.  Comparisons are discussed below.

\begin{figure}[ht!]
\epsscale{1.2}
\plotone{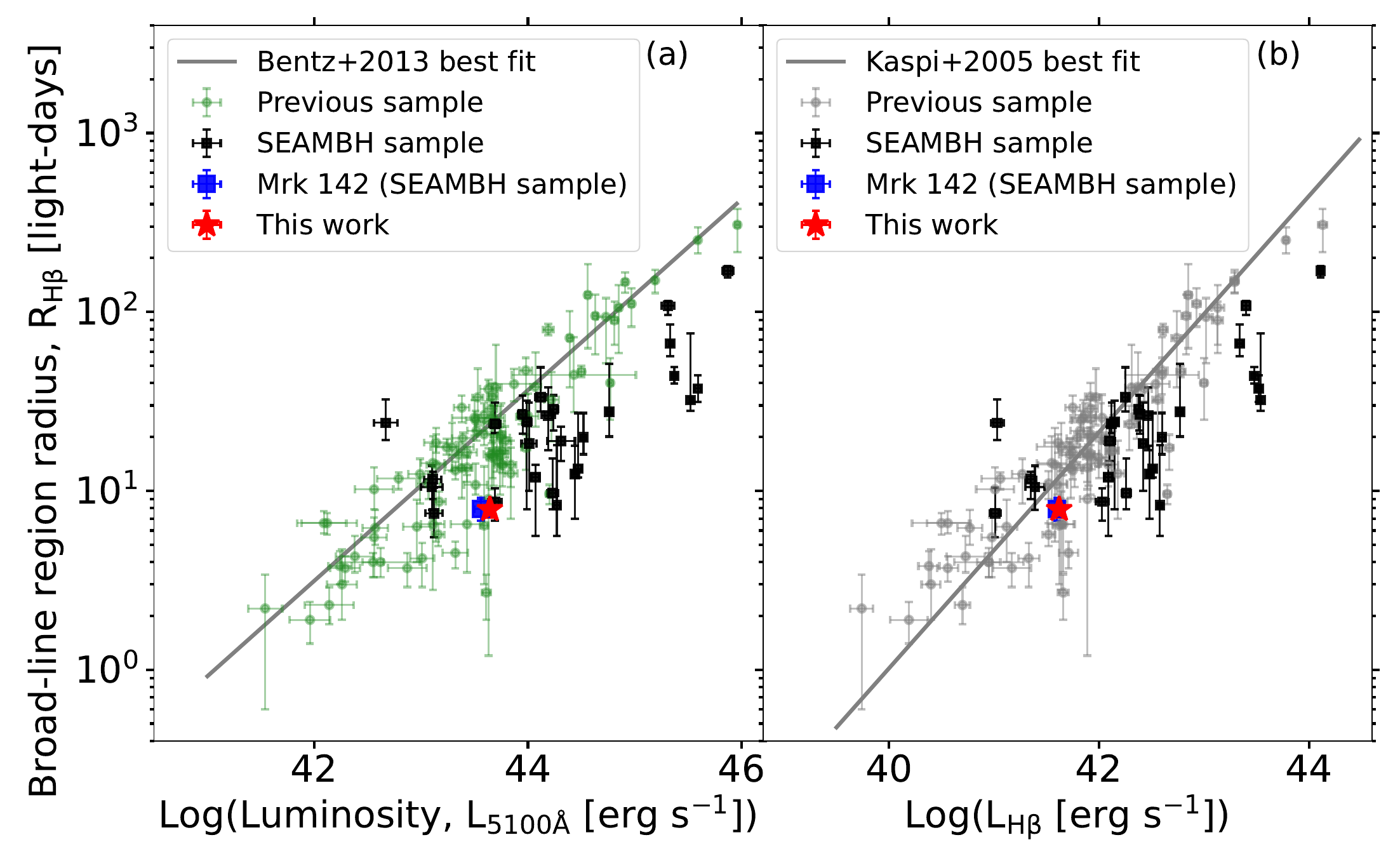}
\caption{Radius-luminosity ({\em R}--{\em L}) scaling relations for \Hbeta\ in the optical including results from this work (red star).  Panel {\em a}: \rlhbetaoptcont\ relation showing Mrk~142 from this work close to the \citet{du_etal_2015} measurement (blue square).  Panel {\em b}: \rlhbetahbeta\ relation showing Mrk~142 from this work overlapping with the measured value in \citet{du_etal_2015}.  The SEAMBH objects \citep[black squares;][]{du_etal_2014, wang_etal_2014_2, hu_etal_2015, du_etal_2015, li_etal_2018, zhang_etal_2019, li_etal_2021} appear on the lower right of the grey solid lines, which represent the best-fit {\em R}--{\em L} relations from \citet{bentz_etal_2013} with $L_{5100}$ (panel {\em a}) and \citet{kaspi_etal_2005} with $L_{\Hbeta}$ (panel {\em b}), indicating a smaller size for the broad-line region in highly accreting AGN compared to the more typical, sub-Eddington AGN mapped in previous studies \citep[green circles, panel {\em a} and grey circles, panel {\em b};][and references therein]{stirpe_etal_1994, santos-lleo_etal_1997, collier_etal_1998, dietrich_etal_1998, peterson_etal_1998, kaspi_etal_2000, santos-lleo_etal_2001, kaspi_etal_2005, bentz_etal_2006, collin_etal_2006, denney_etal_2006, bentz_etal_2007, bentz_etal_2009a, bentz_etal_2009b, denney_etal_2010, dietrich_etal_2012, grier_etal_2012, barth_etal_2013, bentz_etal_2013, bentz_etal_2014, pei_etal_2014, peterson_etal_2014, peterson_etal_2002} at same luminosities.  In particular, the departure of Mrk~142 from the \citet{bentz_etal_2013} best-fit relation in panel {\em a} is more apparent than the deviation from the \citet{kaspi_etal_2005} relation in panel {\em b} (see text for additional details).
\label{fig:rl_comparison_optical}}
\end{figure}

Figure~\ref{fig:rl_comparison_optical} shows the \rlhbetaoptcont\ (panel {\em a}) and \rlhbetahbeta\ (panel {\em b}) relations with the red star representing the Mrk~142 measurements from this work.  In agreement with the findings from previous SEAMBH campaigns (black circles in Figure~\ref{fig:rl_comparison_optical}; see references in the figure caption), the red Mrk~142 star appears to depart from the general trend observed for typical RM objects, especially in the \rlhbetaoptcont\ relation (green circles, panel {\em a}; see the entire reference list in the figure caption).  Comparatively, the departure of Mrk~142 from the \citet{kaspi_etal_2005} best-fit relation (panel {\em b}) is less obvious.  We quantified these Mrk~142 departures in terms of time lag (along the vertical axis) relative to the standard deviation of the departures of the typical RM sample from both best-fit relations ($\sigma_\mathrm{departure}$).  While the departure of Mrk~142 from the \citet{bentz_etal_2013} relation (15.7~light-days) is 1.15$\sigma_\mathrm{departure}$ (where $\sigma_\mathrm{departure} = 13.7$~light-days), its departure from the \citet{kaspi_etal_2005} relation (4.08~light-days) is $<<$1$\sigma_\mathrm{departure}$ (where $\sigma_\mathrm{departure} = 26.8$~light-days).  The latter implies that \rlhbetahbeta\ is a tighter relationship for AGN than \rlhbetaoptcont.  The position of Mrk~142 in the \rlhbetaoptcont\ relationship (panel {\em a}r), considerably below the best-fit line, reiterates the characteristic of super-Eddington AGN exhibiting smaller BLR sizes in contrast to the sub-Eddington population at the same luminosities \citep{du_etal_2016_2}.  \citet{du_etal_2015} tested this deviation of high accretion-rate AGN from the \rlhbetaoptcont\ relationship.  Studying the differences in the BLR sizes for AGN with low ($\dot{M}/\dot{M}_{\mathrm{Edd}}<3$) and high ($\dot{M}/\dot{M}_{\mathrm{Edd}}\geq 3$) mass-accretion rates, \citet{du_etal_2015} inferred that $\dot{M}$ influences the size scales observed in super-Eddington AGN while such a correlation is absent in the low mass-accretion rate objects.

\begin{figure}[ht!]
\epsscale{1.2}
\plotone{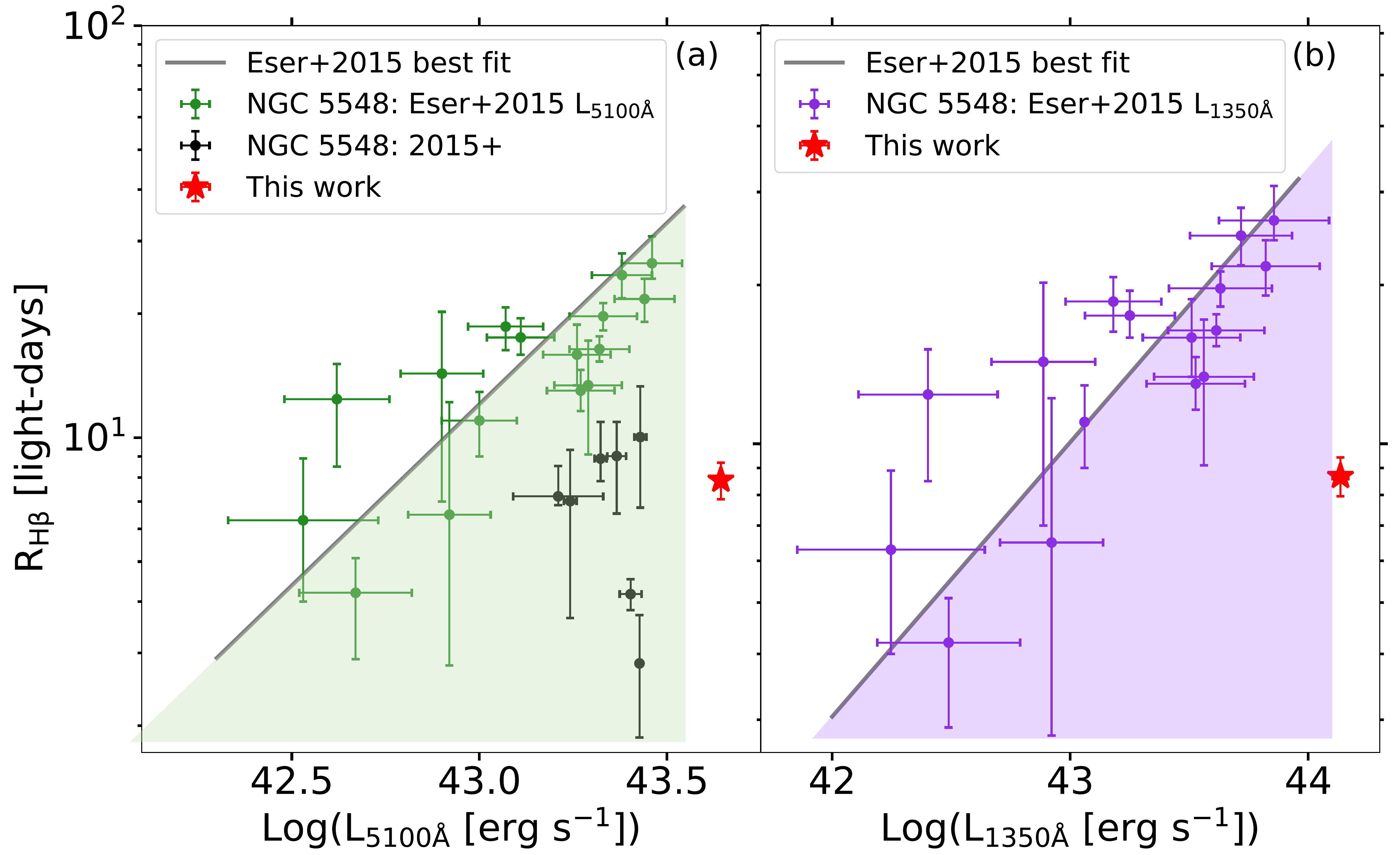}
\caption{Multi-epoch NGC~5548 radius-luminosity ({\em R}--{\em L}) scaling relations for \Hbeta\ in the optical and UV including luminosities from \citet{eser_etal_2015} and results from this work (red star).  Panel {\em a}: NGC~5548 \rlhbetaoptcont\ relation from \citet[][green circles]{eser_etal_2015} on which Mrk~142 from this work is located to the right at higher luminosities.  This follows the similar deviation from the entire RM sample that is observed in Figure~\ref{fig:rl_comparison_optical}, panel {\em a}.  RM measurements for NGC~5548 post 2015 \citep[][black circles]{lu_etal_2016, pei_etal_2017, derosa_etal_2018, lu_etal_2022} are also included for completeness.  Panel {\em b}: \rlhbetauvcont\ relation for NGC~5548, including Mrk~142 from this work.  From its position to the right of the \citet{eser_etal_2015} best-fit {\em R}--{\em L} relation (grey solid line) with $L_{5100}$ in panel {\em a}, Mrk~142 has moved closer to the \citet{eser_etal_2015} best-fit {\em R}--{\em L} relation with $L_{1350}$ in panel {\em b}, suggesting the UV as a better proxy for the driving continuum than the optical.  In general, the offset of the SEAMBHs may reflect a different} spectral energy distribution.
\label{fig:rl_comparison_1350}
\end{figure}

\begin{figure}[ht!]
\epsscale{1.2}
\plotone{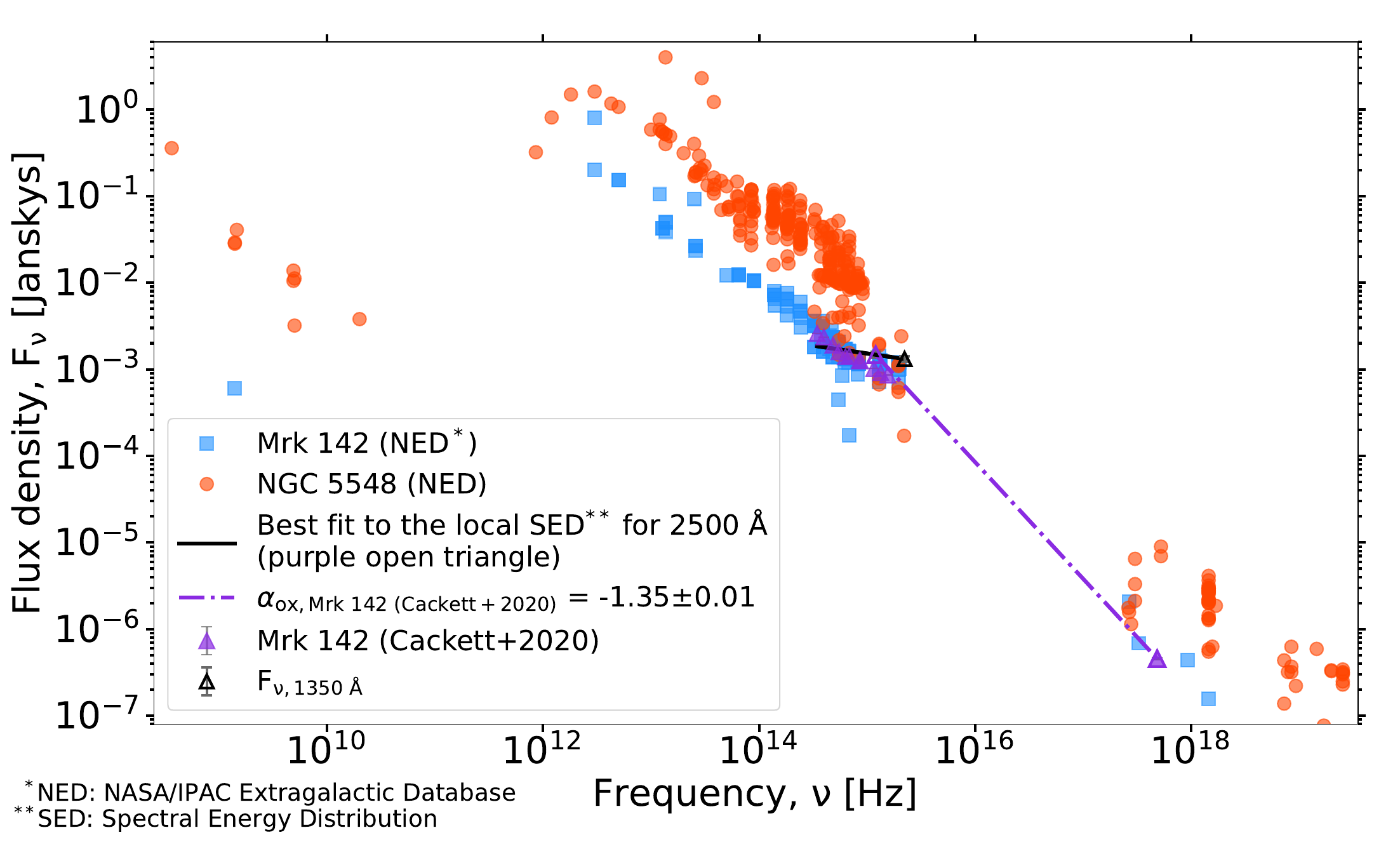}
\caption{Spectral energy distributions (SEDs) of Mrk~142 from the National Aeronautics and Space Administration/Infrared Processing and Analysis Center (NASA/IPAC) Extragalactic Database (NED, blue squares), and from \citet[][mean flux density in each band represented with purple triangles]{cackett_etal_2020}; and NGC~5548 from NED (orange circles) showing similar SED shapes for the two objects.  We derived the optical-to-X-ray slope ($\alpha_\mathrm{ox}$) for the Mrk~142 data from \citet{cackett_etal_2020} regarding a power-law behavior (fit indicated in black solid line) for the SED around the 2500~\AA\ data point (purple open triangle).  Results indicate a softer ionizing SED for Mrk~142 (\(\alpha_\mathrm{ox}=-1.35\)) than NGC~5548 \citep[\(\alpha_\mathrm{ox}=-1.30\);][]{merritt2022}; see text for a detailed discussion.  The fit to the \citet{cackett_etal_2020} SED is extended to indicate the estimated 1350~\AA\ data point (black open triangle) used in Figure~\ref{fig:rl_comparison_1350}.
\label{fig:sed_comparison}}
\end{figure}

Figure~\ref{fig:rl_comparison_1350} shows the \rlhbetaoptcont\ and \rlhbetauvcont\ scaling relations for NGC~5548 over time with optical lag measurements for \Hbeta, and luminosities at 5100~\AA\ and 1350~\AA\ ($L_{1350}$) from \citet{eser_etal_2015}.  Again, the red star in both panels represents the Mrk~142 point from this work.  \citet{eser_etal_2015} formulated a conversion from $L_{5100}$ to $L_{1350}$ for NGC~5548 (see their Equation~4) from all RM campaigns of the object from 1988 to 2008 \citep{peterson_etal_2002, bentz_etal_2007, bentz_etal_2009b, denney_etal_2010}.  We applied that conversion to calculate $L_{1350}$ for NGC~5548 and generated the \rlhbetauvcont\ plot (Figure~\ref{fig:rl_comparison_1350}, panel {\em b}).  Here, we extrapolated the \citet{cackett_etal_2020} UV/optical imaging data -- mean flux densities -- in different bands assuming power-law behavior to estimate a luminosity of \(\log(L_{1350})=44.13\pm0.03\) for Mrk~142 in this work (see Figure~\ref{fig:sed_comparison}).  The shift in the position of Mrk~142 from panel {\em a} to {\em b}, closer to the {\em R}--{\em L} scaling relation in the UV, indicates that the UV emission is a better proxy for the ionizing continuum than the 5100~\AA\ optical emission.

To further understand the comparison between the \rlhbetauvcont\ relations of Mrk~142 (a super-Eddington Seyfert~1 galaxy) and NGC~5548 (a normal Seyfert~1 galaxy), we compared their SEDs from the National Aeronautics and Space Administration/Infrared Processing and Analysis Center (NASA/IPAC) Extragalactic Database (NED) and recent studies, including \citet{cackett_etal_2020}.  Figure~\ref{fig:sed_comparison} shows the SEDs for the above two AGN from NED along with the derived optical-to-X-ray spectral slope (\(\alpha_\mathrm{ox}\)) of $-1.35\pm0.01$ for Mrk~142 from the \citet{cackett_etal_2020} observations (purple triangles representing mean flux densities in different bands).  To derive the $\alpha_\mathrm{ox}$ value, we: (1) fit the local SED around the 2500~\AA\ data point assuming power-law behavior, (2) calculated the dereddened X-ray flux density at 2~keV, and (3) used Equation~1 for $\alpha_\mathrm{ox}$ from \citet{just_etal_2007}.  For NGC~5548, \citet{merritt2022} obtained an $\alpha_\mathrm{ox}$ of \(-1.30\pm0.04\) from simultaneous X-ray/UV/optical observations (see Chapter~2 therein).  A less negative $\alpha_\mathrm{ox}$ for NGC~5548 compared to Mrk~142 suggests a harder ionizing SED for the normal Seyfert~1 galaxy, likely due to the soft X-ray excess observed in the object \citep[e.g.,][]{mehdipour_etal_2015}, than the super-Eddington AGN.  However, the soft excess in NGC~5548 was not evident in 2013, when an obscurer heavily absorbed its soft X-ray flux \citep{mehdipour_etal_2015}.  Most Seyferts/quasars have some sort of soft X-ray excess; the debate for years has been over the origin (ionized disk reflection, warm Comptonized emission, a mixture or both).  The implication from \citet{tortosa_etal_2023} is that for super-Eddington AGN, ionized disk reflection can model the soft excess in this class of AGN well.  It is interesting to witness that despite their distinct types, the SED shapes of Mrk~142 and NGC~5547 are similar as their $\alpha_\mathrm{ox}$ values agree within uncertainties -- a plausible explanation for the tighter \rlhbetauvcont\ correlation than \rlhbetaoptcont\ noted in both the objects.  Future accretion-disk modeling efforts can help understanding such comparisons of {\em R}--{\em L} relationships between normal and super-Eddington AGN.

If UV emission is closer to the driving continuum as seen in Figure~\ref{fig:rl_comparison_1350}, this will affect the black hole mass of Mrk~142 derived from the 5100~\AA\ to \Hbeta\ time lag.  Recently, \citet{cackett_etal_2020} performed the accretion-disk RM analysis for Mrk~142 with data from {\em Swift}, LCO, Zowada, Liverpool, and other ground-based observatories, simultaneous to the Gemini+LJT data taken as a part of the same broader RM campaign.  \citet{cackett_etal_2020} pointed that if the {\em UVW2} band represents the driving continuum, then the black hole mass derived from the \Hbeta\ optical lag is underestimated by $\sim$10\%.  We used the {\em UVW2} to \Hbeta\ time lag result from this work to calculate the black hole mass in Mrk~142.  We obtained a mass of \massbhlog~=~\(6.32\pm0.29\) based on the {\em UVW2} to \Hbeta\ time lag.  This value is $\sim$10\% greater than the black hole mass derived by assuming 5100~\AA\ as the driving continuum band, and thus verifies the discrepancy estimated by \citet{cackett_etal_2020}.  This discrepancy would be as high as $\sim$40\% if X-rays, instead of UV, were the driving continuum \citep{cackett_etal_2020}.  However, our work does not propose any new implications to the accretion-disk structure of Mrk~142.  A robust \Hbeta\ lag measurement with reference to the X-ray continuum from future studies will help understand how X-rays play a role in driving the continuum variability in Mrk~142 and shape its inner accretion disk.

\section{Conclusion} \label{sec:conclusion}

We performed BLR RM analysis of Mrk~142 with medium- and low-resolution optical spectra from Gemini and LJT, simultaneous to the {\em Swift} and LCO+Zowada+Liverpool photometric campaigns reported by \citet{cackett_etal_2020} to measure the UV lag for \Hbeta\ emission line.  With {\tt\string PrepSpec} analysis, we corrected calibration discrepancies for both the Gemini and the LJT spectra individually.  From spectral modeling in {\tt\string Sherpa}, we measured FWHM and fluxes for  \OIIIlam; \Hbetalam; and \HeIlam\ emission lines.  To combine the 5100~\AA\ and \Hbeta\ light curves from Gemini and LJT, we inter-calibrated the respective light curves from the two telescopes in {\tt\string PyROA}.  Applying {\tt\string PyROA} for time-lag analysis, we measured a UV time lag for \Hbeta\ and further derived refined black hole masses.  Placing our results on various {\em R}--{\em L} scaling relations, we verified that our results are consistent with previously published values for Mrk~142.  We summarize our main findings below.

\begin{enumerate}
    \item {\tt\string PyROA}, using Bayesian Information Criterion to evaluate model performance along with a rigorous treatment of uncertainties, provided a robust method for measuring cross-correlation time lags.  This project is one of the early works employing {\tt\string PyROA} technique for measuring RM time lags with real data.  In this process, the longer timescale of LJT spectra nicely complemented the gaps in the Gemini observations.
    \item We measured, for the first time, a UV time lag of \(8.68_{-0.72}^{+0.75}\) days for \Hbeta\ in Mrk~142, with simultaneous photometry in the {\em Swift}/{\em UVW2} band and optical spectroscopy with Gemini and LJT.  Assuming the UV continuum as the primary driver of the observed variability, we derived a black hole mass of \massbhlog~=~\(6.32\pm0.29\).
    \item We obtained a 5100~\AA\ to \Hbeta\ time lag of \(7.89\pm0.80\) days, consistent with the measured value from previous SEAMBH campaigns \citep{du_etal_2015}.  From this lag measurement, we also derived a black hole mass for Mrk~142 of \massbhlog~=~\(6.28\pm0.29\), in agreement with the mass reported by \citet{li_etal_2018}.
    \item We placed the 5100~\AA\ to \Hbeta\ time lag with measured $L_{1350}$ on the \rlhbetauvcont\ relation for NGC~5548 \citep{eser_etal_2015}.  Mrk~142 falls closer to the \rlhbetauvcont\ scaling relation than the \rlhbetaoptcont\ relation indicating that the UV is closer to the ``true'' driving continuum as opposed to 5100~\AA\ band.
\end{enumerate}

In addition, we also recorded supplementary results.  Our spectral analysis indicated blueshifted, broad components for the \HeI\ and \HeII\ emission lines suggestive of wind components in these higher-ionization lines.  To infer the cause of such disk+wind components, we need more higher-resolution data and BLR modeling efforts for super-Eddington AGN.  We intend to study the \HeI\ and \HeII\ lines in further detail in our future work.  Furthermore, BLR RM analysis with the concurrent X-ray data available from {\em Swift} can better inform our understanding of the measured \Hbeta\ time lags with respect to the UV continuum.  We aim to explore this in our future study.

\section{acknowledgments} \label{sec:acknowledgements}
We recognize the support of Rick Edelson in taking the {\em g}-band data from Las Cumbres Observatory.  V.C.K. acknowledges the support of Joel Roediger and the Gemini Observatory staff during the planning of observations and reduction of the Gemini data.  V.C.K. thanks T. A. Boroson, who provided the \FeII\ template to C.H. for the spectral modeling process.  We acknowledge the support of the Natural Sciences and Engineering Research Council of Canada (NSERC), Discovery Grant RGPIN/04157.  V.C.K. acknowledges the support of the Ontario Graduate Scholarships.  E.M.C. gratefully acknowledges support for analysis of the {\em Swift} data from NASA through grant 80NSSC19K0150, and support for analysis of the Zowada Observatory data from the NSF through grant AST-1909199.  C.H. acknowledges support from the National Science Foundation of China (12122305).  The research of V.C.K. was partially supported by the New Technologies for Canadian Observatories, an NSERC CREATE program.  V.C.K. also acknowledges Jonathan Trump, Martin Houde, Stanimir Metchev, and Charles McKenzie for their valuable feedback and discussions.

This research was based on observations obtained at the Gemini Observatory (processed using the Gemini IRAF package), which is managed by the Association of Universities for Research in Astronomy (AURA) under a cooperative agreement with the National Science Foundation on behalf of the Gemini Observatory partnership: the National Science Foundation (United States), National Research Council (Canada), Agencia Nacional de Investigaci\'{o}n y Desarrollo (Chile), Ministerio de Ciencia, Tecnolog\'{i}a e Innovaci\'{o}n (Argentina), Minist\'{e}rio da Ci\^{e}ncia, Tecnologia, Inova\c{c}\~{o}es e Comunica\c{c}\~{o}es (Brazil), and Korea Astronomy and Space Science Institute (Republic of Korea).  This work was enabled by observations made from the Gemini North telescope, located within the Maunakea Science Reserve and adjacent to the summit of Maunakea.  We are grateful for the privilege of observing the Universe from a place that is unique in both its astronomical quality and its cultural significance.

This research used observations from the Lijiang 2.4-meter Telescope funded by the Chinese Academy of Sciences (CAS) and the People’s Government of Yunnan Province.  We acknowledge the support of the National Key R\&D Program of China No. 2021YFA1600404, and the National Science Foundation of China (11833008, 11973029, 11991051, 11991054).

The Liverpool Telescope is operated on the island of La Palma by Liverpool John Moores University in the Spanish Observatorio del Roque de los Muchachos of the Instituto de Astrofisica de Canarias with financial support from the UK Science and Technology Facilities Council.

\vspace{5mm}

\facilities{Gemini North Telescope (GMOS),
            Lijiang 2.4-meter Telescope (YFOSC),
            {\em The Neil Gehrels Swift Observatory},
            Las Cumbres Observatory,
            Dan Zowada Memorial Observatory,
            Liverpool Telescope \citep{steele_etal_2004}}

\software{{\tt\string IRAF},
          Gemini {\tt\string IRAF}, 
          {\tt\string Python}, 
          {\tt\string Astropy} \citep{astropy_collaboration_etal_2013}, 
          {\tt\string PrepSpec}, 
          {\tt\string Sherpa} \citep{freeman_etal_2001, doug_burke_2018_1245678}, 
          {\tt\string PyROA} \citep{donnan_etal_2021},
          {\tt\string ICCF \citep{gaskell_sparke_1986, gaskell_peterson_1987},
          {\tt\string JAVELIN} \citep{zu_etal_2011, zu_etal_2013}}
          }

\clearpage

\appendix

\section{Gemini Spectral Reduction -- Special Cases} \label{app:A}
This section describes the the special cases from the spectral reduction of the Mrk~142 Gemini Spectra that were either treated differently or discarded due to calibration issues.

\begin{itemize}
    \item Epoch 11 narrow-slit standard star spectra: The extracted spectrum from exposure 1 appeared to drop in flux and flatten shorter than (blueward of) $\sim$4740~\AA, whereas the exposure 2 spectrum was flat on both ends.  While it was possible to recover the flat region of exposure 1 spectrum spectrum, the one from exposure 2 was not suitable for further analysis and hence was discarded.  Because the shape of the spectrum blueward of $\sim$4740~\AA\ was not evident from the exposure 2 spectrum, we recovered the region from $\sim$4520~\AA\ to $\sim$4740~\AA\ with reference to the mean wide-slit standard star spectrum (used as the reference to correct for slit losses).  Consequently, only the region longer than (redward of) $\sim$4740~\AA\ was corrected for slit losses with the spline fitting procedure.  The former and the latter were then concatenated to obtain {\it only one} slitloss-corrected spectrum for epoch 11 from exposure 1.  Note that the standard star spectrum from exposure 1 was later used to calibrate the narrow-slit science spectrum from exposure 2.
    
    \item Epoch 11, exposure 1 wide-slit standard star spectrum: Two bumps with bad data were recovered redward of $\sim$5900~\AA\ with reference to the spectrum from exposure 2.  However, the recoverey is not reliable as the two spectra had slightly different count levels.
    
    \item Discarded wide-slit standard star spectra: The wide-slit standard star spectra from exposures 1, 1, and 2 of epochs 21, 22, and 25, respectively, showed bump-like features, which could not be recovered as the standard star spectra from the other exposure of the same epochs had different count levels.  Therefore, the epochs/exposures listed here were discarded.
    
    \item Correction of affected pixels or recovery of bump-like regions on or close to the emission lines of interest: A few science spectra showed bumpy features on or close to the emission lines of interest, \HeIlam\ and \Hbetalam.  We attempted to recover the ``true'' shape of such regions with reference to the other exposure taken with the same slit on the same night.  In most cases, the two exposures had similar count levels.  However, spectral measurements from the recovered spectra with different count levels are less reliable and must be considered carefully.  Following is the list of the specific cases.
    \vspace{0.2cm}
    \newline{Epoch 13, exposure 2 narrow-slit spectrum: A large, downward bump with bad data from $\sim$5545~\AA\ till the red wing of the \HeI\ emission line was recovered with reference to the exposure 1 spectrum.  Spectra from both exposures had similar count levels.}
    \vspace{0.2cm}
    \newline{Epoch 16 narrow-slit spectra: Spectra from both exposures showed partially overlapping bump-like features in the region blueward of the \Hbeta\ emission line.  Because neither of the exposures could be used to recover the shape of the spectrum in the affected region, we recovered the shapes of both the spectra with reference to the mean mean wide-slit science spectrum (used as the reference to correct for slit losses).  In the exposure 2 spectrum, the recovery extended till the blue wing of \Hbeta.  Spectra from both exposures had similar count levels.}
    \vspace{0.2cm}
    \newline{Epoch 21, exposure 2 narrow-slit spectrum: The bumpy region from $\sim$4325~\AA\ to $\sim$4740~\AA\ was recovered with reference to exposure 1 with similar count levels.  The recovered region extended till the tail end of the blue wing of \Hbeta.}
    \vspace{0.2cm}
    \newline{Epoch 27, exposure 1 narrow-slit spectrum: A small bumpy region from till the blue wing of \Hbeta\ was recovered with reference to exposure 2 with similar count levels.}
    \vspace{0.2cm}
    \newline{Epoch 28, exposure 1 narrow-slit spectrum: A 17-pixel wide region residual from sky subtraction on the blue wing of \Hbeta\ was replaced by simulated data values after linear interpolation in that region.  However, this correction was affecting \Hbeta\ line measurements and hence was excluded from the analysis.}
    \vspace{0.2cm}
    \newline{Epoch 30 narrow-slit spectra: Spectra from both exposures showed overlapping bump-like features in the region from $\sim$4370~\AA\ to $\sim$4685~\AA\ that were recovered with reference to the mean wide-slit science spectrum.  In the exposure 1 spectrum, another bumpy feature from the red wing of the \OIII\ emission line at $\sim$5008~\AA\ was recovered with reference to the exposure 2 spectrum with similar count levels.}
    \vspace{0.2cm}
    \newline{Epoch 3, exposure 1 wide-slit spectrum: A bumpy feature on the blue wing of the \HeI\ line was recovered with reference to the exposure 2 spectrum, where both spectra had similar count levels.  This is one of the brighter wide-slit spectra.  However, it was not used to generate the mean wide-slit spectrum for slitloss correction.}
    \vspace{0.2cm}
    \newline{Epoch 11, exposure 1 wide-slit spectrum: A spike, possibly residual of sky subtraction, on the blue side of the \OIII\ peak at $\sim$5008~\AA\ was replaced by simulated data after linear interpolation in the affected region.}
    \vspace{0.2cm}
    \newline{Epoch 14, exposure 2 wide-slit spectrum: A huge bump-like feature with bad data was replaced by simulated data after linear interpolation till the blue wing of \Hbeta.}
    \vspace{0.2cm}
    \newline{Epoch 24, exposure 2 wide-slit spectrum: A huge bump with bad data extended from the region blueward of the \HeI\ line till the red end of the line.  This affected region was recovered with reference to the exposure 1 spectrum, which was at slightly lower count levels than the exposure 2 spectrum.}
    \vspace{0.2cm}
    \newline{Epoch 32, exposure 2 wide-slit spectrum: A bump-like feature from blueward region of the \HeI\ line till the blue side of the \HeI\ peak was recovered with reference to the exposure 1 spectrum.  Spectra from both exposures had similar count levels.}
    
    \item Epoch 25, exposure 2 narrow-slit science spectrum: The region on the blue side of \HeIlam\ emission line shows a large bumpy feature with bad data.  Because the region redward of $\sim$5320~\AA\ has lower count levels overall as compared to the spectrum from exposure 1, the affected region in spectrum 2 was not recovered.  Therefore, this exposure was discarded.
    
    \item Epoch 22, exposure 2 wide-slit science spectrum: The red end of the spectrum from $\sim$6320~\AA\ to $\sim$6400~\AA\ was recovered with reference to the spectrum from exposure 1.  However, the exposure 2 spectrum had slightly higher count levels than exposure 1.
\end{itemize}

\bibliography{manuscript}{}
\bibliographystyle{aasjournal}

\allauthors

\end{document}